\documentclass[12pt, reqno]{amsart}

\makeatletter
\g@addto@macro{\endabstract}{\@setabstract}
\makeatother

\makeatletter
\def\@seccntformat#1{%
	\ifstrequal{#1}{subsection}
	{\bfseries\csname the#1\endcsname.}
	{\csname the#1\endcsname.} 
}
\makeatother

\usepackage{graphics, stackrel}
\usepackage{amsmath, amssymb, amsthm}
\usepackage{graphicx}
\usepackage{verbatim}
\usepackage{amsfonts}
\usepackage{filecontents}
\usepackage{natbib}
\usepackage{fancyvrb}
\usepackage{color}
\usepackage[inline]{enumitem}

\usepackage[citecolor=blue, colorlinks=true, linkcolor=blue]{hyperref}

\usepackage{mathrsfs}
\usepackage{bbm}

\usepackage[left=1.25in, right=1.25in, top=1.0in, bottom=1.15in, includehead, includefoot]{geometry}

\usepackage{multirow}

\DeclareMathOperator{\Var}{Var}
\DeclareMathOperator{\Cov}{Cov}

\newcommand{\tow}{\stackrel { w } {\to} }

\newcommand{\1}{\mathbf 1}

\newcommand*\diff{\mathop{}\!\mathrm{d}}
\renewcommand{\epsilon}{\varepsilon}
\renewcommand{\phi}{\varphi}
\renewcommand{\mid}{\,|\,}

\newcommand{\bB}{\mathscr B}

\newcommand{\fF}{\mathscr F}

\newcommand{\XX}{\mathsf X}

\newcommand{\ZZ}{\mathsf Z}

\newcommand{\CC}{\mathsf C}

\newcommand{\TT}{\mathsf \Theta}
\newcommand{\KK}{\mathsf K}
\renewcommand{\AA}{\mathsf A}
\newcommand{\G}{\mathsf \Gamma}

\newcommand{\pP}{\mathscr P}

\newcommand{\vV}{\mathcal V}

\newcommand{\RR}{\mathbbm R}

\newcommand{\NN}{\mathbbm N}
\newcommand{\PP}{\mathbbm P}

\newcommand{\EE}{\mathbbm E}

\newcommand{\uz}{\underline z}
\newcommand{\uw}{\underline w}
\newcommand{\ua}{\underline a}
\newcommand{\uR}{\underline R}

\newcommand{\tz}{\tilde z}

\newcommand{\rR}{\mathcal R}

\theoremstyle{plain}

\newtheorem{theorem}{Theorem}[section]
\newtheorem{corollary}{Corollary}[section]
\newtheorem{lemma}{Lemma}[section]
\newtheorem{proposition}{Proposition}[section]

\theoremstyle{definition}
\newtheorem{definition}{Definition}[section]

\newtheorem{remark}{Remark}[section]

\newtheorem{assumption}{Assumption}[section]

\renewcommand{\underline}[1]{\text{\b{$#1$}}} 
  
\usepackage{caption}

\usepackage[flushleft]{threeparttable}

\newcommand{\me}{\mathrm{e}}

\usepackage{ dsfont }

\newcommand{\vertiii}[1]{{\left\vert\kern-0.25ex\left\vert\kern-0.25ex\left\vert #1 
    \right\vert\kern-0.25ex\right\vert\kern-0.25ex\right\vert}}
    
\usepackage{subcaption}
\usepackage{graphicx}  

\makeatletter
\newcommand*{\rom}[1]{\expandafter\@slowromancap\romannumeral #1@}
\makeatother

\usepackage{stackengine}

\begin{document}

\title{}
\date{\today}

\begin{center}
    \LARGE 
    Risk Sensitivity and Stationary Equilibrium \\ in Incomplete Markets\footnote{Qingyin Ma gratefully acknowledges the financial support from NSFC (No.72573111). \\
    \emph{Email addresses:} 
    \texttt{qingyin.ma@cueb.edu.cn},\, 
    \texttt{j-stachurski@grips.ac.jp},\, 
    \texttt{wubin@cueb.edu.cn}} 
    
    \vspace{1em}
    \normalsize
    Qingyin Ma\textsuperscript{a} 
    \qquad 
    John Stachurski\textsuperscript{b} 
    \qquad 
    Bin Wu\textsuperscript{c} \par 
    \vspace{1em}
    \textsuperscript{a,\,c}
    ISEM, Capital University of Economics and Business\\ 
    \smallskip 
    \textsuperscript{b}
    National Graduate Institute for Policy Studies
    
    \medskip 
    \normalsize{\today}
\end{center}

\begin{abstract} 
	We study stationary equilibrium and comparative statics in an incomplete market economy with risk-sensitive preferences. We provide conditions under which a stationary equilibrium exists for every nonnegative degree of risk sensitivity. Under independent income, greater risk sensitivity strictly raises aggregate saving at fixed prices unless both economies have zero saving, and strictly lowers the smallest and the largest equilibrium interest rates. Under stochastically monotone Markov income, these comparisons hold whenever one of the two levels of risk sensitivity being compared is sufficiently low or sufficiently high. For pairs of intermediate values, however, greater risk sensitivity can strictly lower saving, contrary to an existing result in the literature, and can strictly raise the smallest and largest equilibrium rates. An exact decomposition identifies the forces driving these reversals.
	
    \vspace{1em}
    
\end{abstract}


\section{Introduction}\label{s:intro}

Concerns over model misspecification feature prominently in modern macroeconomics. Robust control allows households, investors, and policymakers to evaluate decisions against adverse departures from a benchmark probability model. This approach has deeply informed research on consumption and saving, asset pricing, business cycles, welfare, and policy design \citep{hansen1995discounted,hansen2008robustness,hansen1999robust,tallarini2000risk,anderson2005dynamics,bidder2012robust,bidder2018doubts,ellison2015welfare,bianchi2018uncertainty,kasa2018risk}. Recent research connects pessimistic beliefs to survey evidence and documents how these beliefs shape aggregate fluctuations \citep{bhandari2025survey}. In incomplete markets, such concerns operate through households' demand for self-insurance, the distribution of wealth, and the equilibrium prices at which households save. 

We study these interactions within the risk-sensitive specification that underlies much of this literature \citep{hansen1995discounted,tallarini2000risk,anderson2005dynamics,luo2010risk,bauerle2018stochastic,bommier2019risk}, under which households evaluate uncertain future utility via the recursion
\begin{equation*}
	V_t = u(c_t)-\frac{\,\beta\,}{\gamma}
	\log\EE_t\,\me^{-\gamma V_{t+1}},
\end{equation*}
where $\gamma\ge 0$ is the risk-sensitivity parameter, and $\gamma=0$ recovers expected utility.\footnote{Here, $u$ denotes
	the period utility function, $c_t$ consumption, $V_t$ the utility from period-$t$ onward, $\beta$ the discount factor, and $\EE_t$ the expectation conditioned on period-$t$ information.}
For $\gamma>0$, this recursion is equivalent to the multiplier robust preferences of \citet{hansen2008robustness}: the household evaluates future outcomes under the worst-case income law within an entropy neighborhood of the physical law, with $\gamma$ equal to the inverse of the robustness penalty. A larger $\gamma$ lowers this penalty and thereby raises aversion to dispersion in continuation utility, while leaving preferences over deterministic consumption streams unchanged.

Our setting is an incomplete market production economy with borrowing-constrained households, idiosyncratic income risk, and a risk-free asset. Our central concern is how risk sensitivity affects household and aggregate saving, the stationary distribution of wealth, and equilibrium interest rates. This analysis calls for a stationary equilibrium theory for risk-sensitive preferences, which has not been established. Since the tilted probabilities in the Euler equation depend on continuation values, standard existence arguments for expected-utility economies do not apply directly. We provide conditions under which a stationary recursive competitive equilibrium exists for every $\gamma \ge 0$ and the equilibrium interest rate correspondence is upper hemicontinuous in $\gamma$. These conditions accommodate strictly positive, stochastically monotone Markov income processes, including independent income as a special case.

Answering the saving question also requires resolving an existing claim in the literature. In their study of risk-sensitive household saving, \citet[][Proposition 1]{bommier2019risk} state that saving rises with risk sensitivity whenever income is stochastically monotone. We show that this proposition is false as stated. A counterexample satisfies all their maintained assumptions, with time-homogeneous, monotone income, yet saving falls strictly between two positive risk-sensitivity parameters.\footnote{Section~\ref{ss:3_period} isolates the mechanism in a three-period economy. Online Appendix~\ref{oa:counterexamples} provides an infinite-horizon counterexample for the stated household comparison, alongside a general equilibrium economy in which policy, aggregate, and equilibrium orderings all reverse strictly.}
This issue extends beyond household choices: stationary aggregate saving can fall at fixed prices, and the extremal equilibrium interest rates can rise. These findings make it necessary to identify the conditions under which monotone comparisons hold and to characterize how these household effects translate into stationary equilibrium, while establishing equilibrium without assuming global saving monotonicity.

Our first contribution characterizes the effects of risk sensitivity on saving, stationary wealth, and equilibrium interest rates. Under independent income, a higher $\gamma$ weakly raises saving and lowers consumption at all wealth levels, holding prices fixed. Both comparisons are strict wherever saving under the lower parameter is positive. The stationary wealth distribution shifts upward in first-order stochastic dominance, and aggregate saving and wealth rise strictly unless both economies have zero aggregate saving. Since equilibrium capital is positive, this exception does not arise at market clearing: both the smallest and largest equilibrium interest rates fall strictly.

Under stochastically monotone Markov income, the same comparisons hold whenever the lower risk-sensitivity parameter is sufficiently small or the upper one sufficiently large. The policy comparisons apply on a common invariant wealth
domain, and the stationary wealth distribution shifts upward conditional on each productivity state. Thresholds uniform over equilibrium prices ensure that both extremal equilibrium interest rates fall strictly under the same
two-region comparison. In particular, every positive $\gamma$ gives strictly lower extremal rates than the expected-utility benchmark. This does not imply monotonicity between arbitrary positive parameters. We show that the ordering of individual saving, stationary aggregates, and equilibrium rates can all reverse strictly between intermediate values. A numerical illustration further shows that saving reversals can recur across a wide range of $\gamma$.

The mechanism works through valuation of future states and the marginal value of wealth in those states. Raising $\gamma$ alters decision weights directly, while also shifting continuation values and future consumption choices. Under Markov income, these changes differ across productivity states and can overturn the standard precautionary saving comparison. We derive an \emph{exact decomposition} of the change in the marginal return to saving into three components: the direct weight change holding continuation values fixed, the additional reweighting caused by changes in continuation values, and the change in future marginal values. At a state with positive baseline saving, their sum is negative if and only if saving strictly falls. Our decomposition thus provides an exact criterion for individual saving reversals, while our threshold results deliver primitive sufficient conditions for household and equilibrium comparisons.

Our second contribution is a stationary general equilibrium theory for this economy. We prove that a stationary recursive competitive equilibrium exists for every $\gamma \ge 0$. Moreover, the equilibrium interest rate correspondence is nonempty, compact-valued, and upper hemicontinuous in $\gamma$. Thus, whenever risk-sensitivity parameters converge, every limit of corresponding equilibrium rates constitutes an equilibrium rate of the limiting economy. In particular, as $\gamma$ tends to zero, equilibrium rates approach the expected-utility equilibrium set. If equilibrium is unique at the limiting parameter, all nearby equilibrium rates converge to that rate. 

The equilibrium construction and comparative statics reinforce each other tightly. A fixed-price comparison with expected utility, established independently of equilibrium, transfers the divergence of capital supply as gross returns approach $1/\beta$ to every $\gamma\ge0$. Together with continuity and the lower-bound argument, this establishes existence and places all equilibrium rates in a common compact interval. That interval then makes the Markov comparison thresholds uniform across equilibrium prices and maps saving comparisons into orderings of extremal market clearing rates.


Two further results support this construction. First, we prove existence, uniqueness, and parametric continuity of the value function and optimal policy, strict concavity of the value function in wealth, and a unified Euler equation. Our theory covers bounded and unbounded period utilities. Second, under impatience, vanishing absolute risk aversion, and positive income transitions, we establish a unique, globally attracting stationary distribution at fixed prices. Access to a binding state ensures convergence. Uniform bounds on stationary wealth then allow policy continuity to pass to aggregate saving and wealth. These results enable us to unify the endogenous decision weights, stationary wealth distributions, and market clearing within a coherent analysis.

\subsection*{Related literature} 

Our analysis connects to a broad literature on robustness and risk sensitivity in macroeconomics. \cite{tallarini2000risk} and \cite{bidder2012robust} study their implications for business cycles. In environments with idiosyncratic risk, \cite{ellison2015welfare} examine how misspecification concerns affect the welfare cost of exogenous consumption fluctuations, while \cite{kasa2018risk} study robust portfolio choices and wealth inequality in a continuous-time Blanchard--Yaari economy. By contrast, we study the interaction of risk-sensitive consumption and saving decisions, stationary wealth distributions, and endogenously determined factor prices in an incomplete market production economy. 

Our household analysis contributes to the literature on risk-sensitive dynamic programming. \cite{bauerle2018stochastic} study stochastic optimal growth, while \cite{luo2010risk} study risk-sensitive consumption and saving in a linear-quadratic setting with rational inattention. We study a borrowing-constrained income fluctuation problem and link its solution to stationary distributions and competitive equilibrium. Section~\ref{s:motivation} also illustrates why the observational equivalence between robustness and patience in the linear-quadratic setting of \cite{hansen1999robust} does not survive in a nonlinear consumption-saving problem.

On the equilibrium side, a large literature studies incomplete market economies under expected utility \citep{miao2006competitive,kuhn2013recursive,accikgoz2018existence,cao2020recursive,zhu2020existence,light2023general,prohl2025existence}. We retain this benchmark at $\gamma=0$ and use a comparison against it as one ingredient in establishing existence through the entire risk-sensitive family. 
In continuous time, \citet{achdou2026continuous} study recursive equilibrium via mean field game methods. Their Epstein--Zin setting with late resolution of uncertainty excludes the risk-sensitive preferences studied here. We establish upper hemicontinuity in risk sensitivity and strict comparative statics for extremal equilibrium rates.

Finally, our results complement work on precautionary saving and stochastic comparative statics in Markov decision problems \citep{light2018precautionary,light2021stochastic}, and on aggregate and distributional comparative statics in large economies \citep{acemoglu2015robust,jensen2018distributional}. Whereas these studies adopt expected-utility frameworks, we study the equilibrium response to changes in risk sensitivity, establish strict comparisons of extremal equilibrium interest rates without requiring equilibrium uniqueness, and identify when household, aggregate, and equilibrium comparisons can reverse.

\subsection*{Organization}
The paper proceeds as follows. Section~\ref{s:motivation} 
builds intuition via three stylized finite-horizon economies. 
Section~\ref{s:ifp_rsp} solves the risk-sensitive income fluctuation problem
and establishes stochastic stability. Section~\ref{s:GE} proves the existence
and upper hemicontinuity of equilibrium. Section~\ref{s:comp_stats}
derives the comparative statics and the exact decomposition for saving 
comparisons. Proofs and counterexamples are collected in the appendices 
and online appendices.

\section{Three Motivating Examples}\label{s:motivation}

Before turning to the general economy, we illustrate the underlying mechanism 
with three stylized finite-horizon economies. The first shows how risk 
sensitivity generates precautionary saving and lowers equilibrium interest 
rates. The second shows that this monotonicity can fail once income is 
persistent. The third shows that risk sensitivity is not observationally 
equivalent to patience outside the linear-quadratic setting. 

\subsection{A Two-Period Economy: Precautionary Saving and Equilibrium} 
\label{ss:2_period}
Consider a household that lives for two periods: $t=0,1$. In period $0$, 
it has initial wealth $a_0>0$, consumes $c_0$, and saves $s=a_0-c_0$ at 
gross return $R$. In period $1$, it receives labor income $wz$ and consumes 
$c_1(z)=Rs+wz$. Here $w>0$, and $z\in\{z_L,z_H\}$ with $0<z_L<z_H$, 
probabilities $\pi_L,\pi_H>0$, and $\EE(z)=1$. Period utility $u$ 
is increasing, strictly concave, and continuously differentiable on 
$(0, \infty)$ with $u'(0)=\infty$. 

The household evaluates uncertain future utility via the entropic 
certainty equivalent
\begin{equation}\label{eq:entropic}
	\rho_\gamma(X) := 
	-\dfrac1\gamma\log\EE\,\me^{-\gamma X} 
	\quad (\gamma>0) \qquad \text{and}\qquad 
	\rho_0(X) := \EE X
\end{equation}
For a nonconstant $X$, Jensen's inequality gives $\rho_\gamma(X)<\EE X$ 
when $\gamma>0$, hence an uncertain utility is valued below its expected 
utility. As $\gamma\downarrow0$, the certainty equivalent converges to 
the expectation; as $\gamma\to\infty$, it converges to the worst 
realization. Thus $\gamma$ measures sensitivity to adverse outcomes. 
The associated robustness interpretation is developed in 
Section~\ref{ss:preference}. 
Given a discount factor $\beta\in(0,1)$, the household solves
\begin{align*}
	& \text{maximize}  
	&& u(c_0)+\beta\rho_\gamma\bigl[u(c_1(z))\bigr] \\
	&\text{subject to}   
	&& c_0+s=a_0,\quad c_1(z)=Rs+wz,\quad s\ge0. \notag
\end{align*}
The objective is strictly concave, so its unique optimum satisfies
\begin{equation*}
	u'(c_0)= \max\left\{
	\beta R \EE \bigl[u'(c_1(z)) f_s(z)\bigr], u'(a_0)
	\right\},
	\quad \text{where} \quad
	f_s(z):=
	\frac{\me^{-\gamma u(c_1(z))}}{\,\EE\,\me^{-\gamma u(c_1(z))}\,}.
\end{equation*}
Since $\EE f_s = 1$, the probabilities used in the marginal saving 
calculation are $\pi_j f_s(z_j)$. They put more weight on low consumption, 
where an additional unit of wealth has greater marginal utility. 
At $\gamma = 0$, $f_s = 1$ and the first-order condition becomes the usual 
expected-utility Euler condition.

\subsubsection{Precautionary saving}
At any fixed $s$ and $\gamma>0$, both $u'(c_1(z))$ and 
$\me^{-\gamma u(c_1(z))}$ are strictly decreasing in $z$. Hence, 
for all $\gamma>0$,
\begin{equation*}
	\EE \bigl[u'(c_1(z)) f_s(z)\bigr]
	=\frac{\EE \bigl[u'(c_1(z))\me^{-\gamma u(c_1(z))}\bigr]}
	{\EE \me^{-\gamma u(c_1(z))}}
	>\EE u'(c_1(z)).
\end{equation*}
More generally, raising $\gamma$ increases this tilted marginal utility 
at every saving level $s$. Strict concavity therefore implies that optimal 
saving is increasing in $\gamma$, and increases strictly when the household 
with the lower $\gamma$ saves a positive amount. Both households may choose 
zero saving when current resources are sufficiently low. This mechanism 
requires no change in the income dynamics: risk sensitivity raises the 
decision weight on states in which saving is most valuable.

\subsubsection{Equilibrium}
Take a unit mass of households, each endowed 
with $a_0$ units of the period-$0$ good. Their saving supplies capital for 
period-$1$ production. A competitive firm operates the Cobb-Douglas 
technology $F(K,L)=K^\alpha L^{1-\alpha}$ with full depreciation, where $K$ 
is capital, $L$ is labor, and $\alpha\in(0,1)$. With labor normalized to unity,
profit maximization given $R$ yields the capital 
demand and wage schedules
\begin{equation}\label{eq:ex_firm}
	K(R)= (\alpha/ R )^{1/(1-\alpha)}
	\qquad \text{and} \qquad 
	w(R)=(1-\alpha)K(R)^\alpha. 
\end{equation}
Because the productivity shock arrives after saving, all households are
identical in period $0$ and choose the same saving $s$. With log
utility, an interior choice satisfies
\begin{equation}
	\frac{1}{a_0-s}
	= \frac{\beta R\,\EE (Rs+w(R)z)^{-\gamma-1}}
	{\EE (Rs+w(R)z)^{-\gamma}}. \label{eq:ee-log}
\end{equation}
A competitive equilibrium clears the capital market: $s=K(R)$.
Substituting this condition into \eqref{eq:ee-log} and using \eqref{eq:ex_firm}, the equilibrium capital $K$ satisfies
\begin{equation*}
	\frac{1}{a_0-K}
	=\frac{\alpha\beta C_\gamma}{K},
	\qquad \text{where} \quad
	C_\gamma:=\frac{\EE (\alpha+(1-\alpha)z)^{-\gamma-1}}
	{\EE (\alpha+(1-\alpha)z)^{-\gamma}}.
\end{equation*}
There is a unique solution
\begin{equation*}
	K^*=\frac{\alpha\beta C_\gamma}{1+\alpha\beta C_\gamma}\,a_0
	\qquad \text{and} \qquad 
	R^*=\alpha(K^*)^{\alpha-1}.
\end{equation*}
The quantity $C_\gamma$ is a tilted mean of $(\alpha+(1-\alpha)z)^{-1}$. 
It is strictly increasing in $\gamma$ because larger $\gamma$ places more 
weight on low $z$ states. Therefore, the equilibrium capital $K^*$ increases 
and the equilibrium interest rate $R^*$ decreases with risk sensitivity.

\subsection{A Three-Period Economy: When Saving Falls with Risk Sensitivity}
\label{ss:3_period}
In the two-period example, terminal utility is independent of $\gamma$, 
and raising risk sensitivity only reweights the fixed future 
outcomes. With another period, continuation values themselves depend on 
$\gamma$. If current income predicts future income, then this dependence
differs across states and can reverse the saving response.

\begin{figure}[h!]
	\centering
	\includegraphics[width=\textwidth]{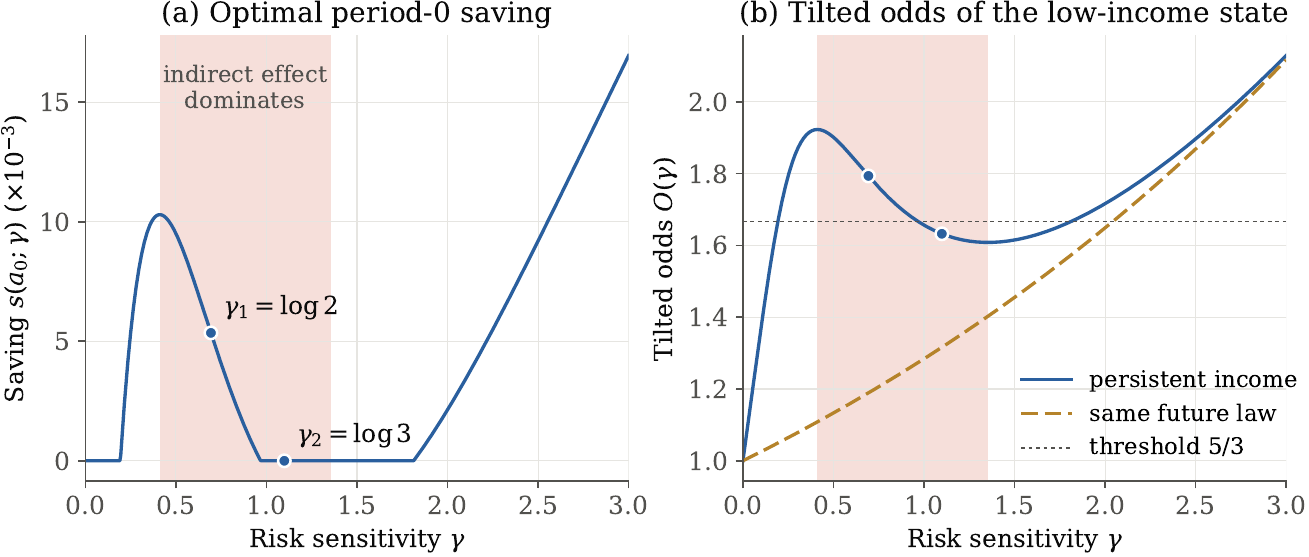}
	\caption{The three-period economy at initial cash $a_0=(160/111)^2$}
	\label{fig:three_period}
\end{figure}

Consider periods $t=0,1,2$ with $u(c)=\sqrt c$, $\beta=3/4$, and $R=w=1$. 
The household saves in periods $0,1$ and consumes all remaining wealth 
in period $2$. Period-1 income is $z_L=1$ or $z_H=25/16$, each with 
probability $1/2$. Conditional period-2 income is 
\begin{equation*}
\begin{array}{c|cc}
	& \Pr(z_2=1) & \Pr(z_2=\hat z)\\
	\hline
	z_1=z_L & 1/4 & 3/4,\ \hat z=9\\
	z_1=z_H & 1/4 & 3/4,\ \hat z=81
\end{array}
\end{equation*}
Work backward. At zero period-0 saving, period-1 cash satisfies 
$a_1 =z_1\in \{z_L,z_H\}$. Since terminal income is at least $1$, 
the marginal benefit of period-1 saving is at most $\beta Ru'(1)=3/8$. 
The marginal cost of saving at zero is $u'(z_L)=1/2$ and $u'(z_H)=2/5$; 
both exceed $3/8$, hence, for every $\gamma$, zero saving is optimal 
in period 1 at both nodes. The corresponding continuation values are 
\begin{equation*}
	v_L(\gamma)=1-\frac{3}{4\gamma}\log\Bigl[
	    \frac14\me^{-\gamma} + \frac34 \me^{-3\gamma}
	\Bigr]
	\quad \text{and} \quad 
	v_H(\gamma)=\frac54-\frac{3}{4\gamma}\log\Bigl[
		\frac14 \me^{-\gamma} + \frac34 \me^{-9\gamma}
	\Bigr],
\end{equation*}
which are computed by substituting the period-2 rewards
into the certainty equivalent and adding the period-1 rewards. 
Thus, at zero period-0 saving, the ratio of the tilted probability of 
the low state to that of the high state is
\begin{equation*}
	O(\gamma):=\me^{\gamma[v_H(\gamma)-v_L(\gamma)]} 
	\quad \Longrightarrow \quad 
	O(\gamma)^4=\me^{\gamma}
	\left[\frac{1+3\me^{-2\gamma}}{1+3\me^{-8\gamma}}\right]^3.
\end{equation*}
At $\gamma_1=\log2$ and $\gamma_2=\log3$,
direct computation gives $O(\gamma_1)>5/3>O(\gamma_2)$: 
greater risk sensitivity reduces the relative weight on the low-income state,
as illustrated in panel (b) of Figure \ref{fig:three_period}.
To see why, write $\Delta v(\gamma) = v_H(\gamma)-v_L(\gamma)$. Then 
\begin{equation*}
	\log\frac{O(\gamma_2)}{O(\gamma_1)}
	= \underbrace{(\gamma_2 - \gamma_1)\Delta v(\gamma_1)}_{
		\text{direct change in weights}}
	+ \; \underbrace{
		\gamma_2\big[\Delta v(\gamma_2) - \Delta v(\gamma_1)\big]}_{
		\text{change in the value gap}}.
\end{equation*}
The first term is positive, as holding continuation values fixed, greater risk 
sensitivity favors the worse state. But the second term is negative, because 
the value gap $\Delta v(\gamma)$ shrinks as $\gamma$ rises, and its magnitude 
dominates the first term between $\gamma_1$ and $\gamma_2$. The high state's 
distant upside loses more value as $\gamma$ rises, narrowing the continuation 
value gap enough to overturn the direct effect.

The envelope identity converts this change in weights into a saving reversal.
At the two income nodes, the marginal values of wealth are $u'(z_L)=1/2$ and
$u'(z_H)=2/5$, both independent of $\gamma$ as zero saving is optimal there for all $\gamma$. 
Their tilted mean is $M(\gamma):= \left[O(\gamma)/2+2/5\right]\big/ \left[1+O(\gamma)\right],$
which is strictly increasing in $O(\gamma)$. Choose initial cash
$a_0=(160/111)^2$, hence $u'(a_0)=111/320$. The derivative of the period-0
objective $J_\gamma$ at zero saving is
$J'_\gamma(0+)=-u'(a_0)+\beta R M(\gamma)
=-111/320+ 3M(\gamma)/4.$
It is zero when $O(\gamma)=5/3$, positive at $\gamma_1$, and negative at 
$\gamma_2$. Strict concavity of the period-0 objective therefore gives 
\begin{equation*}
	s(a_0;\gamma_1)>0=s(a_0;\gamma_2).
\end{equation*}
Increasing risk sensitivity from $\gamma_1$ to $\gamma_2$ therefore eliminates
saving at this state. 
This reversal in saving arises entirely from the change in 
future continuation values, as illustrated in panel (a) of 
Figure~\ref{fig:three_period}. If $(z_L,z_H)$ instead shared the same future 
income law, their value gap $\Delta v(\gamma)$ would be constant and 
equal to $1/4$, so $O(\gamma) = \me^{\gamma/4}$ would be increasing in 
$\gamma$. Persistent income adds the indirect effect, which overturns 
this direct effect for intermediate values of $\gamma$. 

\subsection{Distinguishing Risk Sensitivity from Patience}
In the linear-quadratic permanent income economy of \cite{hansen1999robust}, 
higher robustness can be offset by a lower discount factor, leaving the
consumption--investment allocation unchanged; the two parameters are
observationally equivalent from quantity data. We show that
this equivalence need not extend beyond the linear-quadratic specification.

Return to the two-period problem with $u(c)=\log c$, $\gamma=1$,
$\beta=1/2$, $R=w=1$, and $z\in\{1,2\}$ with equal probabilities. The 
risk-sensitive household solves
\begin{equation*}
	\max_{0\le s\le a_0} \big\{
	    \log(a_0-s)-\beta\log \EE (s+z)^{-1}
	\big\}.
\end{equation*}
At an interior optimum, 
$1/(a_0-s) =\beta \EE(s+z)^{-2} / \EE (s+z)^{-1}$, while an expected-utility 
household choosing the same saving satisfies 
$1/(a_0-s) = \beta'\,\EE (s+z)^{-1}$. The matching discount factor 
is therefore $\beta'(s)=\beta \EE (s+z)^{-2}/[\EE (s+z)^{-1}]^2$,
which varies with $s$. The risk-sensitive optimum is
$s=1$ at $a_0=73/13$ and $s=2$ at $a_0=218/25$, with
compensating discount factors $13/25$ and $25/49$, respectively. 
Strict concavity of each objective delivers a unique optimum for its
Euler equation, so no single discount factor can replicate the 
risk-sensitive decision rule. Hence, the observational equivalence 
of \cite{hansen1999robust} does not hold in this nonlinear, 
borrowing-constrained setting. 

\subsection{Lessons and Roadmap}\label{ss:roadmap}
The examples suggest three lessons. First, when risk sensitivity 
acts only by reweighting future outcomes, as in the two-period economy, 
it raises precautionary saving and lowers the equilibrium interest rate. 
Second, once income is persistent, risk sensitivity also changes 
continuation values, and changes them unevenly across states; 
this indirect effect can reverse the saving comparison. Third, 
risk sensitivity is a distinct preference parameter, not a relabeling 
of patience.

The rest of the paper asks how far these lessons extend to an 
infinite-horizon economy with a stationary wealth distribution and 
endogenous prices. After Section~\ref{s:ifp_rsp} solves the household 
problem, Section~\ref{s:comp_stats} shows that the comparison of the 
two-period economy survives globally under independent income. 
Under persistent income, it survives whenever the lower risk-sensitivity 
parameter is sufficiently small or the upper one sufficiently large, 
but can fail in between. Because these comparisons concern equilibrium 
interest rates, they require a stationary equilibrium theory for 
risk-sensitive preferences. Section~\ref{s:GE} provides it, 
establishing existence for every $\gamma\ge0$ and upper hemicontinuity 
of equilibrium rates in $\gamma$.

\section{An Income Fluctuation Problem with Risk-Sensitive Preferences}
\label{s:ifp_rsp}

This section studies the income fluctuation problem with risk-sensitive 
preferences that serves as the household-side building block of the 
general equilibrium economy. 
We establish the existence, uniqueness, and properties of the 
optimal policy, the Euler equation, and convergence to a unique 
stationary distribution.

\subsection{Preferences and Economic Environment}\label{ss:preference}

Time is discrete and indexed by $t\in\{0,1,\dots\}$. Uncertainty is
driven by a labor productivity shock $\{z_t\}_{t\ge0}$ taking values
in a finite set $\ZZ=\{z_1, \dots, z_N\}\subset(0,\infty)$, 
where $N\ge2$. It is a time-homogeneous Markov chain with transition 
matrix $Q$, and $Q(z,\hat z)$ denotes the probability of moving from 
$z$ to $\hat z$ in one step. Throughout, denote $\RR_+:=[0,\infty)$
and $\RR_{++} := (0,\infty)$.

\subsubsection{Preferences}
The household evaluates consumption recursively: for a consumption plan 
$\{c_t\}_{t\ge0}$, the utility from period $t$ onward, $V_t$, satisfies
\begin{equation}\label{eq:rs_utility}
	V_t=u(c_t)+\beta\rho_t(V_{t+1}),
	\quad \text{where} \quad 
	\rho_t(X):= \begin{cases}
		-\dfrac{\,1\,}{\gamma}\log \EE_t \,\me^{-\gamma X}, 
		& \gamma>0, \\
		\EE_t X, & \gamma=0.
	\end{cases}
\end{equation}
where $u:\RR_+ \to \RR_+$ is the period utility function, which is 
continuous, increasing, strictly concave, twice continuously 
differentiable on $\RR_{++}$, and satisfies $u'(0)=\infty$.\footnote{The condition 
	$u:\RR_+\to\RR_+$ is a presentational convention, rather than
	a substantive restriction. Any period utility that is bounded below 
	on $\RR_+$ is covered: adding a constant makes it nonnegative and, 
	by cash invariance, leaves the optimal policy unchanged. Utilities 
	unbounded below near zero, such as $u(c)=\log c$ or 
	$u(c)=c^{\sigma}/\sigma$ with $\sigma<0$, are also admissible. 
	For these, the Bellman operator is analyzed directly on the
	space of bounded below value functions over the absorbing set
	$[\ua,\infty)\times\ZZ$, where $\ua>0$ is the lowest attainable 
	next-period wealth. 
    See Online Appendix~\ref{ss:opt_utility_ext} for the detailed 
    extension. \label{fn:utility}} 
Moreover, $\beta\in (0,1)$ is the discount factor, and $\EE_t$ 
conditions on period-$t$ information. 

The preference recursion \eqref{eq:rs_utility} is the one introduced in
Section~\ref{s:motivation}, now extended to the infinite horizon: as 
$\gamma\downarrow0$, it approaches the expected utility 
$V_t=\EE_t\sum_{s\ge0}\beta^s u(c_{t+s})$. 
Raising $\gamma$ increases aversion to dispersion in continuation value.

\begin{remark}[Relation to robustness]
	For $\gamma>0$, the certainty equivalent also has the \emph{robustness}
	representation:
	\begin{equation}\label{eq:robust_representation}
		\rho_t(X) = \min_{p \ll Q(z_t,\cdot)} \left\{
		\sum_j p_j X_j + 
		\frac{\,1\,}{\gamma}\sum_j p_j\log\frac{p_j}{Q(z_t,z_j)}
		\right\},
	\end{equation}
	where $X_j$ is the value of $X$ on 
	$\{z_{t+1}=z_j\}$, and the minimum is over probability vectors 
	$p \ll Q(z_t,\cdot)$ (i.e., $Q(z_t,z_j)=0$ implies $p_j=0$), with 
	$0\cdot\log0=0$. The first term is the expectation of $X$ under a 
	distorted income law. The second term penalizes departures from the 
	income law by relative entropy: a larger $\gamma$ lowers this penalty, 
	permitting a more adverse assessment of continuation prospects. 
	Therefore \eqref{eq:rs_utility} is the multiplier form of 
	the robust preferences of \citet{hansen2008robustness}: the household
	evaluates continuation utility under the worst-case income law within an
	entropy neighborhood of the physical one, with $\gamma$ controlling the
	size of that neighborhood. 
	As $\gamma\downarrow0$, the penalty on the departure grows without bound, 
	so the minimizing law converges to $Q$ itself and the certainty equivalent 
	collapses to the expectation.
\end{remark}

\begin{remark}[Relation to Epstein--Zin parameterizations]
	\label{r:mapping}
	When $u(c)=\log c$, the recursion \eqref{eq:rs_utility} is the
	Epstein--Zin recursion with unit elasticity of intertemporal
	substitution and relative risk aversion $1 + \gamma/(1-\beta)$
	\citep{tallarini2000risk}. Thus $\gamma$ is not itself a coefficient of
	relative risk aversion: with an annual discount factor $\beta = 0.96$,
	$\gamma = 1$ corresponds to relative risk aversion of $26$. The mapping
	also depends on the length of the model period through $\beta$. In
	quantitative work, $\gamma$ is therefore usually disciplined through
	detection-error probabilities
	\citep{anderson2003quartet,barillas2009doubts}.
\end{remark}

%

\subsubsection{The household's problem}
A household that enters period $t$ with wealth $a_t$ chooses consumption 
$c_t$ and carries the remainder $a_t-c_t$ into the next period, where it 
earns gross return $R>0$ and receives labor income $w z_{t+1}$ with $w>0$. 
Therefore, the next period's wealth is $a_{t+1}=R(a_t-c_t)+w z_{t+1}$. 
Each period, consumption must satisfy the no-borrowing constraint 
$0\le c_t\le a_t$: the household cannot consume more than its current
resources, so it can never borrow against future labor income. 

A policy $\pi=(\pi_t)_{t\ge0}$ assigns to each history
$h_t=(a_0,z_0,c_0,\dots,a_{t-1},z_{t-1},c_{t-1},a_t,z_t)$ a feasible
consumption choice $c_t\in[0,a_t]$; the set of all policies is
denoted by $\Pi$. A \emph{stationary policy} is a constant sequence
$\pi_t\equiv \varphi$ for all $t\ge0$, with $\Phi$ denoting the set of
all stationary policies. Given a policy $\pi$ and an initial state $(a,z)$, 
the lifetime utility $J(a,z,\pi)$ is defined as the limit of
finite-horizon truncations: for each horizon $n$, impose a zero
continuation value after period $n$, and compute the utility of the
truncated stream $(c_0,\dots,c_n)$ by backward recursion of
\eqref{eq:rs_utility}. This yields $J_n(a,z,\pi)$, the recursive utility 
of the first $n+1$ periods. The lifetime utility is then
\begin{equation*}
	J(a,z,\pi):=\lim_{n\to\infty}J_n(a,z,\pi).
\end{equation*}
Under Assumption~\ref{a:impatience} below, the sequence $\{J_n\}_{n\ge0}$
is increasing and bounded above, so the limit is well defined and
finite; the formal construction is in Appendix~\ref{s:pr_opt}. 
For unbounded below period utilities, the limit remains well defined
but may equal $-\infty$ for suboptimal policies 
(Appendix~\ref{ss:opt_utility_ext}).
Given $(a_0,z_0) = (a,z)$, the household solves 
\begin{align}
	& \sup_{\pi \in \Pi} \; J(a,z,\pi) \label{eq:object} \\
	\text{subject to:}  \qquad  
	& a_{t+1}=R(a_t-c_t)+w z_{t+1},
	\;\; 0\le c_t\le a_t, \;\; t \geq 0. \label{eq:budget}
\end{align}

\subsection{Optimality Results}\label{ss:or}

Throughout, $\EE_z:=\EE[ \, \cdot \mid z_0 = z]$ denotes conditional
expectations, and $\hat X$ denotes the next-period value of a random
variable $X$. 

\begin{assumption}[Impatience]\label{a:impatience}
	The discount factor and gross return satisfy $\beta R<1$.
\end{assumption}

Assumption~\ref{a:impatience} controls the growth of discounted 
continuation utility and is standard in incomplete market models. 
It suffices for the optimality results below. 

The state of the household is the pair $(a,z)$ with 
state space $\XX:=\RR_+\times \ZZ$.
Let $\vV$ be the set of continuous functions $v:\XX\to\RR_+$ that are
increasing in $a$, concave in $a$, and grow at
most linearly in $a$. The precise weighted norm under which
$\mathcal V$ is complete---and under which the Bellman operator
is a contraction---is specified in Appendix~\ref{s:pr_opt}.
For $\gamma\ge0$, $z\in \ZZ$, and $h:\ZZ\to\RR$, define the 
entropic risk expectation:
\begin{equation}\label{eq:entropic_expectation}
	\rho_{\gamma,z}(h)
	:=\rho_{\gamma,z}\bigl(h(\hat z)\bigr)
	:= \begin{cases}
		-\dfrac{\,1\,}{\gamma}\log\EE_z\me^{-\gamma h(\hat z)}, 
		& \gamma>0, \vspace{1mm}\\
		\EE_z h(\hat z), & \gamma=0.
	\end{cases}
\end{equation}
When applied to a continuation value, the argument includes the wealth 
implied by current saving and next-period productivity.


\begin{theorem}[Existence, uniqueness, and fundamental properties of optimal policy]
	\label{t:os_rs_opt}
	If Assumption~\ref{a:impatience} holds, then for all $\gamma\ge 0$:
	\begin{enumerate}
		\item\label{i:vf_be_pol} There exist a unique function $v^* \in \vV$ 
		and a unique stationary policy $c^*\in \Phi$ that satisfy
		\begin{align}\label{eq:os_rs_BE}
			v^*(a,z) &= \max_{c \in [0, a]} \left\{
			    u(c) + \beta \rho_{\gamma,z} \big( 
			        v^*(R(a-c) + w \hat z, \hat z)
			    \big)
			\right\} \\
			&= u(c^*(a,z)) + \beta \rho_{\gamma,z} \big( 
			    v^* (R (a-c^*(a,z)) + w\hat z, \hat z)
			\big). \notag 
		\end{align}
		\item\label{i:v*_prop} The mapping $v^*(a,z)$ is strictly concave 
		in $a$ and continuous.
		\item\label{i:c*_prop} Both $c^*(a,z)$ and $s^*(a,z):= a - c^*(a,z)$ 
		are increasing in $a$ and continuous.
		\item\label{i:vc*_opt} The mappings $(v^*,c^*)$ coincide with 
		the value function and optimal policy, i.e.,
		\begin{equation*}
			v^*(a,z) = \sup_{\pi \in \Pi} J(a,z, \pi) = J(a,z, c^*)
			\quad \text{for all } \; (a,z) \in \XX.
		\end{equation*}
	\end{enumerate}
\end{theorem}

Theorem~\ref{t:os_rs_opt} shows that the income fluctuation problem is well 
posed for all $\gamma\ge0$. At $\gamma=0$, \eqref{eq:os_rs_BE} is the 
standard expected-utility Bellman equation, and the value function drops out 
of the first-order condition; existence can then be proved directly for
the consumption policy. For $\gamma>0$, the tilting weight depends on
the continuation value, so the value function itself becomes the central object
of analysis, and the contraction argument must discipline discounting and
wealth growth jointly. 
Claims~\eqref{i:v*_prop}--\eqref{i:c*_prop} retain the qualitative properties
of the classical income fluctuation problem---concavity, monotonicity in
wealth, and continuity---which are the key inputs for the subsequent
stochastic stability and general equilibrium analyses.

\begin{remark}[A general parametric optimality result]
	Theorem~\ref{t:os_rs_opt} is a special case of the parametric optimality 
	result, Theorem~\ref{t:opt_general}, derived in Appendix~\ref{s:pr_opt}. 
	There, the gross return $R$ varies over an arbitrary compact set 
	$\CC\subset(0,1/\beta)$, and risk sensitivity $\gamma$ over any 
	nonempty $\G\subset[0,\infty)$, with parameter space $\TT=\CC\times\G$. 
	The value function $v^*(a,z;R,\gamma)$ and optimal policy 
	$c^*(a,z;R,\gamma)$ are shown to be jointly continuous on 
	$\XX\times\TT$. No compactness of $\G$ is needed. If $0\in\G$, 
	the solution is right-continuous at $\gamma=0$. This extension 
	delivers the parametric continuity property used in the general 
	equilibrium and comparative statics analyses of 
	Sections~\ref{s:GE}--\ref{s:comp_stats}. Theorem~\ref{t:os_rs_opt} 
	follows immediately by fixing a single pair $(R,\gamma)$.
\end{remark}


\begin{proposition}[Euler equation]
	\label{pr:euler}
	If Assumption~\ref{a:impatience} holds, then for all $\gamma\ge0$, 
	the unique optimal policy $c^*$ satisfies the Euler equation
	\begin{equation}\label{eq:euler}
		u'(c^*(a,z)) = \max \big\{
		\beta R \EE_z u'(c^*(\hat a,\hat z)) \,
		f_\gamma(\hat a, \hat z\mid z), \, u'(a)
		\big\},
	\end{equation}
	where, for all $(a,z,\hat z) \in \XX\times \ZZ$,  
	\begin{equation}\label{eq:tilt_prob}
		\hat a := R (a - c^*(a,z)) + w \hat z
		\quad \text{and} \quad 
		f_\gamma(\hat a, \hat z\mid z) 
		:= \frac{\me^{-\gamma v^*(\hat a,\hat z)}}{
			\;\EE_z\me^{-\gamma v^*(\hat a,\hat z)}\;}.
	\end{equation}
\end{proposition}

Under risk sensitivity, the marginal benefit
of saving is the tilted expectation of future marginal utility. At a fixed
continuation value, the tilted probability $f_\gamma$ places more  
weight on states with lower continuation value; it satisfies $\EE_zf_\gamma=1$
and reduces to  uniform weight at $\gamma=0$. The marginal benefit of
saving therefore depends jointly on future marginal utility and on the
value-dependent weights. For this reason, a higher $\gamma$ does not by 
itself guarantee a saving comparison. When saving is positive, 
current marginal utility equals the discounted tilted marginal benefit. 
At zero saving, it is weakly greater, as reflected by the maximum in 
\eqref{eq:euler}.

\subsection{Stochastic Stability}\label{ss:stability}

Under the optimal policy, the state dynamics are
\begin{align}
	a_{t+1} &= R \left[a_t - c^*(a_t,z_t) \right] + w z_{t+1}, \label{eq:evol_w} \\
	z_{t+1} &\sim Q(z_t, \cdot\,). \label{eq:evol_z}
\end{align}
Let $P$ denote the induced stochastic kernel, 
which governs the transition probabilities of the state process 
$\{(a_t,z_t)\}$. Appendix~\ref{s:prelm} provides its formal definition.
We impose the following condition to control wealth growth at high levels.

\begin{assumption}[Vanishing absolute risk aversion]\label{a:bip_cond}
	The period utility satisfies
	\begin{equation*}
	\lim_{a\to\infty}\eta(a)=0,
	\qquad \text{where} \quad 
	\eta(a):= - u''(a) \big/ u'(a).
	\end{equation*}
\end{assumption}

This condition ensures that marginal utility becomes insensitive 
to bounded changes at sufficiently high wealth (in relative terms): 
$\lim_{a\to\infty}u'(a+h)/u'(a)=1$ for all $h>0$. 
Assumption~\ref{a:bip_cond} clearly holds for the CRRA period utility 
function. More generally, bounded relative risk aversion in the tail 
is sufficient. Together with impatience, it implies mean reversion 
in wealth, as the next result demonstrates.



\begin{proposition}[Mean reversion and existence of a stationary distribution]
	\label{pr:compact}
	If Assumptions~\ref{a:impatience}--\ref{a:bip_cond} hold, then:
	\begin{enumerate}
		\item\label{i:invariance}\textup{(Invariance).} 
		There exists $a^*_R < \infty$ such that the state process 
		$\{(a_t,z_t)\}_{t\geq 0}$ lies in 
		$\XX_R:=[0, a^*_R] \times \ZZ$ almost surely whenever 
		$(a_0,z_0) \in \XX_R$.
		\item\label{i:absorption}\textup{(Absorption).} 
		From any initial state $(a,z)\in \XX$, the state process reaches 
		$\XX_R$ almost surely in finite time.
		\item\label{i:existence}\textup{(Stationary distribution and support).} 
		The stochastic kernel $P$ admits at least one stationary 
		distribution. Moreover, any stationary distribution $\mu_R$ 
		satisfies $\mu_R (\XX_R) = 1$.
	\end{enumerate}
\end{proposition}

Despite market incompleteness and risk sensitivity, wealth does not 
drift to infinity. Intuitively, although the endogenous tilted probability 
redistributes weights across the income states, it neither adds new states 
nor relaxes the budget constraint, so it cannot by itself induce unbounded 
accumulation. The compact absorbing set $\XX_R$, together with smooth 
transitions, ensures existence of a stationary distribution.

To state the next result, we need a little notation, all defined formally in 
Appendix~\ref{s:prelm}. Let $P^n$ be the $n$-step stochastic kernel, 
which measures the transition probability over $n$ periods, and 
$\|\cdot\|_{TV}$ be the total variation norm, which measures the distance 
between two distributions. If households are currently distributed 
according to $\mu$, then $\mu P^n$ is the cross-sectional distribution $n$ 
periods later.

\begin{theorem}[Global stability]
	\label{t:gs_rs}
	If Assumptions~\ref{a:impatience}--\ref{a:bip_cond} hold and $Q$ has positive entries, then the following statements are true.
	\begin{enumerate}
		\item\label{i:gs_XR} \textup{(Uniform ergodicity on $\XX_R$).}
		The state process restricted to $\XX_R$ is uniformly ergodic:
		it admits a unique stationary distribution $\mu^*$ supported on
		$\XX_R$, and there
		exist constants $C<\infty$ and $\vartheta\in(0,1)$ such that
		\begin{equation*}
		\sup_{x\in \XX_R}\|P^n(x,\cdot)-\mu^*\|_{TV}
		\le C\vartheta^n
		\quad \text{for all } \; n\ge0.
		\end{equation*}
		\item\label{i:gs_full} \textup{(Global stability on $\XX$).}
		$\mu^*$ is the unique stationary distribution of $P$ on $\XX$, 
		and for all initial distributions $\mu$,
		\begin{equation*}
		\|\mu P^n-\mu^*\|_{TV} \to 0
		\quad\text{as }\; n\to\infty.
		\end{equation*}
	\end{enumerate}
\end{theorem}

Theorem~\ref{t:gs_rs} shows that the cross-sectional distribution 
converges to a unique limit $\mu^*$ from any initial condition, at a 
geometric rate if the initial state lies in $\XX_R$. The key point is 
that the endogenous tilt does not jeopardize ergodicity: although it
reshapes the optimal policy, the induced transition still assigns 
positive probability to reaching the borrowing constraint in finite 
time (see Lemma~\ref{lm:binding_ft} in Appendix~\ref{s:pr_stability}). 
The theorem justifies interpreting 
$\mu^*$ as the unique \emph{long-run distribution} of the economy, 
and it provides the stationary distribution used to clear the capital 
market in Section~\ref{s:GE}.


\section{General Equilibrium}\label{s:GE}

This section embeds the risk-sensitive household problem into a production
economy, establishes the existence of a stationary competitive equilibrium, 
and the upper hemicontinuity of the equilibrium correspondence in $\gamma$. 
The key challenge is that the tilted probability is endogenously 
determined. The continuity and boundary properties of the
classical existence argument therefore have to be verified with the tilt
varying jointly with prices and the household's policy, rather than held
fixed.

Throughout, $\bar R:=1/\beta$. At prices $(R,w(R))$ and 
risk sensitivity $\gamma$, write $c_{R,\gamma}$ for the optimal 
consumption policy, $P_{R,\gamma}$ for the stochastic kernel in 
\eqref{eq:evol_w}--\eqref{eq:evol_z}, and $\mu_{R,\gamma}$ for its 
unique stationary distribution. The latter is well defined by 
Theorem~\ref{t:gs_rs} when $Q$ has positive entries.
For the general equilibrium analysis, the economy is populated by a unit
mass of households, each facing the problem of Section~\ref{s:ifp_rsp}. 
We normalize the productivity process so that its stationary mean is
$\EE z=1$, with $\EE$ the expectation under the stationary distribution
of $Q$. 

\subsection{Production}\label{ss:firm}
A representative firm uses capital $K$ and effective labor $L$ to produce
output $F(K,L)$. Capital depreciates at rate $\delta \in (0,1)$.

\begin{assumption}[Production technology]\label{a:prod_func}
	The production function $F:\RR_+^2\to\RR_+$ is continuous, concave, and has constant returns to scale. 
	It is strictly increasing and continuously differentiable 
	on $\RR_{++}^2$, $K\mapsto F(K,1)$ is strictly concave, and
	\begin{equation*}
		\lim_{K\downarrow0} F(K,1)=0,\qquad
		\lim_{K\downarrow0}F_1(K,1)=\infty,\qquad
		\ell:=\lim_{K\to\infty}F_1(K,1)<\delta.
	\end{equation*}
\end{assumption}

The Inada conditions in Assumption~\ref{a:prod_func} guarantee that 
capital demand is positive and finite at every relevant interest rate, 
while $\ell<\delta$ ensures that the marginal product of capital 
eventually falls below the depreciation rate.
The firm chooses capital and labor to maximize static profits:
\begin{equation*}
	\max_{K,L\ge0}\Bigl\{F(K,L)-(R-1+\delta)K-wL\Bigr\}.
\end{equation*}
Labor market clearing requires $L=1$: each household supplies $z$ units
of effective labor, and the normalization $\EE z=1$ makes aggregate
effective labor supply equal to $L=\EE z=1$. Hence, the
first-order conditions are evaluated at $L=1$ and given by
\begin{equation}\label{eq:pro_opt}
	R-1+\delta = F_1(K,1)
	\quad\; \text{and} \;\quad 
	w = F_2(K,1).
\end{equation}
Define the capital demand schedule $K(R):=F_1^{-1}(R-1+\delta,1)$, 
the induced wage schedule $w(R):=F_2(K(R),1)$, and the lower bound 
interest rate $\underline R:=1-\delta+\ell<1$. By
Assumption~\ref{a:prod_func}, $K$ is continuous and strictly
decreasing, with
\begin{equation}\label{eq:K_lim}
	\lim_{R\downarrow \uR} K(R)=\infty 	
	\quad\; \text{and} \quad \;
	\lim_{R\uparrow\bar R}K(R) = K(\bar R)\in(0,\infty),
\end{equation}
and the wage satisfies
\begin{equation}\label{eq:wage_capital_prop}
	\lim_{R\uparrow\bar R} w(R)= w(\bar R)>0
	\quad\; \text{and} \quad\;
	\lim_{R\downarrow \uR} \, w(R) / K(R)=0.
\end{equation}
These properties are standard and are formally derived in
Online Appendix~\ref{s:pf_firm_opt}.

\subsection{Stationary Equilibrium}\label{ss:equilibrium}
Given an interest rate $R\in(\underline R,\bar R)$ and risk-sensitivity 
parameter $\gamma$, the household side is summarized by the stationary 
distribution $\mu_{R,\gamma}$ obtained in Section~\ref{ss:stability}. 
The aggregate wealth $A(R,\gamma)$ and aggregate saving $S(R,\gamma)$, 
which finance the firm's capital stock, are respectively 
\begin{equation}\label{eq:agg_wealth}
	A(R,\gamma):=\int a \mu_{R,\gamma}\left[\diff (a,z)\right]
	\quad \text{and} \quad 
	S(R,\gamma):=\int_{\XX}\bigl[a-c_{R,\gamma}(a,z)\bigr]
	\mu_{R,\gamma}[\diff (a,z)].
\end{equation}
In a stationary state, wealth equals the return on saving plus labor
income. Therefore, 
\begin{equation*}
	A(R,\gamma)=R\,S(R,\gamma)+w(R).
\end{equation*}

\begin{definition}\label{d:RCE}
	A \emph{recursive competitive equilibrium} consists of prices $(R,w)$, a
	value function $v^*:\XX\to\RR$, a policy function $c^*:\XX\to\RR_+$,
	and a probability measure $\mu^*$ on $\XX$ such that:
	\begin{enumerate}[label=(\roman*)]
		\item Given $(R,w)$, $v^*$ and $c^*$ solve the household problem 
		\eqref{eq:object}--\eqref{eq:budget}.
		\item Given $(R,w)$, the firm maximizes profits, so its capital
		demand $K$ satisfies the first-order condition \eqref{eq:pro_opt}.
		\item Markets clear: the labor and capital markets clear, satisfying
		\begin{equation*}
			L = \EE z = 1
			\quad\; \text{and} \quad\;
			\int_{\XX}\bigl[a-c^*(a,z)\bigr]\,\mu^*[\diff (a,z)]=K.
		\end{equation*}
		\item\label{i:rce_stat} 
		$\mu^*$ is a stationary distribution of the transition kernel $P$ 
		induced by \eqref{eq:evol_w}--\eqref{eq:evol_z}:
		\begin{equation*}
			\mu^*(B)=\int_{\XX}P((a,z),B)\,\mu^*[\diff (a,z)]
			\quad \text{for all Borel sets } \; B \subset \XX.
		\end{equation*}
	\end{enumerate}
\end{definition}
By part~\ref{i:rce_stat} and the uniqueness statement of
Theorem~\ref{t:gs_rs}, any equilibrium measure must
coincide with $\mu_{R,\gamma}$ at the equilibrium interest rate. The
equilibrium conditions therefore collapse to a single equation
in the interest rate: $S(R,\gamma)=K(R)$.

\subsection{Existence and Upper Hemicontinuity of Equilibrium}\label{ss:existence}
Existence of an equilibrium is equivalent to the existence of a
zero of \emph{excess capital supply}, defined by 
\begin{equation*}
	\Psi(R,\gamma):=S(R,\gamma)-K(R).
\end{equation*}
In what follows, we 
show that $\Psi$ is continuous and changes sign on $(\underline R,\bar R)$.

\begin{lemma}[Continuity of aggregate objects]\label{lm:agg_cont}
	If Assumptions~\ref{a:impatience}, \ref{a:bip_cond},
	and \ref{a:prod_func} hold and $Q$ has positive entries,
	then $A(R,\gamma)$, $S(R,\gamma)$, and $\Psi(R,\gamma)$ are 
	continuous on $(\uR,\bar R)\times [0,\infty)$. 
\end{lemma}

Lemma~\ref{lm:agg_cont} follows from two facts. First, by
Theorem~\ref{t:opt_general}, the value function and the optimal policy are
jointly continuous in $(a,z,R,\gamma)$, so the transition kernel varies
continuously with the parameters. Second, by Proposition~\ref{pr:com_sup},
all stationary distributions for interest rates in a compact set share a
common compact absorbing set, so their supports are uniformly bounded. As
in \citet[][Lemmas~G.2--G.3]{accikgoz2018existence} and
\citet[][Lemma~7 and Theorem~6]{zhu2020existence} in the expected-utility 
case, weak continuity of the stationary distributions alone does not
deliver convergence of their moments; the uniform bound provides the
needed control.


\begin{assumption}[Stochastic monotonicity]
	\label{a:mlr}
	$Q$ has positive entries and is monotone.
\end{assumption}

Here, monotonicity of $Q$ means that higher current productivity
shifts the distribution of future productivity upward in the sense of 
first-order stochastic dominance; see Appendix~\ref{s:prelm} for 
its formal definition. 

\begin{proposition}[Divergence of aggregate wealth]
	\label{pr:agr_cap}
	If Assumptions~\ref{a:impatience}--\ref{a:bip_cond} 
	and~\ref{a:prod_func}--\ref{a:mlr} hold, then 
	$\lim_{R\uparrow\bar{R}} A(R,\gamma)=\infty$ 
	for all $\gamma \ge 0$.
\end{proposition}

Proposition~\ref{pr:agr_cap} shows that the classical divergence of
\citet{huggett1993risk} persists under risk-sensitive preferences:
as $\beta R \uparrow 1$, impatience no longer provides a uniform
bound on wealth accumulation, and aggregate asset demand diverges. 
The argument links two channels. 
The standard \emph{incomplete-markets channel} generates
precautionary saving through the borrowing constraint, even
without convex marginal utility.
The \emph{risk-sensitivity channel} determines whether this
divergence carries over from expected utility.
Its effect on aggregate saving is not immediate: 
higher risk sensitivity shifts weight toward low continuation 
value states, but also changes continuation values unevenly across states.

The proof consists of two steps. When $\gamma=0$, a hypothetical sequence of 
bounded aggregate wealth produces a limiting policy and a stationary 
distribution with finite mean at $\beta R=1$ (Lemma~\ref{lm:c_ubdd}), 
and the limiting Euler inequality forces consumption to be constant along 
stationary paths, contradicting nondegenerate income risk. For $\gamma>0$,
Proposition~\ref{pr:markov_small_gamma} below, 
applied with $\gamma_1=0$, yields $A(R,\gamma)\ge A(R,0)$,
extending divergence to every $\gamma\ge 0$. This fixed-price comparison requires no monotonicity between
arbitrary positive values of $\gamma$. It is established on a common invariant 
wealth set constructed from household primitives and therefore does not presuppose existence of equilibrium.


\begin{remark}[Why the expected-utility proof does not apply directly] 
	The expected-utility arguments of both \citet[][Lemma~1 and 
	Proposition~6]{accikgoz2018existence} 
	and \citet[][Theorem~7 and Proposition~13]{zhu2020existence} 
	exploit Euler restrictions under the law governing realized 
	wealth. With risk sensitivity, the Euler equation instead 
	uses endogenous tilted probabilities.
	Neither does stationarity cancel the tilted continuation term as in 
	\cite{accikgoz2018existence}, nor does the Euler inequality directly 
	produce a marginal utility supermartingale under $Q$ as in 
	\cite{zhu2020existence}.\footnote{With 
		endogenous labor, Zhu’s divergent object is the 
		capital--labor ratio: either wealth diverges or labor supply
		vanishes. The wealth divergence comparison here concerns his 
		exogenous labor case.} 
\end{remark}

\begin{remark}[Role of monotone $Q$]\label{rm:agg_reversal}
	Stochastic monotonicity of $Q$ supports the comparison
	$A(R,\gamma)\ge A(R,0)$ used in
	Proposition~\ref{pr:agr_cap} to extend aggregate wealth
	divergence from expected utility to risk-sensitive preferences.
	This comparison can fail without monotonicity, as shown
	by a counterexample in Online Appendix~\ref{oa:counterexamples}.
\end{remark}

Define the  equilibrium interest rate correspondence by
\begin{equation}\label{eq:equi_set}
	\mathcal R: [0,\infty) \rightrightarrows (\uR,\bar R) 
	\quad \text{s.t.} \quad 
	\mathcal R(\gamma) = \left\{
	    R \in (\uR,\bar R) : \Psi(R,\gamma) = 0
	\right\}.
\end{equation}

\newpage 

\begin{theorem}[Existence and upper hemicontinuity of recursive competitive equilibrium]\label{t:RCE}
	If Assumptions~\ref{a:impatience}--\ref{a:bip_cond} and
	\ref{a:prod_func}--\ref{a:mlr} hold, then:
	\begin{enumerate}
		\item\label{i:exist} 
		A recursive competitive equilibrium exists at every $\gamma\ge0$;
		equivalently, $\mathcal R(\gamma)$ is nonempty for all $\gamma\ge0$.
		\item\label{i:uhc}
		The correspondence $\mathcal R$ is compact-valued and upper 
		hemicontinuous on $[0,\infty)$. 
	\end{enumerate}
\end{theorem}

Theorem~\ref{t:RCE} covers $\gamma=0$, so the expected-utility 
economy is nested as a special case rather than treated separately. 
Moreover, equilibrium interest rates are bounded away from 
$\bar R=1/\beta$: as $R\uparrow\bar R$, aggregate capital supply 
diverges (Proposition~\ref{pr:agr_cap}) while capital demand remains 
finite, so no equilibrium rate can accumulate at the boundary.  
Upper hemicontinuity of the equilibrium correspondence---a classical 
desideratum in general equilibrium theory 
\citep[][]{hildenbrand1972upper}---is delivered by the joint 
continuity of $\Psi$ in $(R,\gamma)$. Corollary~\ref{cor:uhc} 
collects its implications under multiplicity.

\begin{corollary}[Limits and continuity of equilibrium rates]
	\label{cor:uhc}
	If the assumptions of Theorem~\ref{t:RCE} hold, then:
	\begin{enumerate}
		\item If $\gamma_n\to\gamma_0$ and $R_n \in \rR(\gamma_n)$, then 
		every cluster point of $\{R_n\}$ lies in $\rR(\gamma_0)$.
		\item If $\gamma_n \to \gamma_0$, $R_0$ is the unique equilibrium 
		interest rate at $\gamma_0$, and $R_n \in \rR(\gamma_n)$, 
		then $R_n \to R_0$.
		\item If the equilibrium is unique for all $\gamma$, then the unique
		equilibrium interest rate is continuous in $\gamma$ on $[0,\infty)$.
	\end{enumerate} 
\end{corollary}

The proof of Theorem~\ref{t:RCE} proceeds in 
two steps. First, the firm's side delivers continuous capital demand 
and wage schedules. For any fixed $\gamma$, 
saving falls short of capital demand near $\uR$ and exceeds it 
near $\bar R$; continuity of excess supply then yields an equilibrium via 
the intermediate value theorem. Second, the same two endpoint bounds hold 
uniformly in $\gamma$, so all equilibrium rates lie in a common compact 
interval. Since excess supply is jointly continuous in $(R,\gamma)$, 
a limit of equilibrium rates is again an equilibrium rate, which gives 
upper hemicontinuity.

\section{Comparative Statics}\label{s:comp_stats}
This section asks how risk sensitivity affects saving, the stationary
wealth distribution, and equilibrium interest rates. The comparison is
global under independent income and holds in two regions under Markov
income; an exact decomposition then identifies the forces behind reversals.

Let $v_{R,\gamma}$, $c_{R,\gamma}$, $s_{R,\gamma}(a,z):=a-c_{R,\gamma}(a,z)$, 
and $\mu_{R,\gamma}$ be the value function, consumption and saving policies, 
and stationary distribution at prices $(R,w)$ and risk sensitivity
$\gamma$. Aggregate wealth $A(R,\gamma)$, aggregate saving $S(R,\gamma)$, 
excess capital supply $\Psi(R,\gamma)$, and the equilibrium set 
$\mathcal R(\gamma)$ are as in Section~\ref{s:GE}. 
By Theorem~\ref{t:RCE}, $\mathcal R$ is nonempty, compact-valued, 
and upper hemicontinuous, so the extremal interest rates
\begin{equation}\label{eq:iid_extremal_rates}
	R_{\min}(\gamma):=\min\mathcal R(\gamma)
	\qquad \text{and} \qquad 
	R_{\max}(\gamma):=\max\mathcal R(\gamma)
\end{equation}
are well-defined.
In what follows, policy and distributional comparisons are held at fixed 
prices, while equilibrium comparisons allow prices to adjust.

\subsection{Independent Productivity}
\label{ss:iid_comparative}
We first consider {\sc iid} productivity, under which a clean comparative 
static can be established.

\begin{assumption}[Independent productivity with full support]
	\label{a:iid_income}
	The productivity process $\{z_t\}$ is {\sc iid} on $\ZZ$, 
	and every state has positive probability.
\end{assumption}

In this case, current productivity carries no information about future 
income, so the value function and optimal policy depend only on wealth. 
We write them as $v_{R,\gamma}(a)$, $c_{R,\gamma}(a)$, and $s_{R,\gamma}(a)$. 
Assumption~\ref{a:iid_income} implies Assumption~\ref{a:mlr}: the rows 
of $Q$ coincide and all entries are positive. 

\begin{proposition}[Strict comparative statics under {\sc iid} income]
	\label{pr:iid_gamma_saving}
	Suppose Assumptions~\ref{a:impatience}--\ref{a:bip_cond}
	and \ref{a:iid_income} hold.
	Fix $R\in(0,\bar R)$ and $0\le \gamma_1<\gamma_2$.
	Then:
	\begin{enumerate}
		\item 
		Higher risk sensitivity raises individual saving and lowers consumption:
		\begin{equation}\label{eq:iid_policy_order}
			s_{R,\gamma_2}(a)\ge s_{R,\gamma_1}(a)
			\quad \text{and} \quad 
			c_{R,\gamma_2}(a)\le c_{R,\gamma_1}(a)
			\quad\text{for all }a\ge0.
		\end{equation}
		Both inequalities are strict whenever $s_{R,\gamma_1}(a)>0$.
		
		\item 
		Higher risk sensitivity raises aggregate saving and wealth:
		\begin{equation}\label{eq:iid_aggregate_order}
			S(R,\gamma_2)\ge S(R,\gamma_1)
			\quad \text{and} \quad 
			A(R,\gamma_2)\ge A(R,\gamma_1).
		\end{equation}
		Both inequalities are strict unless
		$S(R,\gamma_1)=S(R,\gamma_2)=0$.
		
		\item 
		The stationary wealth distribution under $\gamma_2$
		first-order stochastically dominates that under $\gamma_1$:
		\begin{equation*}
			\int h\,\diff\mu_{R,\gamma_2}
			\ge \int h\,\diff\mu_{R,\gamma_1}
			\;\;\; \text{for all bounded increasing
				$h:\RR_+\to\RR$.}
		\end{equation*}
	\end{enumerate}
\end{proposition}

Under {\sc iid} income, successor states differ only in wealth:
poorer states have lower continuation values and higher marginal
values of wealth. Greater risk sensitivity favors these states,
strengthening the incentive to save. Saving rises strictly wherever
it is initially positive; the borrowing constraint can leave
households with low wealth at zero saving. Coupling the two economies
along a common income path preserves the wealth ordering, which
passes to their stationary distributions. Aggregate saving and
wealth rise strictly unless both economies have zero aggregate saving.


The strict aggregate comparison induces strictly lower 
extremal equilibrium interest rates, as the next result demonstrates.

\begin{theorem}[Strict equilibrium comparison under {\sc iid} income]
	\label{t:iid_gamma_equilibrium}
	If Assumptions~\ref{a:impatience}, \ref{a:bip_cond},
	\ref{a:prod_func}, and \ref{a:iid_income} hold, then for all $0\le\gamma_1<\gamma_2$,
	\begin{equation}\label{eq:iid_equilibrium_order}
		R_{\min}(\gamma_2)< R_{\min}(\gamma_1)
		\qquad \text{and} \qquad 
		R_{\max}(\gamma_2)< R_{\max}(\gamma_1).
	\end{equation}
\end{theorem}

At any equilibrium under $\gamma_1$, aggregate saving equals
positive capital demand. Hence, Proposition~\ref{pr:iid_gamma_saving}
implies that higher risk sensitivity strictly raises
capital supply at that interest rate, while capital demand is
unchanged. Together with the boundary signs of excess supply,
this shifts both extremal equilibrium rates strictly downward,
without requiring equilibrium uniqueness or monotonicity of
capital supply in interest rates.


\subsection{Markov Productivity}\label{ss:mc_prod}

We return to Assumption~\ref{a:mlr} and Markov income.
Current productivity now affects future income prospects, so
continuation values vary with both wealth and productivity.
The {\sc iid} argument no longer applies, but strict comparisons
remain valid when the lower risk-sensitivity parameter is
sufficiently small or the upper one sufficiently large.

For each fixed $R$, let $\XX_R:=[0,a_R^*]\times\ZZ$ be the
common invariant set supplied by Proposition~\ref{pr:compact}
(or Proposition~\ref{pr:com_sup} with $\CC=\{R\}$).
This set is uniform across $\gamma$ and supports every
stationary distribution.

\begin{proposition}[Strict comparative statics under Markov income]
	\label{pr:markov_small_gamma}
	Suppose Assumptions~\ref{a:impatience}, \ref{a:bip_cond},
	and \ref{a:mlr} hold.
	For every $R\in(0,\bar R)$, there exist
	$0<\gamma_R^-<\gamma_R^+$ such that, for all
	$0\le\gamma_1<\gamma_2$ with either
	$\gamma_1\le\gamma_R^-$ or $\gamma_2\ge\gamma_R^+$:
	\begin{enumerate}
		\item Higher risk sensitivity raises individual
		saving and lowers consumption:
		\begin{equation}\label{eq:markov_policy_order}
			s_{R,\gamma_2}(a,z)\ge s_{R,\gamma_1}(a,z)
			\quad \text{and} \quad
			c_{R,\gamma_2}(a,z)\le c_{R,\gamma_1}(a,z)
			\quad\text{on }\;\XX_R.
		\end{equation}
		Both inequalities are strict whenever
		$s_{R,\gamma_1}(a,z)>0$.
		
		\item Higher risk sensitivity raises aggregate
		saving and wealth:
		\begin{equation}\label{eq:markov_aggregate_order}
			S(R,\gamma_2)\ge S(R,\gamma_1)
			\quad \text{and} \quad
			A(R,\gamma_2)\ge A(R,\gamma_1).
		\end{equation}
		Both inequalities are strict unless 
		$S(R,\gamma_1)=S(R,\gamma_2)=0$.
		
		\item Conditional on each productivity state, the stationary
		wealth distribution under $\gamma_2$ first-order
		stochastically dominates that under $\gamma_1$:
		\begin{equation*}
			\int_\XX h\,\diff\mu_{R,\gamma_2}
			\ge \int_\XX h\,\diff\mu_{R,\gamma_1}
		\end{equation*}
		for every bounded $h:\XX\to\RR$ increasing in wealth.
	\end{enumerate}
\end{proposition}

The two thresholds draw on different benchmarks.
Near expected utility, bounds on changes in continuation values
preserve the saving comparison with any higher $\gamma$.
At large $\gamma$, the household's problem approaches the
deterministic benchmark in which income remains at its minimum.
This limit is common across productivity states because each
assigns positive probability to the lowest income.
Bounds on marginal values and tilted probabilities then establish
the comparison with any lower $\gamma$.
In both regions, saving rises strictly wherever it is initially
positive, and stationary aggregates rise strictly unless both
economies have zero saving. Between the thresholds, changes in
continuation values and future consumption can reverse these
comparisons.

Appendix~\ref{s:pr_comparative} constructs the thresholds from
the primitives and the common wealth bound, and makes them
uniform over equilibrium prices. Since equilibrium capital is
positive, the zero-saving exception is excluded, yielding
strict comparisons of extremal equilibrium rates.


\begin{theorem}[Strict equilibrium comparison under Markov income]
	\label{t:markov_gamma_equilibrium}
	Suppose Assumptions~\ref{a:impatience}--\ref{a:bip_cond} and
	\ref{a:prod_func}--\ref{a:mlr} hold. Then there exist 
	$0<\gamma_*^-<\gamma_*^+$ such that, for all $0\le\gamma_1<\gamma_2$
	with either $\gamma_1\le\gamma_*^-$ or $\gamma_2\ge\gamma_*^+$,
	\begin{equation}\label{eq:markov_equilibrium_order}
		R_{\min}(\gamma_2)< R_{\min}(\gamma_1)
		\qquad\text{and}\qquad
		R_{\max}(\gamma_2)< R_{\max}(\gamma_1).
	\end{equation}
\end{theorem}

In particular, every $\gamma>0$ yields strictly lower extremal
equilibrium interest rates than the expected-utility benchmark.
Theorem~\ref{t:markov_gamma_equilibrium} also orders every 
$\gamma_2\ge\gamma_*^+$
against any smaller parameter. Only comparisons with both
parameters in $(\gamma_*^-,\gamma_*^+)$ remain unordered.

\begin{remark}[Strict reversals between positive parameters]\label{rm:eq_reversal}
	In Online Appendix~\ref{oa:counterexamples}, we construct 
	an economy with $\gamma_*^-<\gamma_1<\gamma_2<\gamma_*^+$
	in which higher risk sensitivity strictly lowers saving
	at some states and strictly lowers aggregate saving and wealth,
	while strictly raising both extremal equilibrium interest rates.
\end{remark}

The proofs are in Appendix~\ref{s:pr_comparative}.
As under {\sc iid} productivity, policy comparisons pass to stationary
distributions and aggregates, and then to extremal equilibrium 
rates. The additional step controls how changes in continuation
values affect tilted probabilities under Markov income.

\subsection{Decomposing the Saving Comparison}\label{ss:mech}

Why can saving fall with risk sensitivity in the intermediate
range? We answer this question by decomposing the change in
the marginal return to saving at the baseline choice.


Fix $R \in (0,\bar R)$, $w>0$, $(a,z)\in\XX$, and $0\le\gamma_1<\gamma_2$. 
Suppose 
$s_1:=s_{R,\gamma_1}(a,z)>0$.\footnote{If 
	$s_1=0$, saving cannot fall because borrowing is prohibited.} 
Let $x_j:=Rs_1+w z_j$. 
For each $k\in\{1,2\}$, define continuation values and marginal values 
at these common successor states $V_k$ and $m_k$:
\begin{equation*}
	V_{kj}:=v_{R,\gamma_k}(x_j,z_j)
	\quad \text{and} \quad
	m_{kj}:=\partial v_{R,\gamma_k}(x_j,z_j)/\partial x_j
	    \equiv u'[c_{R,\gamma_k}(x_j, z_j)],
\end{equation*}
where the last equality follows from
the envelope identity (Lemma~\ref{lm:T_diff}).
Define the tilted probability vectors $p_k$ and $\tilde p$:
\begin{equation*}
	p_{kj}\propto Q(z,z_j)\me^{-\gamma_k V_{kj}} \quad (k=1,2)
	\qquad\text{and} \qquad
	\tilde p_j\propto Q(z,z_j)\me^{-\gamma_2 V_{1j}}.
\end{equation*}
Thus, $\tilde p$ varies risk sensitivity while holding the baseline 
continuation value function fixed. Moreover, let $L_j:=V_{1j}-V_{2j}$. 

\begin{proposition}[Exact decomposition of the saving comparison]
	\label{pr:decomp}
	If Assumption~\ref{a:impatience} holds, 
	then $L_j \ge 0$ for all $j$, and
	\begin{equation*}
		s_{R,\gamma_2}(a,z) \ge s_{R,\gamma_1}(a,z)
		\quad \iff \quad 
		T+W+B \ge 0,
	\end{equation*}
	where
	\begin{equation*}
		T=\frac{\Cov_{p_1}\bigl(m_1,\me^{-(\gamma_2-\gamma_1)V_1}\bigr)
		}{\EE_{p_1} \me^{-(\gamma_2-\gamma_1)V_1}},
		\quad
		W=\frac{\Cov_{\tilde p}\bigl(m_1,\me^{\gamma_2 L}\bigr)
		}{\EE_{\tilde p\,}\me^{\gamma_2 L}}
		\quad \text{and} \quad 
		B=\EE_{p_2}(m_2-m_1).
	\end{equation*}
	A strict reversal $s_{R,\gamma_2}(a,z)<s_{R,\gamma_1}(a,z)$ 
	occurs if and only if $T+W+B<0$.
\end{proposition}


The term $T$ is the direct tilt. Holding continuation values
at $V_1$, higher risk sensitivity shifts weight toward low continuation 
value states. This weakly raises the marginal benefit of saving when
$\Cov_{p_1}(m_1,\me^{-(\gamma_2-\gamma_1)V_1})\ge0$.
Under {\sc iid} income, strict increase and strict concavity
of the value function imply $T>0$: poorer states have lower
continuation values and higher marginal values of wealth.
Under Markov income, however, values and derivatives also depend on
productivity, so ordering continuation values alone does not
determine the sign of $T$.

The term $W$ captures the additional reweighting caused by
changes in continuation values. Relative to $\tilde p$,
the actual tilt favors states with larger value losses $L_j$.
Therefore $W<0$ when these losses are associated with lower
baseline marginal values, as measured by
$\Cov_{\tilde p}(m_1,\me^{\gamma_2L})<0$.
Finally, $B$ captures changes in future marginal values,
holding the weights at $p_2$ fixed. A negative $B$ means
that expected marginal utility falls under these weights;
it need not imply higher expected consumption.

In the two-period economy of Section~\ref{ss:2_period},
terminal consumption equals wealth, so continuation values
and marginal values are independent of $\gamma$.
Hence $W=B=0$, leaving only $T>0$.
In the three-period economy of Section~\ref{ss:3_period},
the calculation at $s=0$ holds successor marginal values
fixed, so $m_1=m_2$ and $B=0$. The shrinking continuation value gap
produces a negative $W$ that outweighs the direct tilt.
This calculation concerns marginal incentives at the
borrowing constraint; Proposition~\ref{pr:decomp} evaluates
them at positive baseline saving.

In the general economy, all three terms can operate. When $T \ge 0$, a 
reversal requires $W+B<-T$. The decomposition holds exactly for a given state 
and parameter pair, while the thresholds in Section~\ref{ss:mc_prod} 
provide sufficient comparisons from the primitives, without computing
these endogenous objects.

\begin{figure}[h!]
	\centering
	\includegraphics[width=\textwidth]{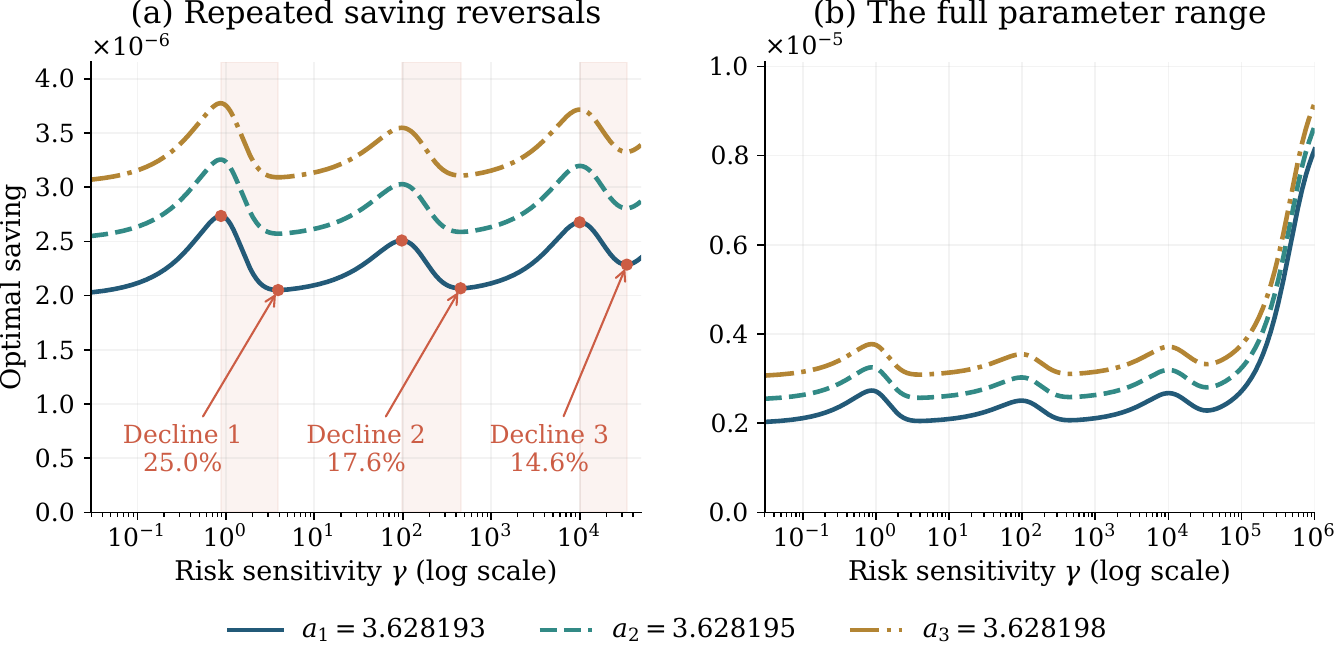}
	\caption{Optimal saving at fixed prices and productivity $z_1$.} 
	\label{fig:markov_saving_reversals}
\end{figure} 

Figure~\ref{fig:markov_saving_reversals} illustrates repeated
saving reversals at fixed prices and productivity $z_1$,
for three nearby wealth levels.\footnote{We set $u(c)=\sqrt{c}$,
	$\beta=0.5$, $R=1.05$, $\Delta=p=10^{-5}$, $\epsilon=10^{-8}$,
	with income
	$y:=wz=(1,1+\Delta,1+2\Delta,1+30\Delta,1+60\Delta,1.03,1.06,9,25)$.
	Let $Q_{ij}=\epsilon+p\1_{\{j=1\}} +(1-p-9\epsilon) \bar Q_{ij}$, 
	where the only nonzero entries of $\bar Q$ are 
	$(\bar Q_{12},\bar Q_{13})=(0.2,0.8)$, 
	$(\bar Q_{24},\bar Q_{26},\bar Q_{28})=(0.005,0.045,0.95)$, and
	$(\bar Q_{i5},\bar Q_{i7},\bar Q_{i9})=(0.005,0.045,0.95)$ for 
	$i=3,\dots,9$.
	Hence, $Q$ is strictly positive and monotone.
	For its stationary distribution $\pi$, set
	$w=\sum_j\pi_jy_j$ and $z_j=y_j/w$.
	Both panels use $a_k=(\beta R)^{-2}+\Delta b_k$
	and $(b_1,b_2,b_3)=(7.5,7.75,8)$, all above the
	zero-saving region $[0,(\beta R)^{-2}]$.}
Panel~(a) shows three successive declines spanning more than
four orders of magnitude in $\gamma$.
At the lowest wealth level, the peak-to-trough declines are
approximately $25\%$, $18\%$, and $15\%$. The other two wealth 
levels trace the same pattern, shifted in level, confirming that this 
nonmonotonicity is not knife-edge in wealth.
Panel~(b) extends the range to $\gamma=10^6$ and shows
a pronounced recovery at large $\gamma$. Yet even at $\gamma=10^5$, 
saving remains slightly below its first peak, attained at $\gamma<1$.
A strict reversal thus occurs between parameters more than
five orders of magnitude apart, illustrating how wide the
intermediate region in Proposition~\ref{pr:markov_small_gamma} can be.
A local recovery alone does not establish the upper-threshold
comparison, which orders saving against every smaller parameter.

\appendix

\section{Preliminaries}\label{s:prelm}

Let $(\Omega, \fF, \PP)$ be a fixed probability space hosting all random 
variables. The productivity process $\{z_t\}$ and its natural filtration 
$\{\fF_t\}$ are defined thereon. For $z\in \ZZ$, $\PP_z$ denotes the 
probability conditional on $z_0 = z$, with conditional expectations written
\begin{equation*}
	\EE_z:=\EE[ \, \cdot \mid z_0 = z] 
	\qquad \text{and} \qquad 
	\EE_{a,z}:= \EE[ \,\cdot \mid (a_0,z_0) = (a, z)].
\end{equation*}
Expectation without a subscript $\EE$ refers to the stationary process, 
where $z_0$ is drawn from its stationary distribution.
Moreover, we denote $\uz := \min \ZZ$ and $\bar z := \max \ZZ$.

Recall that $Q$ is the probability transition matrix of $\{z_t\}$, with 
$Q(z,\hat z)$ denoting the probability of transitioning from $z$ to $\hat z$ in 
one step. We say that $\{z_t\}$ or $Q$ is \textit{monotone} if for all 
$z_1,z_2 \in \ZZ$ with $z_1 \leq z_2$ and all bounded increasing functions 
$h:\ZZ \to \RR$, we have
$\sum_{\hat z\in \ZZ} h(\hat z) Q(z_1, \hat z) 
\leq \sum_{\hat z\in \ZZ} h(\hat z) Q(z_2,\hat z)$.

More generally, for a given topological space $\XX$, let $\bB(\XX)$ be its 
Borel sets and $\pP(\XX)$ denote the set of probability measures on $\bB(\XX)$. 
A \emph{stochastic kernel} $P$ on $\XX$ is a map $P \colon \XX \times \bB(\XX) 
\to [0, 1]$ such that $x \mapsto P(x, B)$ is $\bB(\XX)$-measurable for each $B 
\in \bB(\XX)$, and $B \mapsto P(x, B)$ is a probability measure on $\bB(\XX)$ 
for each $x \in \XX$. Let $P^1 := P$. For all $t \in \NN$, $x, y \in \XX$, 
$B \in\bB(\XX)$, and $\mu \in \pP(\XX)$, define 
\begin{equation*}
	P^t (x, B) := \int P^{t-1}(y, B) P(x, \diff y)
	\quad \text{and} \quad 
	(\mu P^t)(B) := \int P^t (x, B) \mu(\diff x).
\end{equation*}
We call $\mu^* \in \pP(\XX)$ a \emph{stationary distribution} of $P$ if $\mu^* 
P = \mu^*$. $P$ is called \emph{Feller} if $x \mapsto \int h(y) P(x, \diff y)$ 
is continuous on $\XX$ whenever $h$ is bounded and continuous on $\XX$. 

A sequence $\{ \mu_n \} \subset \pP(\XX)$ is called \emph{tight}, if, for all 
$\epsilon>0$, there exists a compact $K \subset \XX$ such that 
$\mu_n(\XX \backslash K) \leq \epsilon$ for all $n$. We say a sequence 
$\{\mu_n\} \subset \pP(\XX)$ \emph{converges weakly to} $\mu\in \pP(\XX)$, 
denoted as $\mu_n \tow \mu$, if $\int h\diff \mu_n \to \int h \diff \mu$ 
for all bounded continuous functions $h:\XX \to \RR$. For a finite signed 
measure $\mu$ on $\bB(\XX)$, define its total variation norm 
$\|\mu\|_{TV} := \sup_{g: |g| \leq 1} \left| \int g \diff \mu \right|$. We say 
that a stochastic kernel $P$ is \textit{globally stable} if $P$ admits a unique 
stationary distribution $\mu^*\in \pP(\XX)$ and, for all $\mu\in \pP(\XX)$, we 
have $\|\mu P^t - \mu^*\|_{TV} \to 0$ as $t\to\infty$.

\begin{lemma}[Properties of the entropic risk measure]
	\label{lm:entropic}
	For $\gamma\in[0,\infty)$, define the entropic risk measure of a bounded
	random variable $X$ by
	\begin{equation*}
	\rho_\gamma(X):=
	\begin{cases}
		-\dfrac1\gamma\log\EE\,\me^{-\gamma X}, & \gamma>0,\\
		\EE X, & \gamma=0.
	\end{cases}
	\end{equation*}
	Then the following properties hold for all $\gamma\in[0,\infty)$:
	\begin{enumerate}
		\item\label{i:mono} (Monotonicity) $X\le Y$ implies 
		$\rho_\gamma(X)\le\rho_\gamma(Y)$.
		\item\label{i:concave} (Concavity)
		$\rho_\gamma(\lambda X+(1-\lambda)Y)
		\ge\lambda\rho_\gamma(X)+(1-\lambda)\rho_\gamma(Y)$
		for all $\lambda\in[0,1]$.
		\item\label{i:ge0} (Non-negativity) 
		$0=\rho_\gamma(0)\le\rho_\gamma(X)$ for all $X\ge0$.
		\item\label{i:jensen} (Jensen's inequality) 
		$\rho_\gamma(X)\le\EE X$, with equality if $\gamma=0$.
		\item\label{i:drift_inv} (Cash invariance) 
		$\rho_\gamma(X+d)=\rho_\gamma(X)+d$ for all $d\in\RR$.
		\item\label{i:lips} (Lipschitz continuity)
		$\lvert\rho_\gamma(X)-\rho_\gamma(Y)\rvert\le\lVert X-Y\rVert_\infty$.
		\item\label{i:conti_gam} (Continuity in $\gamma$) 
		For every bounded $X$, the map $\gamma\mapsto\rho_\gamma(X)$ is 
		continuous on $[0,\infty)$, and the convergence
		$\rho_\gamma(X)\to\EE X$ as $\gamma\downarrow0$ is uniform 
		over all random variables bounded by a common constant.
		\item\label{i:mono_gam} (Monotonicity in $\gamma$) 
		$0\le\gamma_1\le\gamma_2$ implies
		$\rho_{\gamma_1}(X)\ge\rho_{\gamma_2}(X)$.
	\end{enumerate}
\end{lemma}

\begin{proof}
	Properties~\eqref{i:mono}--\eqref{i:lips} follow from 
	\citet[][Section~4.1]{follmer2011stochastic} for $\gamma>0$; the case 
	$\gamma=0$ is trivial. For \eqref{i:conti_gam}, fix $K<\infty$ 
	and $X$ with $\lvert X\rvert\le K$ almost surely. Since
	$\me^{-t}=1-t+\mathcal O(t^{2})$ uniformly on bounded intervals,
	$
		\EE\,\me^{-\gamma X}=
		1-\gamma\,\EE X+\mathcal O\bigl(\gamma^{2}K^{2}\me^{\gamma K}\bigr).
	$ 
	As $|\EE X| \le \EE|X| \le K$,  
	$\delta := \EE\,\me^{-\gamma X} -1 = \mathcal O (\gamma K)$, then
	$\log (1+\delta) = \delta + \mathcal O (\delta^2)$, yielding
	\begin{equation*}
		-\frac1\gamma\log\EE\,\me^{-\gamma X}
		= -\frac1\gamma \log (1+\delta) 
		=\EE X+O\bigl(\gamma K^{2}\me^{\gamma K}\bigr)\to\EE X
		\quad \text{as }\; \gamma\downarrow0,
	\end{equation*}
	uniformly over $\lvert X\rvert\le K$. Continuity for $\gamma>0$ 
	follows directly from continuity of the exponential and logarithm.
	Regarding claim~\eqref{i:mono_gam}, set
	$\phi(\gamma):=\log\EE\,\me^{-\gamma X}$ for $\gamma>0$.
	Differentiating under the tilted measure gives
	$\phi''(\gamma)=\Var_\gamma(X)\ge0$, where
	$\operatorname{Var}_\gamma$ is variance under the reweighted 
	probability $\me^{-\gamma X}/\EE\,\me^{-\gamma X}$. So $\phi$ is
	convex with $\phi(0)=0$. Convexity implies that the secant
	slopes $\phi(\gamma)/\gamma$ are increasing in $\gamma>0$. Accordingly,
	$\rho_\gamma(X)=-\phi(\gamma)/\gamma$ is decreasing, and letting
	$\gamma\downarrow0$ gives $\rho_\gamma(X)\le\EE X=\rho_0(X)$.
\end{proof}


\section{Proof of Theorem~\ref{t:os_rs_opt} and Parametric Optimality}
\label{s:pr_opt}
In this appendix, we prove a parametric extension of 
Theorem~\ref{t:os_rs_opt}, allowing variation of both gross return $R$ and 
risk sensitivity $\gamma$. Set $\bar R:=1/\beta$. The parameter space is
$\TT:=\CC\times\G$, where $\CC\subset(0,\bar R)$ is nonempty compact, 
$\G\subset[0,\infty)$ is nonempty, and wage follows a
continuous map $R\mapsto w(R) > 0$. We establish uniform existence, 
uniqueness, and core properties of the value function and 
optimal policy for $\theta=(R,\gamma)\in\TT$.


\subsection{Setup}\label{ss:opt_setup}
Fix a compact set $\CC\subset(0,\bar R)$, a nonempty subset
$\G\subset[0,\infty)$, and set
\begin{equation*}
	\TT:=\CC\times\G
	\qquad \text{and} \qquad 
	\theta=(R,\gamma).
\end{equation*}
Let $w(R)$ be a continuous, strictly positive wage schedule on 
$(0,\bar R)$. For a candidate
value function $v$ and  consumption policy $c$, write
$v_\theta$ and $c_\theta$ for their sections at the parameter $\theta$,
i.e., $v_\theta(a,z):=v(a,z;\theta)$ and $c_\theta(a,z):=c(a,z;\theta)$.
We now construct a weighting function and a contraction modulus that are
common to all $\theta\in\TT$. Define
\begin{equation}\label{eq:params}
	R_+:=\max_{R\in\CC} R
	\qquad \text{and} \qquad 
	b_\CC:=\bar z\max_{R\in\CC}w(R)<\infty.
\end{equation}
The constant $b_\CC$ is finite because $\CC$ is compact and $w(R)$ is
continuous; it does not depend on $\gamma$. Since $\beta\in(0,1)$ and 
$R_+<\bar R=1/\beta$, we can choose $M_\CC$ such that
\begin{equation}\label{eq:MC_choice_G}
	\max\{1,R_+\}<M_\CC<1/\beta
	\qquad\Longrightarrow\qquad \beta M_\CC<1.
\end{equation}
Choose $L_\CC$ and define the weight function 
$\kappa_\CC:\RR_+\to[1,\infty)$ by
\begin{equation}\label{eq:kappa_C_G}
	L_\CC\ge\max\{1,b_\CC/(M_\CC-1)\}
	\qquad\text{and}\qquad 
	\kappa_\CC(a):=a+L_\CC.
\end{equation}
For every $\theta=(R,\gamma)\in\TT$, $(a,\hat z)\in\XX$, and $c\in[0,a]$,
we have
\begin{equation}\label{eq:pointwise_drift_C_G}
	\kappa_\CC\bigl(R(a-c)+w(R)\hat z\bigr)
	\le R_+a+b_\CC+L_\CC
	\le M_\CC(a+L_\CC)=M_\CC\kappa_\CC(a),
\end{equation}
where the last inequality follows from $R_+<M_\CC$ and
$b_\CC\le(M_\CC-1)L_\CC$. In particular,
\begin{equation}\label{eq:expected_drift_C_G}
	\max_{c\in[0,a]}\EE_z\kappa_\CC\bigl(R(a-c)+w(R)\hat z\bigr)
	\le M_\CC\kappa_\CC(a)
	\quad\text{for all }(a,\theta)\in\RR_+\times\TT.
\end{equation}
We next verify that $u$ is bounded by the common weight function. If
$0\le a\le1$, monotonicity gives $u(a)\le u(1)$. If $a\ge1$, concavity of
$u$ gives $u(a)\le u(1)+u'(1)(a-1)\le u(1)+u'(1)a$.
Thus $u(a)\le u(1)+u'(1)a$ for all $a\ge0$, which implies
\begin{equation}\label{eq:u_kappa_C_G}
	\lVert u\rVert_{\kappa_\CC}
	:=\sup_{a\ge0} [u(a)/\kappa_\CC(a)]
	\le u'(1)+u(1)/L_\CC<\infty.
\end{equation}
By Lemma~\ref{lm:entropic}, for every $\gamma\ge0$ the map 
$h\mapsto\rho_{\gamma,z}(h)$ is monotone, concave, cash-invariant, and 
$1$-Lipschitz in the sup norm, and the map 
$(\gamma,h)\mapsto\rho_{\gamma,z}(h)$ is jointly continuous on
$[0,\infty)\times\RR^{N}$ (recall that $N$ is the number of states in $\ZZ$).

Let $b_{\kappa_\CC}(\XX\times\TT)$ be the space of continuous functions
$v:\XX\times\TT\to\RR$ such that
\begin{equation*}
	\lVert v\rVert_{\kappa_\CC}
	:=\sup_{(a,z,\theta)\in\XX\times\TT}
	\frac{\lvert v_\theta(a,z)\rvert}{\kappa_\CC(a)}<\infty.
\end{equation*}
Let $\vV_\TT$ be the subset of $b_{\kappa_\CC}(\XX\times\TT)$ consisting
of functions that are concave in $a$, increasing in $a$, and take 
nonnegative values. Then $(\vV_\TT,\lVert\cdot\rVert_{\kappa_\CC})$ is 
complete, because $b_{\kappa_\CC}(\XX\times\TT)$ is a Banach space for 
any index set $\G$ (a uniform limit of bounded continuous functions is 
bounded continuous), and concavity, monotonicity, and nonnegativity are 
preserved under pointwise limits. Let $\Phi_\TT$ denote the set of 
continuous functions $c:\XX\times\TT\to\RR_+$ such that 
$c_\theta(a,z)\in[0,a]$ for all $(a,z,\theta)\in\XX\times\TT$.

A history of length $t$ is
$h_t=(a_0,z_0,c_0,\dots,a_{t-1},z_{t-1},c_{t-1},a_t,z_t)$, where
$h_0=(a_0,z_0)$. Let $H_t$ be the set of histories such that
$c_s\in[0,a_s]$ for all $s<t$. A policy $\pi=(\pi_t)_{t\ge0}$ assigns
to each history $h_t\in H_t$ a feasible action $c_t\in[0,a_t]$; $\Pi$
denotes the set of all policies, and $\Phi\subset\Pi$ the set of stationary
policies. For each $t\ge0$, let $B_{\kappa_\CC}(H_t)$ be the set of
nonnegative Borel measurable functions $v:H_t\to\RR_+$ for which there is a
constant $d_{v,t}<\infty$ such that $v(h_t)\le d_{v,t}\kappa_\CC(a_t)$ for
every $h_t\in H_t$. Fix $\theta=(R,\gamma)\in\TT$ and
$\pi=(\pi_t)_{t\ge0}\in\Pi$. Let
$\hat S^{\theta}_{\pi_t}
:=\bigl(R(a_t-\pi_t(h_t))+w(R)z_{t+1},\,z_{t+1}\bigr)$
be the next-period states.
Moreover, for $v\in B_{\kappa_\CC}(H_{t+1})$, define
\begin{equation*}
	\rho^{\theta}_{\pi_t,h_t}(v)
	:=\rho_{\gamma,z_t}\bigl(
	    v(h_t,\pi_t(h_t),\hat S^{\theta}_{\pi_t})
	\bigr)
	\quad\text{and}\quad 
	(L^{\theta}_{\pi_t}v)(h_t)
	:=u(\pi_t(h_t))+\beta\rho^{\theta}_{\pi_t,h_t}(v).
\end{equation*}
For all $\pi\in\Pi$, let $J_{n,\theta}(a,z,\pi)$ and $J_\theta(a,z,\pi)$
denote respectively the finite and infinite horizon criteria $J_n$ and
$J$ of Section~\ref{s:ifp_rsp}, evaluated at the parameter $\theta$. That is,
\begin{equation}\label{eq:J_theta}
	J_{n,\theta}(a,z,\pi):=(L^{\theta}_{\pi_0}\circ\cdots\circ L^{\theta}_{\pi_n})0(a,z)
	\quad\text{and}\quad 
	J_\theta(a,z,\pi):=\lim_{n\to\infty}J_{n,\theta}(a,z,\pi).
\end{equation}

\subsection{Main Proofs}
We aim to prove the following generalized optimality result.

\begin{theorem}[Parametric existence, uniqueness, and optimality]
	\label{t:opt_general}
	For any compact subset $\CC\subset(0,\bar R)$ and any nonempty subset
	$\G\subset[0,\infty)$ with $\TT:=\CC\times\G$, we have:
	\begin{enumerate}
		\item\label{i:vf-be}
		There exist a unique $v^*\in\vV_\TT$ and a unique
		$c^*\in\Phi_\TT$ such that, for all $(a,z,\theta)\in\XX\times\TT$,
		\begin{align}
			v_\theta^*(a,z) &= \max_{c\in[0,a]} 
			\Big\{
		        u(c) + \beta\rho_{\gamma,z}\bigl(
		            v_\theta^*(R(a-c)+w(R)\hat z,\hat z)
		        \bigr)
			\Big\} \label{eq:be_general}\\
			&= u\bigl(c_\theta^*(a,z)\bigr) 
			+\beta \rho_{\gamma,z}\bigl(
			    v_\theta^*(R(a-c_\theta^*(a,z))+w(R)\hat z,\hat z)
			\bigr).  \notag
		\end{align}
		\item\label{i:v*} 
		The mapping $v_\theta^*(a,z)$ is strictly concave in $a$, increasing in $a$, 
		and continuous.
		\item\label{i:c*} 
		Both $c_\theta^*(a,z)$ and $s_\theta^*(a,z):=a-c_\theta^*(a,z)$ are 
		increasing in $a$ and continuous.
		\item\label{i:v*_opt} 
		The mappings $(v^*_\theta,c_\theta^*)$ coincide with the 
		value function and optimal policy:
		\begin{equation*}
			v_\theta^*(a,z) = \sup_{\pi\in\Pi}J_\theta(a,z,\pi)
			= J_\theta(a,z,c_\theta^*) 
			\quad \text{for all }\; (a,z,\theta)\in\XX\times \TT.
		\end{equation*}
	\end{enumerate}
\end{theorem}

To prove Theorem~\ref{t:opt_general},
we introduce the Bellman operator $T_\TT$. For all $v\in\vV_\TT$, 
\begin{equation}\label{eq:bo_general}
	(T_\TT v)_\theta(a,z)
	:=\max_{c\in[0,a]}
	\Bigl\{
		u(c)+\beta\rho_{\gamma,z}\bigl(
		    v_\theta(R(a-c)+w(R)\hat z,\hat z)
		\bigr)
	\Bigr\}.
\end{equation}

\begin{lemma}[Self-map on the candidate value space]\label{lm:selfmap_C}
	$T_\TT$ maps $\vV_\TT$ into itself. 
\end{lemma}

\begin{proof}
	Fix $v\in\vV_\TT$. Since $\rho_{\gamma,z}(h)\ge0$ for $h\ge0$ and all
	$\gamma\ge0$, we have $(T_\TT v)_\theta(a,z)\ge0$. By Lemma~\ref{lm:entropic}\eqref{i:jensen}, for all 
	$(a,z,\theta)\in\XX\times\TT$,
	\begin{align*}
		(T_\TT v)_\theta(a,z) 
		&\le\max_{c\in[0,a]} \bigl\{
		    u(c)+\beta\,\EE_z v_\theta(R(a-c)+w(R)\hat z,\hat z)
		\bigr\} \\
		&\le u(a)+\beta\max_{c\in[0,a]}
		    \EE_z v_\theta(R(a-c)+w(R)\hat z,\hat z) \\
		&\le\lVert u\rVert_{\kappa_\CC}\kappa_\CC(a)
		    +\beta\lVert v\rVert_{\kappa_\CC}
		    \max_{c\in[0,a]}\EE_z\kappa_\CC(R(a-c)+w(R)\hat z).
	\end{align*}
	From \eqref{eq:expected_drift_C_G}, we then obtain $(T_\TT v)_\theta(a,z) \le \bigl(\lVert u\rVert_{\kappa_\CC} 
	+\beta M_\CC\lVert v\rVert_{\kappa_\CC}\bigr)\kappa_\CC(a)$. Thus
	\begin{equation}\label{eq:T_growth_C_G}
		\lVert T_\TT v\rVert_{\kappa_\CC}
		\le\lVert u\rVert_{\kappa_\CC}
		    +\beta M_\CC\lVert v\rVert_{\kappa_\CC}<\infty.
	\end{equation}
	To show that $T_\TT v$ is jointly continuous, let
	$\hat v_\theta(a,z,c)
	:=\rho_{\gamma,z}\bigl(v_\theta(R(a-c)+w(R)\hat z,\hat z)\bigr)$.
	Since $\ZZ$ is finite and $v$ is jointly continuous, the vector
	$\bigl(v_\theta(R(a-c)+w(R)\hat z,\hat z)\bigr)_{\hat z\in \ZZ}$ is 
	continuous in $(a,z,\theta,c)$. By 
	Lemma~\ref{lm:entropic}\eqref{i:conti_gam}, the map
	$(\gamma,h)\mapsto\rho_{\gamma,z}(h)$ is jointly continuous on
	$[0,\infty)\times\RR^{N}$. Hence $\hat v_\theta(a,z,c)$ is continuous 
	on $\mathrm{Gr}(\Gamma_\TT)$, where
	\begin{equation*}
		\mathrm{Gr}(\Gamma_\TT)
		:=\bigl\{(a,z,\theta,c):
		    (a,z,\theta)\in\XX\times\TT,\ c\in[0,a]
		\bigr\} 
		\;\text{ with }\; \Gamma_\TT(a,z,\theta):=[0,a].
	\end{equation*}
	Since $u$ is continuous, 
	$(a,z,\theta,c)\mapsto u(c)+\beta\hat v_\theta(a,z,c)$ is continuous. 
	Moreover, $\Gamma_\TT$ is continuous, nonempty-valued, 
	and compact-valued. By the theorem of the maximum
	\citep[see, e.g.,][Theorem~3.6]{stokey1989}, $T_\TT v$ is 
	jointly continuous on $\XX\times\TT$.
	
	Next, we show that $T_\TT v(a,z)$ is increasing in $a$. Let $a_2>a_1$. 
	Each $c\in[0,a_1]$ remains feasible at $a_2$, and
	$R(a_2-c)+w(R)\hat z\ge R(a_1-c)+w(R)\hat z$. Since $v_\theta$ is 
	increasing in $a$, the objective at $(a_2,z,\theta)$ evaluated at $c$ 
	is no smaller than the corresponding objective at $(a_1,z,\theta)$. 
	Taking maxima shows that $(T_\TT v)_\theta$ is increasing in $a$.
	
	Finally, we show that $a\mapsto(T_\TT v)_\theta(a,z)$ is concave. Fix
	$\theta\in\TT$ and let $a_1,a_2\in\RR_+$ with $a_1\ne a_2$ and
	$\lambda\in(0,1)$. Define $a:=\lambda a_1+(1-\lambda)a_2$. Let
	$c_j\in[0,a_j]$ attain the maximum in \eqref{eq:bo_general} at $a_j$, 
	$j=1,2$, and define $c:=\lambda c_1+(1-\lambda)c_2$. Then $c\in[0,a]$ and
	$
		R(a-c)+w(R)\hat z
		=\lambda\bigl(R(a_1-c_1)+w(R)\hat z\bigr)
		+(1-\lambda)\bigl(R(a_2-c_2)+w(R)\hat z\bigr).
	$
	Concavity of $v_\theta(\cdot,\hat z)$ then gives
	$
		v_\theta(R(a-c)+w(R)\hat z,\hat z)
		\ge\lambda v_\theta(R(a_1-c_1)+w(R)\hat z,\hat z)
		+(1-\lambda)v_\theta(R(a_2-c_2)+w(R)\hat z,\hat z).
	$
	By Lemma~\ref{lm:entropic}\eqref{i:mono}--\eqref{i:concave}, we obtain
	$\hat v_\theta(a,z,c) \ge\lambda\hat v_\theta(a_1,z,c_1)
	+(1-\lambda)\hat v_\theta(a_2,z,c_2)$.
	Combining this inequality with concavity of $u$,
	\begin{align*}
		(T_\TT v)_\theta(a,z)
		&\ge\lambda\bigl[u(c_1)+\beta\hat v_\theta(a_1,z,c_1)\bigr]
		+(1-\lambda)\bigl[u(c_2)+\beta\hat v_\theta(a_2,z,c_2)\bigr] \\
		&=\lambda(T_\TT v)_\theta(a_1,z)
		+(1-\lambda)(T_\TT v)_\theta(a_2,z).
	\end{align*}
	Thus, $(T_\TT v)_\theta(a,z)$ is concave in $a$. Together with 
	\eqref{eq:T_growth_C_G} and the preceding
	continuity and monotonicity arguments, we obtain 
	$T_\TT\vV_\TT\subset\vV_\TT$.
\end{proof}

\begin{remark}[Convergence to the expected-utility operator as $\gamma\downarrow0$]
	\label{rm:B_conv}
	For fixed $v\in\vV_\TT$ and $R\in\CC$, 
	Lemma~\ref{lm:entropic}\eqref{i:conti_gam} gives
	$(T_\TT v)_{(R,\gamma)}\to(T_\TT v)_{(R,0)}$ as $\gamma\downarrow0$,
	uniformly on compact subsets of $\XX$, where $(T_\TT v)_{(R,0)}$ is 
	the standard expected-utility Bellman operator obtained by setting 
	$\gamma=0$.
\end{remark}

\begin{lemma}
	\label{lm:contr_C}
	$T_\TT$ is a contraction mapping on
	$(\vV_\TT,\lVert\cdot\rVert_{\kappa_\CC})$ with modulus $\beta M_\CC<1$.
\end{lemma}

\begin{proof}
	Fix $v_1,v_2\in\vV_\TT$ and set $d:=\lVert v_1-v_2\rVert_{\kappa_\CC}$.
	By \eqref{eq:pointwise_drift_C_G}, for every feasible $(a,\theta,c,\hat z)$,
	\begin{align*}
		&\lvert (v_1)_\theta(R(a-c)+w(R)\hat z,\hat z)
		      -(v_2)_\theta(R(a-c)+w(R)\hat z,\hat z)
		\rvert \\
		&\le d\,\kappa_\CC(R(a-c)+w(R)\hat z)
		\le d\,M_\CC\kappa_\CC(a).
	\end{align*}
	The last expression is constant with respect to $\hat z$. By 
	Lemma~\ref{lm:entropic}\eqref{i:lips},
	\begin{equation*}
		\left| 
		\rho_{\gamma,z}\bigl((v_1)_\theta(R(a-c)+w(R)\hat z,\hat z)\bigr)
		-\rho_{\gamma,z}\bigl((v_2)_\theta(R(a-c)+w(R)\hat z,\hat z)\bigr)
		\right|
		\le d M_\CC\kappa_\CC(a).
	\end{equation*}
	Since $|\max_{c\in[0,a]}f(c)-\max_{c\in[0,a]}h(c)| 
	\le\max_{c\in[0,a]}|f(c)-h(c)|$ for continuous $f,h$,	
	we obtain
	$|(T_\TT v_1)_\theta(a,z)-(T_\TT v_2)_\theta(a,z)|
	\le\beta M_\CC d \,\kappa_\CC(a)$, which gives
	$\|T_\TT v_1-T_\TT v_2\|_{\kappa_\CC}
	\le\beta M_\CC \|v_1-v_2\|_{\kappa_\CC}$.
	The conclusion then follows from \eqref{eq:MC_choice_G}.
\end{proof}

\begin{lemma}[Admissibility and monotonicity of the policy operator]
	\label{lm:L_ope_C}
	For all $\theta\in\TT$, $\pi\in\Pi$, and $t\ge0$, 
	the following statements hold: 
	\begin{enumerate*}
		\item\label{i:propog_Lpi} 
		$L^{\theta}_{\pi_t}:B_{\kappa_\CC}(H_{t+1})\to B_{\kappa_\CC}(H_t)$;
		\item\label{i:mono_Lpi} 
		$(L^{\theta}_{\pi_t}v_{t+1})(h_t)
		\le(L^{\theta}_{\pi_t}\hat v_{t+1})(h_t)$
		for $h_t\in H_t$ and $v_{t+1}\le\hat v_{t+1}$.
	\end{enumerate*}
\end{lemma}

\begin{proof}
	Fix $v_{t+1}\in B_{\kappa_\CC}(H_{t+1})$. By 
	Lemma~\ref{lm:entropic}\eqref{i:ge0}--\eqref{i:jensen} and
	\eqref{eq:expected_drift_C_G}, we have
	\begin{align*}
		0&\le\rho^{\theta}_{\pi_t,h_t}(v_{t+1})
		\le\EE_t v_{t+1}(h_t,\pi_t(h_t),\hat S^{\theta}_{\pi_t}) \\
		&\le d_{v,t+1}\EE_t\kappa_\CC\bigl(R(a_t-\pi_t(h_t))+w(R)z_{t+1}\bigr)
		\le M_\CC d_{v,t+1}\kappa_\CC(a_t).
	\end{align*}
	Moreover, by monotonicity of $u$ and \eqref{eq:u_kappa_C_G}, we have,
	$u(\pi_t(h_t))\le u(a_t)\le\lVert u\rVert_{\kappa_\CC}\kappa_\CC(a_t)$.
	Combining the inequalities above yields
	$(L^{\theta}_{\pi_t}v_{t+1})(h_t)
	\le\bigl(\lVert u\rVert_{\kappa_\CC}
	+\beta M_\CC d_{v,t+1}\bigr)\kappa_\CC(a_t)$.
	Hence claim~\eqref{i:propog_Lpi} holds. Claim~\eqref{i:mono_Lpi} follows 
	immediately from Lemma~\ref{lm:entropic}\eqref{i:mono}.
\end{proof}

\begin{lemma}[Uniform bounds for the recursive criterion]
	\label{lm:J_ope_C}
	For all $(a,z,\theta)\in\XX\times\TT$ and $\pi\in\Pi$, the sequence
	$\{J_{n,\theta}(a,z,\pi)\}_{n\ge0}$ is increasing and
	\begin{equation}\label{eq:J_bound_C}
		0\le J_{n,\theta}(a,z,\pi)
		\le\frac{\lVert u\rVert_{\kappa_\CC}}{1-\beta M_\CC}
		\kappa_\CC(a)
		\quad\text{for all }\; n\ge0.
	\end{equation}
	Moreover, $J_\theta$ is well defined.
\end{lemma}

\begin{proof}
	Fix $(a,z,\theta)\in\XX\times\TT$ and $\pi\in\Pi$. Since $u$ is 
	nonnegative, the definition of $J_{n,\theta}$ and 
	Lemma~\ref{lm:entropic}\eqref{i:ge0} imply
	$J_{n,\theta}(a,z,\pi)\ge0$. To see that 
	$\{J_{n,\theta}(a,z,\pi)\}_{n\ge0}$ is
	increasing, by Lemma~\ref{lm:L_ope_C},
	\begin{equation*}
		J_{n,\theta}(a,z,\pi)
		=(L^{\theta}_{\pi_0}\circ\cdots\circ L^{\theta}_{\pi_n})0(a,z)
		\le(L^{\theta}_{\pi_0}\circ\cdots\circ L^{\theta}_{\pi_n})
		    (L^{\theta}_{\pi_{n+1}}0)(a,z)
		=J_{n+1,\theta}(a,z,\pi).
	\end{equation*}
	Set $D_\CC:=\lVert u\rVert_{\kappa_\CC}/(1-\beta M_\CC)$. Since
	$\beta M_\CC<1$, $D_\CC\ge\lVert u\rVert_{\kappa_\CC}$. Hence, for all
	$h_n\in H_n$,
	\begin{equation*}
		L^{\theta}_{\pi_n}0(h_n)=u(\pi_n(h_n))
		\le u(a_n)\le\lVert u\rVert_{\kappa_\CC}\kappa_\CC(a_n)
		\le D_\CC\kappa_\CC(a_n).
	\end{equation*}
	The monotonicity of $L_{\pi_{n-1}}^\theta$,
	Lemma~\ref{lm:entropic}\eqref{i:jensen}, and
	\eqref{eq:expected_drift_C_G} then imply that
	\begin{align*}
		&L^{\theta}_{\pi_{n-1}}(L^{\theta}_{\pi_n}0)(h_{n-1})
		\le L^{\theta}_{\pi_{n-1}}(D_\CC\kappa_\CC)(h_{n-1}) \\
		&\le u(\pi_{n-1}(h_{n-1})) +\beta D_\CC
		\EE_{n-1}\kappa_\CC\bigl(
		    R(a_{n-1}-\pi_{n-1}(h_{n-1}))+w(R)z_n
		\bigr)  \\
		&\le\lVert u\rVert_{\kappa_\CC}\kappa_\CC(a_{n-1})
		+\beta D_\CC\max_{c\in[0,a_{n-1}]}
		\EE_{n-1}\kappa_\CC(R(a_{n-1}-c)+w(R)z_n) \\
		&\le\bigl(\lVert u\rVert_{\kappa_\CC}
		+\beta M_\CC D_\CC\bigr)\kappa_\CC(a_{n-1})
		=D_\CC\kappa_\CC(a_{n-1}).
	\end{align*}
	where the last equality follows from the definition of $D_\CC$.
	Continuing backward gives 
	$(L^{\theta}_{\pi_t}\circ\cdots\circ L^{\theta}_{\pi_n})0(h_t)
	\le D_\CC\kappa_\CC(a_t)$ for $t=n,n-1,\dots,0$. In particular,
	\begin{align*}
		&J_{n,\theta}(a,z,\pi)
		= (L_{\pi_0}^\theta\circ\cdots\circ L_{\pi_n}^\theta)0(a,z)
		\leq D_\CC\kappa_\CC(a)
		= \|u\|_{\kappa_\CC} \kappa_\CC(a) / (1-\beta M_\CC) \\
		&\Longrightarrow \qquad 
		0 \leq J_{n,\theta}(a,z,\pi)
		\leq \|u\|_{\kappa_\CC} \kappa_\CC(a) / (1-\beta M_\CC)
		\quad\text{for all }\; n\geq0.
	\end{align*}
	Since $\kappa_\CC,M_\CC,\lVert u\rVert_{\kappa_\CC}$ do not depend on
	$\theta\in\TT$, the bound is uniform over $\theta$. So the increasing 
	sequence is bounded above with a finite limit. 
	Hence $J_\theta$ is well defined.
\end{proof}

\begin{proof}[Proof of Theorem~\ref{t:opt_general}]
	
	By Lemma~\ref{lm:contr_C} and \eqref{eq:MC_choice_G},
	the Banach fixed point theorem gives a unique fixed point
	$v^*\in\vV_\TT$.
	Membership in $\vV_\TT$ implies joint continuity on
	$\XX\times\TT$ and concavity in wealth.
	Define the continuation return
	\begin{equation*}
	H_\theta(s,z):=
	\beta\rho_{\gamma,z}
	\bigl(v_\theta^*(Rs+w(R)\hat z,\hat z)\bigr).
	\end{equation*}
	It is jointly continuous by continuity of $v^*$ and $w$,
	finiteness of $\ZZ$, and
	Lemma~\ref{lm:entropic}\eqref{i:conti_gam}.
	Monotonicity and concavity of the entropic expectation
	also imply concavity in $s$.
	Thus the Bellman objective
	$u(c)+H_\theta(a-c,z)$ is jointly continuous and
	strictly concave in $c$.
	It has a unique maximizer $c_\theta^*(a,z)$ on $[0,a]$.
	The maximum theorem gives joint continuity of $c^*$
	and hence of $s_\theta^*(a,z):=a-c_\theta^*(a,z)$.
	This proves claim~\eqref{i:vf-be}, the continuity part
	of claim~\eqref{i:c*}, and claim~\eqref{i:v*}
	except for strict concavity.
	
	To prove policy monotonicity, fix $(z,\theta)$.
	Consider a unique maximizer $q(a)$ of
	$\phi(q)+\psi(a-q)$ over $q\in[0,a]$, with $\psi$ concave.
	If $a_1<a_2$ but $q_1:=q(a_1)>q_2:=q(a_2)$,
	both choices are feasible at both wealth levels.
	Concavity of $\psi$ makes
	\begin{equation*}
	D(a):=\phi(q_1)-\phi(q_2)
	+\psi(a-q_1)-\psi(a-q_2)
	\end{equation*}
	increasing on $[a_1,a_2]$.
	Unique optimality gives $D(a_1)>0>D(a_2)$,
	a contradiction.
	Applying this argument with
	$(\phi,\psi)=(u,H_\theta(\cdot,z))$ proves
	monotonicity of consumption, and reversing these roles
	proves monotonicity of saving.
	Hence claim~\eqref{i:c*} holds.
	
	Regarding claim~\eqref{i:v*_opt} , fix $\theta\in\TT$ and $\pi\in\Pi$.
	The Bellman inequality, monotonicity of the policy
	operators, and $v_\theta^*\ge0$ imply
	$
	v_\theta^*
	\ge
	(L_{\pi_0}^{\theta}\circ\cdots\circ
	L_{\pi_n}^{\theta})v_\theta^*
	\ge
	(L_{\pi_0}^{\theta}\circ\cdots\circ
	L_{\pi_n}^{\theta})0
	=J_{n,\theta}(\cdot,\pi).
	$
	Taking limits and then the supremum over policies gives
	\begin{equation}\label{eq:value_upper_C}
		v_\theta^*\ge\sup_{\pi\in\Pi}J_\theta(\cdot,\pi).
	\end{equation}
	Let $L_{c_\theta^*}$ be the stationary policy operator
	\begin{equation*}
	(L_{c_\theta^*}f)(a,z)
	:=u(c_\theta^*(a,z))
	+\beta\rho_{\gamma,z}
	\bigl(f(Rs_\theta^*(a,z)+w(R)\hat z,\hat z)\bigr).
	\end{equation*}
	The Bellman equation gives
	$L_{c_\theta^*}v_\theta^*=v_\theta^*$.
	The estimate in Lemma~\ref{lm:contr_C} applies to this
	operator on $b_{\kappa_\CC}(\XX)$ with the same modulus
	$r:=\beta M_\CC<1$.
	Since
	$J_{n,\theta}(\cdot,c_\theta^*)=L_{c_\theta^*}^{\,n+1}0$,
	\begin{equation*}
	\|v_\theta^*-J_{n,\theta}(\cdot,c_\theta^*)\|_{\kappa_\CC}
	\le r^{n+1}\|v_\theta^*\|_{\kappa_\CC}\to 0
	\quad \text{as }\; n\to \infty.
	\end{equation*}
	Hence, $J_\theta(\cdot,c_\theta^*)=v_\theta^*$.
	Together with \eqref{eq:value_upper_C}, this proves
	claim~\eqref{i:v*_opt}.

	It remains to prove strict concavity of $v^*$ in wealth.
	Fix $\theta=(R,\gamma)\in\TT$ and abbreviate
	$v:=v_\theta^*$, $c:=c_\theta^*$, and
	$\rho_z:=\rho_{\gamma,z}$.
    Since additional wealth can be consumed immediately without
    changing saving or future states, $v(\cdot,z)$ is
    strictly increasing for every $z\in\ZZ$.
    Suppose strict concavity fails. Concavity then implies 
    $v(m,z)=[v(a,z)+v(b,z)]/2$
    for some $z\in\ZZ$, $a<b$, and $m=(a+b)/2$.
    Write $c_a=c(a,z)$, $c_b=c(b,z)$, and define
    $x_j:=R(a-c_a)+w(R)z_j$, $y_j:=R(b-c_b)+w(R)z_j$,
    $U_j=v(x_j,z_j)$, $V_j=v(y_j,z_j)$, and
    $M_j=v((x_j+y_j)/2,z_j)$.
    The feasible choice $(c_a+c_b)/2$ at $m$ gives
    \begin{equation*}
    v(m,z)
    \ge u\left(\frac{c_a+c_b}{2}\right)+\beta\rho_z(M)
    \ge \frac{u(c_a)+u(c_b)}2
    +\frac{\beta}{2}\bigl[\rho_z(U)+\rho_z(V)\bigr]
    = v(m,z).
    \end{equation*}
    The second inequality follows from concavity of $u$ and $v$,
    and monotonicity and concavity of $\rho_z$.
    Equality throughout and strict concavity of $u$ imply $c_a=c_b$.
    Moreover, $M\ge(U+V)/2$, so strict monotonicity of $\rho_z$
    on components with positive transition probability implies
    $M_j=(U_j+V_j)/2$ whenever $Q(z,z_j)>0$.
    
    As $y_j-x_j=R(b-a)>0$, the same argument applies at
    each such successor pair.
    Inductively, the optimal wealth paths from $a$ and $b$
    have identical consumption at every date along every
    income history with positive probability.
    Their finite-horizon recursive utilities with zero terminal
    continuation therefore coincide:
    $J_{n,\theta}(a,z,c_\theta^*)
    =J_{n,\theta}(b,z,c_\theta^*)$ for every $n$.
    Taking limits and using the optimal policy verification 
    gives $v(a,z)=v(b,z)$, contradicting strict increase of $v_\theta^*$.
    Thus $v_\theta^*(\cdot,z)$ is strictly concave for every
    $(z,\theta)\in\ZZ\times\TT$.
\end{proof}

\begin{proof}[Proof of Theorem~\ref{t:os_rs_opt}]
	This follows directly from Theorem~\ref{t:opt_general} with $\CC=\{R\}$, $\G=\{\gamma\}$, and $w=w(R)$.
\end{proof}

\begin{lemma}[Monotonicity of the value function in productivity]
	\label{lm:mono_productivity}
	If the transition matrix $Q$ is monotone, then, for all
	$\theta\in\TT$, the value function $v^*_\theta(a,z)$ from 
	Theorem~\ref{t:opt_general} is increasing in $z$.
\end{lemma}

\begin{proof}
	Set $v_0\equiv0$ and $v_{n+1}:=T_\TT v_n$ for $n\geq0$. 
	By Lemma~\ref{lm:contr_C}, these iterates converge
	pointwise to $v^*$, so it suffices to prove the claim
	for each $v_n$. 
	The claim holds for $n=0$.  Suppose it holds for $v_n$.
	Fix $\theta=(R,\gamma)$, $a\ge0$, and $c\in[0,a]$.
	Since $v_{n,\theta}$ is increasing in both wealth
	and productivity,
	$h(\hat z):=v_{n,\theta}(R(a-c)+w(R)\hat z,\hat z)$
	is increasing.
	For $\gamma>0$, monotonicity of $Q$ makes
	$\EE_z\me^{-\gamma h(\hat z)}$ decreasing in $z$,
	hence $\rho_{\gamma,z}(h)$ increasing in $z$.
	For $\gamma=0$, the same conclusion follows directly
	from monotonicity of $\EE_z h$.
	Thus the Bellman objective is increasing in $z$
	for every feasible $c$.
	Maximizing over the common set $[0,a]$ proves the
	claim for $v_{n+1}$ and completes the induction.
\end{proof}

\section{Proof of Proposition~\ref{pr:euler}}\label{s:pr_euler}
We return to the fixed-price setup of 
Sections~\ref{ss:preference}--\ref{ss:or}: gross return $R$, 
wage $w$, and risk sensitivity $\gamma\ge0$ are
all held fixed, and we adopt the main-text notation instead of 
the parametric notation of Appendix~\ref{s:pr_opt}. Formally, this is the 
special case $\CC=\{R\}$ and $\G=\{\gamma\}$
with $w=w(R)$. We thus apply the results of 
Appendix~\ref{s:pr_opt} with $\TT=\CC\times \G$, 
so the parametric objects 
$(\vV_\TT,v^*_\theta,c^*_\theta,v_\theta,c_\theta)$ reduce to
$(\vV,v^*,c^*,v,c)$. We also write the Bellman operator
$T_\TT$ from \eqref{eq:bo_general} simply as
\begin{equation*}
	T v(a,z)
	=\max_{c\in[0,a]} \Bigl\{
		u(c)+\beta\rho_{\gamma,z}\bigl(v(R(a-c)+w\hat z,\hat z)\bigr)
	\Bigr\},
\end{equation*}
which for $\gamma>0$ coincides with
\begin{equation}\label{eq:bo_rs}
	T v(a,z) =\max_{c\in[0,a]} \left\{
		u(c)-\frac{\beta}{\gamma}\log\EE_z
		\me^{-\gamma v(R(a-c)+w\hat z,\hat z)}
	\right\}.
\end{equation}
To derive the Euler equation uniformly for $\gamma\ge0$, define, 
for $z\in \ZZ$, $s\ge0$, and $\hat z\in \ZZ$,
\begin{equation}\label{eq:infh_tilt}
	f_\gamma(s,z,\hat z) :=
	\begin{cases}
		\dfrac{\me^{-\gamma v^*(Rs+w\hat z,\hat z)}}
		{\EE_z\me^{-\gamma v^*(Rs+w\hat z,\hat z)}}, 
		& \gamma>0, \vspace{.5mm}\\
		1, & \gamma=0.
	\end{cases}
\end{equation}
Then $f_\gamma(s,z,\cdot)$ is a probability density with respect to
$Q(z,\cdot)$. The Euler equation is
\begin{equation*}
	u'(c^*(a,z)) =\max\Bigl\{
		\beta R\,\EE_z\bigl[u'(c^*(\hat a,\hat z))\,
		f_\gamma(a-c^*(a,z),z,\hat z)\bigr],\, u'(a)
	\Bigr\},
\end{equation*}
where $\hat a=R(a-c^*(a,z))+w\hat z$. 


\begin{lemma}[Differentiability]
	\label{lm:T_diff}
	For any $v \in \vV$, let $c$ be the unique optimizer that attains the 
	maximum defining $Tv$. Then 
	$\partial Tv(a,z)/\partial a = u'(c(a,z))$ for all $a>0$ and $z\in\ZZ$. 
	In particular, $\partial v^*(a,z)/\partial a = u'(c^*(a,z))$ 
	for all $a > 0$ and $z \in \ZZ$.
\end{lemma}

\begin{proof}
	For each $v\in\vV$, the Bellman operator can be written equivalently as
	\begin{equation}\label{eq:be_equiv}
		Tv(a,z)=\sup_{s\in[0,a]}W(a,z,s)
		\quad\text{with}\quad
		W(a,z,s)=u(a-s)+\beta\rho_{\gamma,z}
		\bigl(v(Rs+w\hat z,\hat z)\bigr).
	\end{equation}
	Let $s(a,z):=a-c(a,z)$ be the unique maximizer
	in~\eqref{eq:be_equiv}. Fix $a>0$ and $z\in\ZZ$.
	Since $u'(0)=\infty$, we have $c(a,z)>0$ and $s(a,z)<a$.
	Choose $0<\varepsilon<c(a,z)$.
	For $|x-a|<\varepsilon$, we have $0\le s(a,z)<x$,
	so $s(a,z)$ is feasible at wealth $x$. Hence
	\begin{equation*}
		W(x,z,s(a,z))\le Tv(x,z)
		\quad \text{and} \quad 
		W(a,z,s(a,z))=Tv(a,z).
	\end{equation*}
	Since $u$ is concave and continuously differentiable on $\RR_{++}$, 
	$x\mapsto W(x,z,s(a,z))$ is concave and continuously
	differentiable on $(a-\varepsilon,a+\varepsilon)$.
	By Lemma~\ref{lm:selfmap_C}, $Tv(\cdot,z)$ is concave,
	hence \citet[Lemma~1]{benveniste1979} yields differentiability  
	at $a$, with $\partial Tv(a,z)/\partial a =u'(a-s(a,z))=u'(c(a,z))$.
	Since $Tv^*=v^*$, applying this identity to $v^*$
	establishes the second claim.
\end{proof}

\begin{proof}[Proof of Proposition~\ref{pr:euler}]
	Define
	$
		\hat v(s,z):=\rho_{\gamma,z}
		\bigl(v^*(Rs+w\hat z,\hat z)\bigr).
	$
	Since $\ZZ$ is finite and $Rs+w\hat z\ge w\underline z>0$,
	Lemma~\ref{lm:T_diff} and direct differentiation give
	\begin{equation}\label{eq:part_vhat}
		\hat v'(s,z)
		:= \partial \hat v(s,z)/\partial s
		=R\EE_z\bigl[
		u'(c^*(Rs+w\hat z,\hat z))
		f_\gamma(s,z,\hat z)
		\bigr],
	\end{equation}
	where the derivative at $s=0$ is taken from the right.
	For $\gamma>0$, this follows by differentiating the
	logarithm of the finite exponential sum; for $\gamma=0$,
	it follows by differentiating the finite expectation,
	with $f_0\equiv1$.
	
	Fix $a>0$ and set $s^*:=a-c^*(a,z)$.
	Optimal consumption is positive, so $s^*<a$.
	If $s^*>0$, the first-order condition for
	$u(a-s)+\beta\hat v(s,z)$ gives
	\begin{equation*}
		u'(c^*(a,z))
		=\beta\hat v'(s^*,z)
		=\beta R\EE_z\bigl[
		u'(c^*(\hat a,\hat z))f_\gamma(s^*,z,\hat z)
		\bigr],
	\end{equation*}
	where $\hat a:=Rs^*+w\hat z$.
	Since $c^*(a,z)<a$, strict concavity of $u$ also gives
	$u'(c^*(a,z))>u'(a)$.
	If $s^*=0$, optimality at the lower boundary gives
	$u'(a)\ge\beta\hat v'_+(0,z)$, and
	$c^*(a,z)=a$.
	Substituting \eqref{eq:part_vhat} in these two cases
	yields \eqref{eq:euler}.
	At $a=0$, feasibility gives $c^*(0,z)=0$, and the
	equation holds with $u'(0)=+\infty$.
\end{proof}

\section{Proof of Proposition~\ref{pr:compact} and Theorem~\ref{t:gs_rs}}
\label{s:pr_stability}

In this appendix, we prove Proposition~\ref{pr:compact} and 
Theorem~\ref{t:gs_rs}. We shall need both the fixed-price objects of 
Section~\ref{s:ifp_rsp} and the parametric objects of 
Appendix~\ref{s:pr_opt}.

\subsection{Key Notation}
Let $\mathsf X_0:=(0,\infty)\times\mathsf Z$. In the
fixed-price and fixed-$\gamma$ setting of Section~\ref{s:ifp_rsp}, 
let $c(a,z):=c^*(a,z)$ be the optimal consumption policy, and 
define
\begin{equation}\label{eq:g_fixed}
	g(a,z,\hat z):=R[a-c(a,z)]+w\hat z,
	\qquad (a,z,\hat z)\in\mathsf X\times\mathsf Z,
\end{equation}
which is increasing in $a$ by Theorem~\ref{t:os_rs_opt}.
For the parametric results, fix a compact subset
$\CC\subset(0,\bar R)$, where $\bar R:=1/\beta$, and let 
$R\mapsto w(R)>0$ be a continuous wage schedule on $\CC$. Set
$\TT:=\CC\times[0,\infty)$ and $\theta=(R,\gamma)$.
For $\theta\in\TT$, let $T_\theta$ denote the Bellman 
operator~\eqref{eq:bo_general} of Appendix~\ref{s:pr_opt}. 
Define the finite-horizon value functions
\begin{equation*}
	v_{0,\theta}:=0
	\qquad \text{and} \qquad 
	v_{n+1,\theta}:=T_\theta v_{n,\theta},\quad n\ge0.
\end{equation*}
For $n\ge1$, let $c_{n,\theta}(a,z)$ denote the unique maximizer in
$v_{n,\theta}=T_\theta v_{n-1,\theta}$, and define
\begin{equation*}
	s_{n,\theta}(a,z):=a-c_{n,\theta}(a,z)
	\quad\text{and}\quad 
	g_{n,\theta}(a,z,\hat z):=R\,[a-c_{n,\theta}(a,z)]+w(R)\hat z.
\end{equation*}
For the infinite-horizon objects, write
\begin{equation*}
	c_\theta(a,z):=c_\theta^*(a,z),\;\;
	s_\theta(a,z):=a-c_\theta(a,z),\;\;
	g_\theta(a,z,\hat z):=R\,[a-c_\theta(a,z)]+w(R)\hat z.
\end{equation*}
Clearly, the fixed-price and fixed-$\gamma$ objects are the special case 
$\CC=\{R\}$ and $\G=\{\gamma\}$ with $w(R)=w$, so that $c=c_\theta$ and 
$g=g_\theta$. Lemma~\ref{lm:finite_horizon} and Proposition~\ref{pr:com_sup}
below are stated in the parametric notation, while Lemmas~\ref{lm:threshold}, 
\ref{lm:binding_ft}--\ref{lm:mono_prob} and the proofs of 
Proposition~\ref{pr:compact} and Theorem~\ref{t:gs_rs} use the fixed-price
and fixed-$\gamma$ notation.

\subsection{Proofs}
As in Appendix~\ref{s:pr_opt}, we prove a uniform parametric extension of 
Proposition~\ref{pr:compact}. The following lemma shows that, except in 
extreme cases, the borrowing constraint is binding if and only if assets 
fall below a certain threshold.

\begin{lemma}[Threshold for saving decision]
	\label{lm:threshold}
	Given $z\in \ZZ$, either (i) $c(a,z) = a$ for all $a\geq 0$, or 
	(ii) there exists a threshold $\bar a(z) \in (0,\infty)$ 
	such that $c(a,z) = a$ if and only if $a\leq \bar a(z)$.
\end{lemma}

\begin{proof}
	Fix $z\in \ZZ$. Suppose (i) does not hold. Notice that $c(a,z) > 0$ 
	for all $(a,z) \in \XX_0$ since $u'(0) = \infty$. Since in addition 
	$\hat a = R (a - c(a,z)) + w\hat z \geq w \hat z \geq w \uz>0$,
	we have $u'(c(\hat a,\hat z))\le\max_{z \in \ZZ}u'(c(w\uz, z))<\infty$. 
	Hence
	\begin{equation*}
		\beta R \EE_z u'(c(\hat a,\hat z)) 
		\me^{-\gamma v^*(\hat a, \hat z)} /
		\EE_z \me^{-\gamma v^*(\hat a, \hat z)}
		\leq \beta R \max_{z \in \ZZ} u'(c(w\underline z, z)) < \infty.
	\end{equation*}
	The Euler equation then implies that $c(a,z) = a$ when $a$ is 
	sufficiently close to $0$. Since in addition $a \mapsto a - c(a,z)$ is 
	increasing and continuous by Theorem~\ref{t:os_rs_opt}, and $c(a,z)<a$ 
	for some $a>0$ when (i) does not hold, we have
	\begin{equation*}
		\bar a(z) := \max \{a\in (0,\infty): c(a,z) = a\} < \infty.
	\end{equation*}
	By definition, $c(a,z) = a$ if and only if $a\leq \bar a(z)$. Hence, 
	we have shown that claim~(ii) holds if claim~(i) does not hold. 
	The proof is now complete.
\end{proof}

\begin{lemma}[Finite-horizon Bellman iterations]
	\label{lm:finite_horizon}
	Let $\CC\subset(0,\bar R)$ be a compact subset, 
	$\G\subset[0,\infty)$ be nonempty, and set $\TT:=\CC\times\G$. 
	For $v\in \vV_\TT$, let $T_\TT$ be as
	in \eqref{eq:bo_general}. Define $v_{0,\theta}:=0$ and
	$v_{n+1,\theta}:=T_\TT v_{n,\theta}$ for $n\ge0$. For $n\ge1$, 
	let $c_{n,\theta}$ denote the unique maximizer in
	$v_{n,\theta}=T_\TT v_{n-1,\theta}$, and set
	$s_{n,\theta}(a,z):=a-c_{n,\theta}(a,z)$. For $n\ge0$, define
	\begin{equation}\label{eq:fh_tilt}
		f_{n,\theta}(s,z,\hat z):=
		\begin{cases}
			\dfrac{\me^{-\gamma v_{n,\theta}(Rs+w(R)\hat z,\hat z)}}
			{\EE_z\me^{-\gamma v_{n,\theta}(Rs+w(R)\hat z,\hat z)}}, 
			& \gamma>0,\vspace{.5mm} \\ 
			1, & \gamma=0.
	\end{cases}
	\end{equation}
	Then, for all $n\ge0$ and $\theta\in\TT$:
	\begin{enumerate}
		\item\label{i:vn}
		$v_{n,\theta}$ is nonnegative, jointly continuous, 
		increasing in $a$, and concave in $a$.
		\item\label{i:csn} 
		For $n\ge1$, $c_{n,\theta}$ and $s_{n,\theta}$ are continuous,
		nonnegative, and increasing in $a$.
		\item\label{i:fh_euler} 
		For $n\ge1$, whenever $0<c_{n+1,\theta}(a,z)<a$, with
		$s:=a-c_{n+1,\theta}(a,z)$,
		\begin{equation*}
		    u'(c_{n+1,\theta}(a,z))
			=\beta R\,\EE_z\Bigl[
			    u'\bigl(c_{n,\theta}(Rs+w(R)\hat z,\hat z)\bigr)\,
			    f_{n,\theta}(s,z,\hat z)
			\Bigr].
		\end{equation*}
		\item\label{i:fh_to_infh}
		$v_{n,\theta}\to v_\theta$ uniformly on
		compact subsets of $\XX$, and
		$c_{n,\theta}(a,z)\to c_\theta(a,z)$ and
		$s_{n,\theta}(a,z)\to s_\theta(a,z)$
		for all $(a,z)\in \XX$, where $v_\theta$ and $c_\theta$ are the 
		fixed point and optimal policy of Theorem~\ref{t:opt_general}.
	\end{enumerate}
\end{lemma}

\begin{proof}
	Since $v_{0,\theta}=0\in \vV_\TT$ and $T_\TT \vV_\TT \subset \vV_\TT$ 
	by Lemma~\ref{lm:selfmap_C}, induction gives 
	$v_{n,\theta}\in \vV_\TT$ for all $n$, which is claim~\eqref{i:vn}. 
	For claim~\eqref{i:csn}, fix $n\ge1$. The maximizer in
	$v_{n,\theta}=T_\TT v_{n-1,\theta}$ exists uniquely by strict concavity 
	of $u$ and concavity of the continuation value in $c$ (as in the proof of
	Theorem~\ref{t:opt_general}); the maximum theorem gives continuity of 
	$c_{n,\theta}$, and the same reasoning from  
	Theorem~\ref{t:opt_general}\eqref{i:c*} implies $c_{n,\theta}$ and
	$s_{n,\theta}$ are increasing in $a$.
	
	For claim~\eqref{i:fh_euler}, fix $n\ge1$ and $\theta\in\TT$. 
	Lemma~\ref{lm:T_diff}, applied at the fixed parameter $\theta$ to the 
	function $v_{n-1,\theta}$, gives
	$\partial v_{n,\theta}(a,z)/\partial a=u'(c_{n,\theta}(a,z))$ for $a>0$.
	The derivative of the map
	$s\mapsto\rho_{\gamma,z}\bigl(v_{n,\theta}(Rs+w(R)\hat z,\hat z)\bigr)$ is
	computed exactly as in~\eqref{eq:part_vhat} of Appendix~\ref{s:pr_euler}, 
	with $v^*$ replaced by $v_{n,\theta}$ and $f_\gamma$ replaced by 
	$f_{n,\theta}$:
	\begin{equation*}
		\frac{\partial \rho_{\gamma,z}\bigl(
			v_{n,\theta}(Rs+w(R)\hat z,\hat z)
		\bigr)}{\partial s}
		= R\,\EE_z\left[
			\frac{\partial v_{n,\theta}(Rs+w(R)\hat z,\hat z)}{\partial a}
			f_{n,\theta}(s,z,\hat z)
		\right].
	\end{equation*}
	The first-order condition for $c_{n+1,\theta}$ in
	$v_{n+1,\theta}=T_\TT v_{n,\theta}$ then gives claim~\eqref{i:fh_euler}.
	
	For claim~\eqref{i:fh_to_infh}, $v_{n,\theta}=T_\TT^n0$. By 
	Lemma~\ref{lm:contr_C}, $T_\TT$ is a contraction with modulus 
	$\beta M_\CC<1$, so $\lVert v_{n,\theta}-v_\theta\rVert_{\kappa_\CC}\to0$, 
	hence $v_{n,\theta}\to v_\theta$ uniformly on compact subsets of $\XX$. 
	For the policy convergence, fix $\theta\in\TT$, $a>0$, $z\in \ZZ$, 
	and write
	\begin{equation*}
		\rho_{n,\theta}(s,z)
		:=\beta\rho_{\gamma,z}\bigl(v_{n,\theta}(Rs+w(R)\hat z,\hat z)\bigr)
		\quad\text{and}\quad 
		\rho_\theta(s,z)
		:=\beta\rho_{\gamma,z}\bigl(v_\theta(Rs+w(R)\hat z,\hat z)\bigr).
	\end{equation*}
	By Lemma~\ref{lm:entropic}\eqref{i:lips}, we have
	\begin{equation*}
		\sup_{s\in[0,a]}\lvert\rho_{n,\theta}(s,z)-\rho_\theta(s,z)\rvert
		\le\beta\sup_{(x,\hat z)\in[0,Ra+w(R)\bar z]\times \ZZ}
		\lvert v_{n,\theta}(x,\hat z)-v_\theta(x,\hat z)\rvert\to0.
	\end{equation*}
	Consider any convergent subsequence 
	$s_{n_k+1,\theta}(a,z)\to\bar s\in[0,a]$. Set
	\begin{equation*}
		\varepsilon_k
		:=\sup_{s\in[0,a]}\lvert
		\rho_{n_k,\theta}(s,z)-\rho_\theta(s,z)
		\rvert\to0.
	\end{equation*}
	By optimality of $s_{n_k+1,\theta}(a,z)$, for all $s\in[0,a]$, 
	we have
	\begin{align*}
		&u(a-s_{n_k+1,\theta}(a,z))
		+\rho_{n_k,\theta}(s_{n_k+1,\theta}(a,z),z)
		\ge u(a-s)+\rho_{n_k,\theta}(s,z) \\
		\Longrightarrow \quad 
		&u(a-s_{n_k+1,\theta}(a,z))+\rho_\theta(s_{n_k+1,\theta}(a,z),z)
		\ge u(a-s)+\rho_\theta(s,z)-2\varepsilon_k.
	\end{align*}
	Letting $k\to\infty$ and using continuity gives
	\begin{equation*}
		u(a-\bar s)+\rho_\theta(\bar s,z)
		\ge u(a-s)+\rho_\theta(s,z)
		\quad\text{for all }\; s\in[0,a].
	\end{equation*}
	Thus $\bar s$ maximizes the limiting objective; uniqueness gives
	$\bar s=s_\theta(a,z)$. Since $[0,a]$ is compact, the whole sequence
	converges, so $s_{n,\theta}(a,z)\to s_\theta(a,z)$ and
	$c_{n,\theta}(a,z)\to c_\theta(a,z)$. At $a=0$ both policies equal zero.
	This proves claim~\eqref{i:fh_to_infh}.
\end{proof}

\begin{proposition}[Common compact support]\label{pr:com_sup}
	Let $\CC\subset(0,\bar R)$ be any compact subset. 
	The following statements hold:
	\begin{enumerate}
		\item\label{i:invariance_para}\textup{(Invariance).}
		There exists an $a_\CC<\infty$, independent of $\gamma\ge0$
		and $n\ge1$, such that $\XX_\CC:=[0,a_\CC]\times\ZZ$
		is invariant under every finite-horizon and infinite-horizon
		optimal policy. That is, for any
		$(a,z,R,\hat z)\in\XX_\CC\times\CC\times\ZZ$,
		$\gamma\ge0$, and $n\ge1$, we have
		$g_{n,\theta}(a,z,\hat z)\le a_\CC$ and 
		$g_{\theta}(a,z,\hat z)\le a_\CC.$
		\item\label{i:absorption_para}\textup{(Absorption).}
		Under each infinite-horizon optimal policy, $\XX_\CC$ is
		reached almost surely in finite time from any initial state
		$(a,z)\in\XX$, for all $R\in\CC$ and $\gamma\ge0$.
		\item\label{i:stationary_support_para}\textup{(Stationary support).}
		Every stationary distribution $\mu_{\theta}$ of the optimal
		state process satisfies $\mu_{\theta}(\XX_\CC)=1$,
		for every $R\in\CC$ and $\gamma\ge0$.
	\end{enumerate}
\end{proposition}

\begin{proof}
	By Assumption~\ref{a:bip_cond}, the absolute risk aversion
	$\eta(a):=-u''(a)/u'(a)$ converges to zero as $a\to\infty$.
	Hence, for every fixed $h>0$,
	\begin{equation*}
		\log \frac{u'(a+h)}{u'(a)}
		=\int_a^{a+h}\frac{u''(t)}{u'(t)}\,\diff t
		=-\int_a^{a+h}\eta(t)\,\diff t\to0,
	\end{equation*}
	\begin{equation}\label{eq:additive_flat}
		\Longrightarrow \qquad 
		\lim_{a\to\infty} u'(a+h)/u'(a)=1
		\qquad\text{for all }\;h>0.
	\end{equation}
	Let $\rho_{\gamma,z}$ be the entropic risk expectation 
	\eqref{eq:entropic_expectation} 
	and $f_{n,\theta}$ be the finite horizon tilt 
	\eqref{eq:fh_tilt}; at $\gamma=0$
	both reduce to their expected-utility limits. 
	Recall $R_+,b_\CC$ in \eqref{eq:params} and define
	\begin{equation*}
		R_-:=\min_{R\in\CC}R>0, \qquad 
		q_\CC:=\beta R_+<1
		\qquad\text{and}\qquad
		\ua:=\min_{R\in\CC}w(R)\underline z>0.	
	\end{equation*}
	\emph{Step 1. Consumption diverges uniformly.}
	We first show that
	\begin{equation}\label{eq:markov_consumption_lower}
		c_{n,\theta}(a,z)\ge\min\{a,\ua\}
		\qquad\text{for all } n\ge1,\ R\in\CC,\ \gamma\ge0,\ (a,z)\in\XX.
	\end{equation}
	The claim holds at $n=1$ because $c_{1,\theta}(a,z)=a$. 
	Suppose it holds at $n$ but $c_{n+1,\theta}(a,z)<\min\{a,\ua\}$. 
	Then $0<c_{n+1,\theta}(a,z)<a$, so the Euler equation of 
	Lemma~\ref{lm:finite_horizon}\eqref{i:fh_euler} applies. 
	Every next-period wealth satisfies
	$\hat a_j=R(a-c_{n+1,\theta}(a,z))+w(R)z_j\ge \ua$.
	By the induction hypothesis,
	$c_{n,\theta}(\hat a_j,z_j)\ge\min\{\hat a_j,\ua\}=\ua$. Hence
	\begin{equation*}
		u'(c_{n+1,\theta}(a,z))
		=\beta R\,\EE_z\Bigl[u'\bigl(c_{n,\theta}(\hat a,\hat z)\bigr)
		f_{n,\theta}(a-c_{n+1,\theta}(a,z),z,\hat z)\Bigr]
		\le\beta R\,u'(\ua)<u'(\ua).
	\end{equation*}
	But $c_{n+1,\theta}(a,z)<\ua$, so $u'(c_{n+1,\theta}(a,z))>u'(\ua)$, a
	contradiction, so \eqref{eq:markov_consumption_lower} holds.
	Next, we verify
	\begin{equation}\label{eq:mc_diverge}
		\lim_{a\to\infty}m_\CC(a)=\infty,
		\qquad\text{where}\quad
		m_\CC(a):=
		\inf_{\substack{R\in\CC,\ z\in\ZZ\\\gamma\ge0,\ n\ge1}}
		c_{n,\theta}(a,z).
	\end{equation}
	In particular, by Lemma~\ref{lm:finite_horizon}\eqref{i:csn} and
	\eqref{eq:markov_consumption_lower}, $m_\CC(a)>0$ for all $a>0$
	and $m_\CC$ is increasing. Suppose to the contrary that
	$m_\CC(a)\to M\in(0,\infty)$.
	Choose $a_k\to\infty$ and indices $(n_k,R_k,\gamma_k,z_k)$ such that
	$m_\CC(a_k)\le c_k:=c_{n_k,R_k,\gamma_k}(a_k,z_k) \le m_\CC(a_k)+1/k$.
	Then $c_k\to M$. For all sufficiently large $k$, we have
	$c_k\le M+1<a_k$, so current consumption is interior and
	$n_k\ge2$ (since $c_{1,\theta}\equiv a$).
	Every next-period wealth satisfies
	\begin{equation*}
		\hat a_{k}
		=g_{n_k,\theta_k}(a_k,z_k, \hat z)
		\ge R_-(a_k-c_k)=:d_k\to\infty.
	\end{equation*}
	By Lemma~\ref{lm:finite_horizon}\eqref{i:csn}
	and the definition of $m_\CC$, we have
	\begin{equation*}
		c_{n_k-1,\theta_k}(\hat a_{k},\hat z)
		\ge c_{n_k-1,\theta_k}(d_k,\hat z)
		\ge m_\CC(d_k)
		\qquad\text{for all }\; \hat z\in\ZZ.
	\end{equation*}
	Since $u'$ is decreasing and $\EE_z f_{n_k-1,\theta_k}=1$,
	the Euler equation gives
	\begin{equation*}
		u'(c_k)
		\le\beta R_k u'(m_\CC(d_k))
		\le\beta R_+ u'(m_\CC(d_k)).
	\end{equation*}
	Letting $k\to\infty$, since $a_k,d_k\to\infty$ and
	$m_\CC(a)\to M$ as $a\to\infty$, we obtain
	$u'(M)\le q_\CC u'(M)$, which contradicts $q_\CC<1$ and
	$u'(M)>0$. Hence \eqref{eq:mc_diverge} holds.
	
	\emph{Step 2. Uniform eventual drift coefficient.}
	Define the return-adjusted saving rate
	\begin{equation*}
		\sigma_{n,\theta}(a,z)
		:= R\bigl(a-c_{n,\theta}(a,z)\bigr) \big/ a,
		\qquad a>0.
	\end{equation*}
	We first show that
	\begin{equation}\label{eq:sigR<1}
		\sigma_{n,\theta}(a,z)<1
		\qquad\text{for all }\;(a,z,R)\in\XX_0\times\CC,\;
		n\ge1,\ \gamma\ge0.
	\end{equation}
	If $R<1$, then $\sigma_{n,\theta}(a,z)\le R<1$.
	If $R=1$, then $\sigma_{n,\theta}(a,z)=1-c_{n,\theta}(a,z)/a<1$
	since $c_{n,\theta}(a,z)>0$.
	For $R>1$, let $\lambda_R:=1-1/R\in(0,1)$ and argue by induction
	on $n$. The claim holds at $n=1$ because $c_{1,\theta}(a,z)=a$.
	Suppose it holds at $n$ but fails at $n+1$ for some $(a,z)\in\XX_0$.
	Then $c:=c_{n+1,\theta}(a,z)\le\lambda_Ra<a$ and $R(a-c) \ge a$, so the 
	choice is interior.	For all $\hat z$, we have
	$\hat a:=R(a-c)+w(R)\hat z \ge a + w(R) \hat z>a$,
	and the induction hypothesis gives
	$c_{n,\theta}(\hat a,\hat z)>\lambda_R\hat a>\lambda_R a\ge c$. 
	So $u'(c_{n,\theta}(\hat a,\hat z))<u'(c)$, and the Euler 
	equation yields $u'(c)\le\beta R\,u'(c)<u'(c)$, a contradiction. 
	This proves \eqref{eq:sigR<1}.
	
	\emph{Step 3. Uniform eventual inward drift.}
	We claim that there exists $A<\infty$ such that
	\begin{equation}\label{eq:interior_a_hat}
		Rs_{n,\theta}(a,z)+b_\CC\le a-1
		\quad\text{for all }\; 
		a\ge A,\ n\ge1,\ R\in\CC,\ \gamma\ge0,\ z\in\ZZ.
	\end{equation}
	Suppose not. Then there exist $a_k\to\infty$ and
	$(n_k,R_k,\gamma_k,z_k)$ such that, writing
	$c_k:=c_{n_k,\theta_k}(a_k,z_k)$, 
	$R_k(a_k-c_k)+b_\CC>a_k-1.$
	Set $\lambda_k:=1-1/R_k$. Rearranging gives
	$c_k<\lambda_ka_k+ (b_\CC+1) / R_k.$
	Since $c_k\ge m_\CC(a_k)\to\infty$ by \eqref{eq:mc_diverge}
	and $(b_\CC+1)/R_k\le (b_\CC+1)/R_-<\infty$, we obtain $\lambda_ka_k\to\infty$. Because $a_k\to\infty$, this forces $\lambda_k>0$ for all sufficiently large
	$k$, hence $R_k>1$ and $0<\lambda_k<1$ eventually. 
	Define $x_k:=\lambda_k(a_k-b_\CC-1)\to\infty$.
	For every $\hat z\in\ZZ$, using
	$1-\lambda_k=1/R_k$, the next-period wealth satisfies
	\begin{equation*}
		\hat a_{k}
		\ge R_k(a_k-c_k)
		>R_k\left[a_k-\lambda_ka_k- (b_\CC+1)/R_k\right]
		=a_k-b_\CC-1,
	\end{equation*}
	and the lower bound is positive for large $k$. Hence, saving is positive,
	$c_k<a_k$, and $n_k\ge2$. Applying \eqref{eq:sigR<1} at horizon $n_k-1$
	and state $(\hat a_{k},\hat z)$ yields
	\begin{equation*}
		c_{n_k-1,\theta_k}(\hat a_{k},\hat z)
		>\lambda_k\hat a_{k}
		>\lambda_k(a_k-b_\CC-1)=x_k.
	\end{equation*}
	From $1/R_k=1-\lambda_k$,
	$c_k<\lambda_ka_k+ (b_\CC+1) / R_k = x_k+b_\CC+1$.
	As $u'$ is decreasing, $u'(x_k+b_\CC+1)<u'(c_k)$ and 
	$u'(c_{n_k-1,\theta_k}(\hat a_k,\hat z))<u'(x_k)$. 
	Then, by the Euler equation,
	\begin{equation*}
		u'(x_k+b_\CC+1)\le u'(c_k)
		\le\beta R_k u'(x_k)\le q_\CC u'(x_k).
	\end{equation*}
	Hence, $u'(x_k+b_\CC+1)/u'(x_k)\le q_\CC<1$.
	But $x_k\to\infty$, so \eqref{eq:additive_flat} with $h=b_\CC+1$
	implies $u'(x_k+b_\CC+1)/u'(x_k)\to1$, a contradiction.
	This proves \eqref{eq:interior_a_hat}.
	
	\emph{Step 4. Passage to the infinite horizon.}
	By Lemma~\ref{lm:finite_horizon}\eqref{i:fh_to_infh}, for each fixed $\theta\in\TT$, $v_{n,\theta}\to v_\theta$
	uniformly on compact subsets of $\XX$, and
	\begin{equation}\label{eq:conver_policy}
		s_{n,\theta}(a,z)\to s_\theta(a,z)
		\quad\text{and}\quad 
		c_{n,\theta}(a,z)\to c_\theta(a,z)
		\quad \text{for all } \; (a,z)\in \XX.
	\end{equation}
	\emph{Step 5. Invariance and absorption.}
	By \eqref{eq:conver_policy}, 
	passing to the limit in \eqref{eq:interior_a_hat} yields
	\begin{equation}\label{eq:bd_ahat}
		g_{\theta}(a,z,\hat z)
		\le Rs_{\theta}(a,z)+b_\CC\le a-1
		\quad \text{for all } \; a\ge A.
	\end{equation}
	The same passage extends \eqref{eq:markov_consumption_lower}
	and the lower bound $c_{n,\theta}\ge m_\CC$ to the
	infinite-horizon policy.
	Choose $a_\CC>\max\{A,b_\CC+1\}$ sufficiently large that
	\eqref{eq:interior_a_hat} and \eqref{eq:bd_ahat} hold for all $a\ge a_\CC$.
	For $a\le a_\CC$, monotonicity of saving and these bounds give
	\begin{equation}\label{eq:markov_strict_saving_cap}
		\begin{aligned}
			Rs_{n,\theta}(a,z)+w(R)\bar z
			&\le Rs_{n,\theta}(a_\CC,z)+b_\CC\le a_\CC-1,\\
			Rs_{\theta}(a,z)+w(R)\bar z
			&\le Rs_{\theta}(a_\CC,z)+b_\CC\le a_\CC-1.
		\end{aligned}
	\end{equation}
	Thus $\XX_\CC:=[0,a_\CC]\times\ZZ$ is invariant for all
	$R\in\CC$, $\gamma\ge0$, and $n\ge1$, proving
	claim~\eqref{i:invariance_para}.
	
	For absorption, fix $a_0>a_\CC$. By \eqref{eq:bd_ahat},
	$a_{t+1}\le a_t-1$ whenever $a_t>a_\CC$.
	Hence $a_t\le a_\CC$ after at most $\lfloor a_0-a_\CC\rfloor+1$
	periods along every productivity path; by invariance the process
	remains in $\XX_\CC$ thereafter. This proves
	claim~\eqref{i:absorption_para}.
	
	\emph{Step 6. Stationary support.}
	Let $\mu_\theta$ be a stationary distribution of the state process under
	the optimal policy at parameter $\theta=(R,\gamma)\in\TT$. By
	stationarity,
	\begin{equation*}
		\mu_\theta(\XX_\CC)
		=\int_\XX P_\theta^t(x,\XX_\CC)\,\mu_\theta(\diff x)
		\qquad\text{for all }t\ge1.
	\end{equation*}
	Let $\tau_x$ be the first hitting time of the state process to
	$\XX_\CC$ conditional on $(a_0,z_0)=x$, with $\tau_x=\infty$ if the
	process never enters $\XX_\CC$. By claim~\eqref{i:invariance_para}, 
	the process never leaves $\XX_\CC$ once it enters, so the events
	$\{\tau_x\le t\}$ and $\{(a_t,z_t)\in\XX_\CC\}$ coincide. 
	Claim~\eqref{i:absorption_para} then gives
	$P_\theta^t(x,\XX_\CC)=\PP_x(\tau_x\le t)\to\PP_x(\tau_x<\infty)=1$
	for every $x\in\XX$. The dominated convergence theorem then gives
	$\mu_\theta(\XX_\CC)=1$, proving
	claim~\eqref{i:stationary_support_para}.
\end{proof}

\begin{proof}[Proof of Proposition~\ref{pr:compact}]
	Claims~\eqref{i:invariance}--\eqref{i:absorption} follow directly from
	Proposition~\ref{pr:com_sup} with $\CC=\{R\}$. The support
	property $\mu_R(\XX_R)=1$ is exactly
	claim~\eqref{i:stationary_support_para} of Proposition~\ref{pr:com_sup}
	under $\CC=\{R\}$. It remains to verify the existence of a stationary
	distribution. Since $\XX_R$ is invariant and absorbing, we may restrict
	the state process to $\XX_R$. We first show that $P$ is Feller on 
	$\XX_R$. Let $h:\XX_R\to\RR$ be bounded and continuous. Then
	\begin{equation*}
		(Ph)(a,z)
		=\int h(\hat a,\hat z)\,P\bigl((a,z),\diff (\hat a,\hat z)\bigr)
		=\sum_{\hat z\in\ZZ}
		h\bigl(g(a,z,\hat z),\hat z\bigr)Q(z,\hat z)
	\end{equation*}
	for all $(a,z)\in \XX_R$, where $g(a,z,\hat z)$ is given by 
	\eqref{eq:g_fixed}. By Theorem \ref{t:os_rs_opt},  
	$a\mapsto c(a,z)$ is continuous for each $z\in\mathsf Z$, so
	$g$ is continuous in $a$. Since $\mathsf Z$ is finite 
	and $h$ is continuous, $Ph$ is continuous on $\XX_R$. 
	Thus $P$ is Feller on the compact state space $\XX_R$. By the 
	Krylov--Bogolyubov theorem, a Feller Markov chain on a compact 
	state space admits at least one stationary distribution. Hence 
	claim~\eqref{i:existence} holds. 
\end{proof}


\begin{lemma}[The borrowing constraint binds with positive probability]
	\label{lm:binding_ft}
	For all $(a,z)\in \XX$, we have 
	$\PP_{(a,z)} \left(\cup_{t \geq 0} \{c_t=a_t\} \right) > 0.$
\end{lemma}

\begin{proof}
	Since the stated claim clearly holds when $a=0$, 
	it suffices to verify that it holds for $(a,z)\in \XX_0$. 
	Suppose, to the contrary, that for some $(a, z) \in \XX_0$, we have
	\begin{equation*}
		\PP_{(a,z)} \left(
			\cap_{t\geq 0} \{c_t < a_t\}
		\right) 
		= 1-\PP_{(a,z)} (\cup_{t\ge0} \{c_t=a_t\}) = 1.
	\end{equation*}
	In other words, the borrowing constraint never binds with probability 
	one. Recall the tilted probability $f(\hat a,\hat z \mid z)$ defined by 
	\eqref{eq:tilt_prob}. By definition, $\EE_z f(\hat a,\hat z\mid z)=1$. 
	Note that $a_t \ge w z_t\ge w \uz$ for all $t\ge 1$ and $c(w\uz,z)>0$ 
	for all $z\in \ZZ$ because $u'(0) = \infty$. By the monotonicity of 
	$c(\cdot,z)$, we have 
	$c(a_t,z_t) \geq c(w\uz, z_t) \geq \min_{z\in\ZZ} c(w\uz, z) =:B > 0$,
	Hence, $u'(c(a_t,z_t))\le u'(B)<\infty$ almost surely for all $t\ge 1$. 
	Since the borrowing constraint never binds with probability one, based on 
	the Euler equation,
	\begin{align*}
		u'(c(a,z)) &= \beta R \EE_0 f(a_1, z_1\mid z) u'(c(a_1, z_1)) \\
		&= \beta R \EE_0 f(a_1, z_1\mid z) \cdots 
		\beta R \EE_{t-1} f(a_t, z_t\mid z_{t-1}) u'(c(a_t, z_t)) \\
		&\leq u'(B) \beta R \EE_0 f(a_1, z_1\mid z) 
		\cdots \beta R \EE_{t-1} f(a_t, z_t\mid z_{t-1}) 
		= (\beta R)^t u'(B)
	\end{align*}
	with probability one for all $t \geq 1$. Letting $t \to \infty$ yields 
	$0 < u'(c(a,z)) \leq 0$ with probability one, which is a contradiction. 
	Hence the stated claim holds.
\end{proof}

In the rest of this section, let $Q$ have positive entries, as in 
Theorem~\ref{t:gs_rs}. Then
\begin{equation}\label{eq:q_min}
	q_{\min} := \min_{z,\hat z\in \ZZ} Q(z,\hat z) > 0.
\end{equation} 

\begin{lemma}[An accessible binding state]
	\label{lm:binding_ft_2}
	There exists $\tilde z \in \ZZ$ such that $w\tz \leq \bar a(\tz)$, 
	and for all $(a,z)\in \XX$, there exists $t\geq 1$ with
	$
		\PP_{(a, z)} \left(
		    \{c_t = a_t\} \cap \{z_t = \tz\}
		\right) > 0.
	$
\end{lemma}

\begin{proof}
	If $c(a,z)=a$ for all $a\ge0$ for some $z\in \ZZ$, the stated claim 
	obviously holds. Now suppose for all $z\in\ZZ$, $c(a,z) < a$ for some 
	$a\ge0$. By Lemma~\ref{lm:threshold}, there exists $\bar a(z) \in 
	(0,\infty)$ such that $c(a,z)=a$ if and only if $a \leq \bar a(z)$. 
	Lemma~\ref{lm:binding_ft} then implies that there exists $\tz\in\ZZ$ 
	such that $w \tilde z \leq \bar a(\tilde z)$, otherwise the borrowing 
	constraint never binds starting from any $(a,z)$ with $a>\bar a(z)$. Note 
	that Lemma~\ref{lm:binding_ft} also implies that for all $(a,z) \in \XX$, 
	there exists some $t$ and $\check z \in\ZZ$ such that 
	$M(a,z) := \PP_{(a,z)} \left(
		\{c_{t-1} = a_{t-1}\} \cap \{z_{t-1}=\check z\}
	\right) > 0$.
	Hence, by Lemma~\ref{lm:threshold}, we have
	\begin{align*}
		&\PP_{(a,z)} \left(
		\{c_t = a_t\} \cap \{z_t=\tilde z\}
		\right)
		= \PP_{(a,z)} \left(
		\{a_t\leq \bar a(z_t)\} \cap \{z_t=\tilde z\}
		\right) \\
		\geq\;& \PP_{(a,z)} \left(
		\{a_t\leq \bar a(z_t)\} \cap \{z_t=\tilde z\}
		\mid  \{c_{t-1} = a_{t-1}\} \cap \{z_{t-1}=\check z\}
		\right) M(a,z)  \\
		=\;& \PP_{(a,z)}
		\{w z_t \leq \bar a(z_t) \mid  z_t=\tilde z\} \,
		Q(\check z, \tilde z) M(a,z)
		\ge q_{\min} M(a,z) > 0.
	\end{align*}
	The stated claim is verified.
\end{proof}

\begin{lemma}[Monotone binding transition]
	\label{lm:mono_prob}
	$\ell_t (a,z) := \PP_{(a, z)} \left(
		\{c_t = a_t\} \cap \{z_t = \tilde z \}
	\right)$
	is decreasing in $a$ for all $z\in \ZZ$ and $t\geq 1$,
	where $\tz$ is defined as in Lemma~\ref{lm:binding_ft_2}.
\end{lemma}

\begin{proof}
	We extend the range of $\bar a(z)$ to $(0,\infty]$ by defining 
	$\bar a(z) = \infty$ if $c(a,z) \equiv a$. By 
	Lemma~\ref{lm:threshold}, there exists $\bar a(z)\in (0,\infty]$ 
	such that $c(a,z) = a$ if and only if $a \leq \bar a(z)$. 
	Hence, for all $t \geq 0$, 
	$\1 \left(\{c_t = a_t\} \cap \{z_t = \tz\} \right)
	= \1 \left(\{a_t \leq \bar a(z_t)\} \cap \{z_t = \tz\}\right)$,
	which is decreasing in $a_t$. Recall function 
	$g$ defined in \eqref{eq:g_fixed}. When $t = 1$, 
	$
		\ell_1 (a,z)
		= \EE_{a,z} \1 \left( 
		\{a_1 \leq \bar a(z_1)\} \cap \{z_1 = \tz\} 
		\right)
		= Q(z, \tilde z) \1\left\{
		g(a,z,\tilde z) \leq \bar a(\tilde z) 
		\right\}.
	$
	Because $g(a,z,\hat z)$ is increasing in $a$, $\ell_1 (a,z)$ is 
	decreasing in $a$. Now suppose the stated claim holds for 
	arbitrary $t$, we show that it holds for $t+1$. Note that
	\begin{align*}
		\ell_{t+1} (a,z) 
		&= \EE_{a,z} \EE_{(a_1,z_1)} \left\{
		\1 \left( 
		\{a_{t+1} \leq \bar a(z_{t+1})\} \cap \{z_{t+1} = \tilde z\} 
		\right) 
		\mid \fF_1
		\right\}  \\
		&= \EE_{a,z} \EE_{(a_1,z_1)} 
		\1 \left( 
		\{a_{t} \leq \bar a(z_{t})\} \cap \{z_{t} = \tilde z\} 
		\right) 
		= \EE_{a,z} \ell_{t}(a_1,z_1).
	\end{align*}
	So $\ell_{t+1} (a,z)=\EE_{a,z} \ell_t \left(g(a,z,z_1), z_1\right)$.
	Since $\ell_t (a_1, z_1)$ is decreasing in $a_1$ and $g(a,z,z_1)$ is 
	increasing in $a$, $\ell_{t+1} (a, z)$ is decreasing in $a$. 
	Hence the stated claim holds.
\end{proof}

\begin{proof}[Proof of Theorem~\ref{t:gs_rs}]
	\emph{\eqref{i:gs_XR} Uniform ergodicity on $\XX_R$.}
	By Proposition \ref{pr:compact}, the set $\XX_R$ is invariant
	and supports a stationary distribution $\mu^*$. Let $\tilde z$ be the state supplied by
	Lemma~\ref{lm:binding_ft_2}, and put
	$x_*:=(w\tilde z,\tilde z)$.
	Fix $\hat z\in\ZZ$. The same lemma gives $t\ge1$ such that
	$q_1:=
	\PP_{(a_R^*,\hat z)}
	\{c_t=a_t,\ z_t=\tilde z\}>0.
	$
	From any $(a,z)\in\XX_R$, the event $z_1=\hat z$
	has probability at least $q_{\min}>0$.
	On this event, invariance gives $a_1\le a_R^*$,
	so Lemma~\ref{lm:mono_prob} implies that the probability
	of binding at productivity $\tilde z$ after a further
	$t$ periods is at least $q_1$.
	One more transition to $\tilde z$ then places the process
	at $x_*$, since zero saving implies next-period wealth
	$w\tilde z$. Thus, with $m:=t+2$ and $q:=q_{\min}^2q_1>0$,
	\begin{equation*}
		P^m(x,\cdot)\ge q\,\delta_{x_*}(\cdot)
		\quad \text{for all }\, x\in\XX_R.
	\end{equation*}
	Hence the whole restricted state space is small and
	the chain is $\delta_{x_*}$--irreducible.
	Moreover, $w\tilde z\le\bar a(\tilde z)$ and
	Lemma~\ref{lm:threshold} imply zero saving at $x_*$,
	so $P(x_*,\{x_*\})\ge Q(\tilde z,\tilde z)>0$.
	The chain is therefore aperiodic.
	Theorem~16.0.2 of \citet{meyn2009markov} yields
	uniform ergodicity on $\XX_R$ and uniqueness of $\mu^*$
	there.
	
	\emph{\eqref{i:gs_full} Global stability on $\XX$.}
	Let $x_t:=(a_t,z_t)$ and
	$\tau:=\inf\{t\ge0:x_t\in\XX_R\}$.
	By Proposition~\ref{pr:compact}, $\tau<\infty$ almost
	surely from every initial state, and $\XX_R$ is absorbing.
	Fix any initial distribution $\mu$ and, for each $n$, set
	$T:=\lfloor n/2\rfloor$.
	For every measurable $f$ with $|f|\le1$, the Markov
	property gives
	\begin{equation*}
		\mu P^nf-\mu^*f
		=\EE_\mu\bigl[P^{n-T}f(x_T)-\mu^*f\bigr].
	\end{equation*}
	On $\{\tau\le T\}$, we have $x_T\in\XX_R$, so
	part~\eqref{i:gs_XR} bounds the integrand in absolute
	value by $C\vartheta^{n-T}$.
	On $\{\tau>T\}$, it is bounded by $2$.
	Taking the supremum over $|f|\le1$ therefore yields
	$
	\|\mu P^n-\mu^*\|_{TV}
	\le C\vartheta^{n-T}
	+2\PP_\mu(\tau>T)\to 0
	$
	as $n \to \infty$ since $\tau<\infty$ almost surely under every initial
	distribution.
	Finally, any stationary $\mu$ satisfies
	$\mu=\mu P^n$, so this convergence implies $\mu=\mu^*$.
	Thus $P$ is globally stable on $\XX$.
\end{proof}

\section{Proof of Proposition~\ref{pr:agr_cap}}\label{s:pr_agg_wealth}

Consider the general equilibrium setting of Section~\ref{s:GE}, where the
household faces the wage schedule $w(R)$ defined in Section~\ref{ss:firm}.
Throughout, Assumption~\ref{a:mlr} is in force. 
Let $\{R_n\}$ be a sequence in $(\underline R,\bar R)$ with
$R_n\uparrow\bar R$, and set
\begin{equation*}
	w_n:=w(R_n) \qquad \text{and} \qquad
	\bar w:=\lim_{R\uparrow\bar R}w(R),
\end{equation*}
which is finite (see Online Appendix~\ref{s:pf_firm_opt}).
Write $v_{R,\gamma}, c_{R,\gamma}$, $\mu_{R,\gamma}$ for the value function,
consumption policy, and stationary distribution at prices $(R,w(R))$ and 
risk-sensitivity parameter $\gamma$. At $\gamma=0$, the entropic risk 
expectation $\rho_{0,z}$ of \eqref{eq:entropic_expectation} is the 
conditional expectation $\EE_z$, so these objects are the expected-utility 
benchmark.

We first establish divergence at $\gamma=0$. For the sequence above, set
\begin{align*}
	c_n(a,z):=c_{R_n,0}(a,z)
	\qquad \text{and} \qquad 
	v_n(a,z):=v_{R_n,0}(a,z).
\end{align*}
Let $A(R_n,0)$ and $S(R_n,0)$ be the aggregate wealth and saving under 
$\mu_{R_n,0}$, as defined in \eqref{eq:agg_wealth} of the main text. 
Define the optimal wealth transition at $(R_n,w_n,0)$ by
\begin{equation}\label{eq:gn_func}
	g_n(a,z,\hat z):=R_n[a-c_n(a,z)]+w_n\hat z,
	\qquad (a,z,\hat z)\in\mathsf X\times\mathsf Z,
\end{equation}
so that $\hat a_n=g_n(a,z,\hat z)$ is next period's assets.   

\begin{lemma}[A finite-mean invariant distribution at the critical return]
	\label{lm:c_ubdd}
	Suppose $\liminf_{n\to\infty}A(R_n,0)<\infty$ for some
	sequence $\{R_n\}$ with $R_n\uparrow \bar R$. Then there exist a probability measure 
	$\mu_0 \in \pP(\XX)$ and a continuous policy $c_0$ such that
	\begin{equation*}
		\int_\XX a\, \diff\mu_0<\infty,
		\qquad 0\le c_0(a,z)\le a,
	\end{equation*}
	and the maps $a\mapsto c_0(a,z), a-c_0(a,z)$ are increasing for each 
	$z\in\ZZ$. Moreover, $\mu_0$ is stationary for the transition kernel 
	$P_{\bar R,0}$ and $c_0$ satisfies the Euler inequality
	\begin{equation}\label{eq:critical_euler_ineq}
		u'(c_0(a,z)) \ge \EE_z u'\left(
		c_0(\bar R(a-c_0(a,z)) + \bar w\hat z,\hat z)
		\right).
	\end{equation}
\end{lemma}

\begin{proof}
	By assumption, there exist
	$M<\infty$ and a subsequence of $\{R_n\}$ along which $A(R_n,0)\le M$.
	Passing to this subsequence, we may assume
	$A(R_n,0)\le M$ for all $n$.
	
	\emph{Step 1. Existence of $\mu_0$ with finite first moment.}
	Let $\mu_n:=\mu_{R_n,0}$. 
	By the definition of $A(R_n,0)$ in \eqref{eq:agg_wealth} and the Markov 
	inequality, $\mu_n([L,\infty)\times \ZZ)\le M/L$ for all $L>0$ and $n$. 
	Since $\ZZ$ is finite, the sequence $\{\mu_n\}$ is tight. By Prokhorov's 
	theorem, there is a weakly convergent subsequence. With slight abuse of 
	notation, we write  $\mu_n \tow \mu_0 \in \pP(\XX)$. For each $L>0$, the 
	map $a\mapsto a\wedge L$ is bounded and continuous. By weak convergence,
	\begin{equation*}
		\int_\XX (a\wedge L)\diff\mu_0
		= \lim_{n\to\infty}\int_\XX (a\wedge L)\diff\mu_n
		\le M.
	\end{equation*}
	Letting $L\to\infty$, the monotone convergence theorem gives 
	$\int_\XX a\diff\mu_0\le M<\infty$. 
	
	\emph{Step~2. Existence and properties of $c_0$.} For all $n$ and 
	$z\in\ZZ$, Theorem~\ref{t:opt_general} applied at $\theta=(R_n,0)$ 
	implies that $a\mapsto c_n(a,z), a-c_n(a,z)$ are increasing. 
	Thus, for $a_2>a_1$, $0\le c_n(a_2,z)-c_n(a_1,z)\le a_2-a_1$.
	So, $\{c_n(\cdot,z)\}_n$ is uniformly Lipschitz continuous. Since 
	additionally $0\le c_n(a,z)\le a$, the family is uniformly bounded 
	on compact sets. Taking the subsequence from Step~1, we repeatedly 
	apply the Arzelà--Ascoli theorem to extract nested subsequences
	$\{c_{n_j^{(k)}}\}_{j\ge1}$ converging uniformly to $c^{(k)}$ on 
	$[0,k]\times\ZZ$. Since $c^{(k+1)}|_{[0,k]\times\ZZ}=c^{(k)}$, 
	we define a continuous policy $c_0$ on $\XX$ by
	$c_0|_{[0,k]\times\ZZ}:=c^{(k)}$ for all $k\ge1$.
	Set $N_j:=n_j^{(j)}$. For each fixed $k$, the tail
	$\{N_j\}_{j\ge k}$ is a subsequence of $\{n_j^{(k)}\}_{j\ge1}$, 
	so $c_{N_j}\to c_0$ uniformly on $[0,k]\times\ZZ$.
	Relabeling this subsequence as $n$, we obtain local uniform 
	convergence $c_n\to c_0$ while retaining $\mu_n\tow\mu_0$. 
	Passing to the limit gives $0\le c_0(a,z)\le a$, with 
	$a\mapsto c_0(a,z), a-c_0(a,z)$ increasing for all $z\in\ZZ$.	
	
	\emph{Step~3. Stationarity of $\mu_0$ for $P_{\bar R,0}$.}
	Since $R_n\uparrow\bar R$ and $c_n\to c_0$ on $\AA \times \ZZ$ for each 
	compact $\AA$, by \eqref{eq:gn_func}, for each 
	$\hat z\in \ZZ$, $g_n(a,z,\hat z)\to g_0 (a,z,\hat z):=\bar 
	R(a-c_0(a,z))+\bar w\hat z$ uniformly on compact subsets of $\XX$. 
	Let $P_n$ be the stochastic kernel induced by $g_n$ at $\gamma=0$ 
	and $P_{\bar R,0}$ the kernel induced by $g_0$. 
	For bounded continuous function $h:\XX\to\RR$, $\mu_nP_n=\mu_n$ gives
	$\int_\XX h\diff\mu_n = \int_\XX h \diff (\mu_n P_n)
	=\int_\XX P_nh\diff\mu_n$. We show that
	\begin{equation}\label{eq:critical_stationarity}
		\int_\XX h\diff\mu_0=\int_\XX P_{\bar R,0}h\diff\mu_0
		\quad \text{for all bounded continuous } h:\XX\to\RR,
	\end{equation}
	which implies $\mu_0 P_{\bar R,0} = \mu_0$, i.e., $\mu_0$ is a 
	stationary distribution of $P_{\bar R,0}$.
	To see~\eqref{eq:critical_stationarity},
	\begin{align}\label{eq:stationarity_decomp}
		\left|\int_\XX h\diff\mu_0-\int_\XX P_{\bar R,0}h\diff\mu_0\right|
		&\le \left|\int_\XX h\diff\mu_0-\int_\XX h\diff\mu_n\right| 
		+ \left|\int_\XX (P_nh-P_{\bar R,0}h)\diff\mu_n \right| \notag \\
		&+\left|
		\int_\XX P_{\bar R,0}h\diff\mu_n-\int_\XX P_{\bar R,0}h\diff\mu_0
		\right|.
	\end{align}
	The first and third terms vanish under weak convergence 
	$\mu_n \tow \mu_0$, since $h$ and $P_{\bar R,0}h$ are bounded 
	continuous. For the middle term, fix $\epsilon>0$. By tightness of 
	$\{\mu_n\}$, pick a compact set $\KK\subset \XX$ with 
	$\sup_n\mu_n(\XX\setminus\KK) < \epsilon/(4\|h\|_\infty)$. Because 
	$\max \{|P_{\bar R,0}h|, |P_n h|\} \le \|h\|_\infty$, we obtain
	$\int_{\XX\setminus\KK} |P_n h - P_{\bar R,0} h| \diff \mu_n 
	    \leq 2 \|h\|_\infty \sup_n \mu_n(\XX\setminus\KK)
		< \epsilon/2$.
	Since $g_n\to g_0$ uniformly on $\KK$ and $h$ is bounded and continuous, 
	for all large $n$,
	\begin{align*}
		&\sup_{(a,z)\in \KK}|P_nh(a,z)-P_{\bar R,0}h(a,z)| \\
		\le \,& \sup_{(a,z)\in \KK} \sum_{\hat z \in \ZZ} \left|
		h(g_n(a,z,\hat z), \hat z) -
		h(g_0(a,z,\hat z), \hat z) 
		\right| Q(z,\hat z)
		< \epsilon / 2.
	\end{align*}
	Therefore, for all large $n$,
	\begin{equation*}
		\left|\int_\XX (P_nh-P_{\bar R,0}h)\diff\mu_n \right| 
		\leq \int_{\XX\setminus\KK} |P_n h - P_{\bar R,0} h| \diff \mu_n 
		+ \int_{\KK} |P_n h - P_{\bar R,0} h| \diff \mu_n 
		< \epsilon/2 + \epsilon/2 = \epsilon.
	\end{equation*}
	Letting $n\to\infty$ in \eqref{eq:stationarity_decomp} yields 
	\eqref{eq:critical_stationarity}, hence $\mu_0$ is stationary 
	for $P_{\bar R,0}$.
	
	\emph{Step~4. Euler inequality.}
	Proposition~\ref{pr:euler} applied at $\gamma=0$ gives
	\begin{equation}\label{eq:rs_euler_ineq_n}
		u'(c_n(a,z))
		\ge \beta R_n\EE_z[u'(c_n(\hat a_n,\hat z))]
		\quad \text{for all } n \text{ and } (a,z) \in \XX,
	\end{equation}
	with $u'(0):=\infty$.
	It remains to prove \eqref{eq:critical_euler_ineq}. 
	Fix $(a,z)\in \XX$. If $c_0(a,z)=0$, then \eqref{eq:critical_euler_ineq} 
	holds trivially since $u'(c_0(a,z))=\infty$. Suppose $c_0(a,z)>0$. By 
	\eqref{eq:rs_euler_ineq_n},
	\begin{equation*}
		u'(c_n(a,z))\ge \beta R_n \sum_{\hat z\in \ZZ} Q(z,\hat z)
		u'(c_n(g_n(a,z,\hat z),\hat z)).
	\end{equation*}
	Since $c_n(a,z)\to c_0(a,z)>0$, we have $u'(c_n(a,z))\to u'(c_0(a,z))$. 
	Also, $\beta R_n\to \beta\bar R=1$. For each $\hat z\in \ZZ$, 
	$g_n(a,z,\hat z)\to g_0(a,z,\hat z)$. By the uniform convergence of 
	$\{c_n\}$ to $c_0$ on compact sets, the continuity of $c_0$, and the 
	triangle inequality, we obtain
	\begin{align*}
		\left|c_n(g_n(a,z,\hat z),\hat z)
		- c_0(g_0(a,z,\hat z),\hat z)\right|
		&\leq \left|c_n(g_n(a,z,\hat z),\hat z)
		- c_0(g_n(a,z,\hat z),\hat z)\right|  \\
		&+ \left|c_0(g_n(a,z,\hat z),\hat z)
		- c_0(g_0(a,z,\hat z),\hat z)\right| \to 0.
	\end{align*}
	Moreover, $c_0(g_0(a,z,\hat z),\hat z)>0$ for every
	$\hat z\in\mathsf Z$: otherwise $c_n(g_n(a,z,\hat z),\hat z)\to0$
	for some $\hat z$, then $u'(c_n(g_n(a,z,\hat z),\hat z))\to\infty$,
	which would force the right-hand side of \eqref{eq:rs_euler_ineq_n} 
	to diverge and contradict the boundedness of the left-hand side, since
	$u'(c_n(a,z))\to u'(c_0(a,z))<\infty$. Hence each limit is finite.
	Taking limits then verifies \eqref{eq:critical_euler_ineq}.
\end{proof}

\begin{proof}[Proof of Proposition~\ref{pr:agr_cap}]
	
	We first show that $A(R,0)\to\infty$ as $R\uparrow\bar R$.
	Suppose this fails. Then there exist
	$M<\infty$ and a sequence $R_n\uparrow\bar R$ such that
	$A(R_n,0)\le M$ for every $n$.
	Lemma~\ref{lm:c_ubdd} therefore yields a pair $(\mu_0,c_0)$
	such that
	(i) $\int_\XX a\diff\mu_0<\infty$,
	(ii) $\mu_0$ is invariant for $P_{\bar R,0}$, and
	(iii) $c_0$ satisfies the Euler inequality
	\eqref{eq:critical_euler_ineq}.
	
	\emph{Step 1. Martingale $\{u'(c_t)\}$ process.} Let 
	$\{(a_t,z_t)\}_{t\ge 0}$ have transition kernel $P_{\bar R,0}$ and 
	invariant distribution $\mu_0$, with $(a_0,z_0) \sim  \mu_0$. Define 
	$c_t:=c_0(a_t,z_t)$ and $\phi_t:=u'(c_t)\in(0,\infty]$. Evaluating 
	\eqref{eq:critical_euler_ineq} at $(a_t,z_t)$ and using the 
	Markov property gives
	\begin{equation}\label{eq:phi_supermartingale}
		\phi_t\ge \EE[\phi_{t+1}\mid\fF_t]
		\qquad\text{almost surely.}
	\end{equation}
	Thus $\{\phi_t\}$ is a supermartingale. For $K>0$, define 
	$\phi_t^K:=\phi_t\wedge K$. Since $x\mapsto x\wedge K$ is 
	increasing and concave, 
	$\EE[\phi_{t+1}^K\mid\fF_t]
	\le \EE[\phi_{t+1}\mid\fF_t]\wedge K
	\le \phi_t\wedge K = \phi_t^K$.
	Hence $\{\phi_t^K\}$ is a bounded supermartingale. Because 
	$\{(a_t,z_t)\}_{t\ge 0}$ is stationary, $\EE[\phi_t^K]=\EE[\phi_{t+1}^K]$. 
	Since the random variable $\phi_t^K-\EE[\phi_{t+1}^K\mid\fF_t]$ is 
	nonnegative and has zero expectation, 
	$\EE[\phi_{t+1}^K\mid\fF_t]=\phi_t^K$ almost surely, implying that 
	$\{\phi_t^K\}$ is indeed a bounded martingale.
	
	\emph{Step 2. Almost sure constancy of $\{c_t\}$.}
	Let $\Delta_t:=\phi_{t+1}^K-\phi_t^K$. Since $\{\phi_t^K\}$ 
	is a martingale, $\EE[\Delta_t\mid\fF_t]=0$. Thus
	$\EE[\phi_t^K\Delta_t] 
	= \EE[\phi_t^K \EE[\Delta_t\mid\fF_t]] = 0$, which gives
	$\EE[(\phi_{t+1}^K)^2]
	= \EE[(\phi_t^K+\Delta_t)^2] 
	= \EE[(\phi_t^K)^2]+\EE[\Delta_t^2].$
	By stationarity, $\EE[(\phi_{t+1}^K)^2]=\EE[(\phi_t^K)^2]$, hence 
	$\EE[\Delta_t^2]=0$ and $\phi_{t+1}^K=\phi_t^K$ almost surely. Letting 
	$K\to\infty$ through a countable sequence gives $\phi_{t+1}=\phi_t$ 
	almost surely. Since $u'$ is strictly decreasing, $c_{t+1}=c_t$ almost 
	surely. As $\EE_{\mu_0} a_t < \infty$ and $c_t \leq a_t$, we have 
	$c_t<\infty$ almost surely. Intersecting over all $t$, there is a finite 
	random variable $\bar c$ with $c_t=\bar c$ for all $t\ge 0$ almost surely. 
	Thus consumption is constant along almost every realized path starting 
	from $(a_0,z_0)\sim \mu_0$.
	
	\emph{Step 3. Constructing zero conditional variance.}
	Define $s_t:=a_t-c_t\ge 0$. At $R=\bar R$, $a_{t+1}=\bar R s_t+\bar wz_{t+1}$.
	Since $c_{t+1}=c_t=\bar c$, 
	$s_{t+1} = a_{t+1}-c_{t+1} = \bar R s_t+\bar wz_{t+1}-\bar c$, so
	$s_t=\bar R^{-1}s_{t+1}+\bar R^{-1}(\bar c-\bar wz_{t+1})$. 
	Iterating $N$ periods gives
	\begin{equation}\label{eq:s_iterate_N}
		s_t = \bar R^{-N}s_{t+N} + 
		\sum_{j=1}^{N} \bar R^{-j}(\bar c-\bar wz_{t+j}).
	\end{equation}
	Since $\mu_0$ has finite first moment, $\EE_{\mu_0}s_t<\infty$. By 
	Tonelli's theorem and stationarity,
	$\EE_{\mu_0}\left[\sum_{N=1}^{\infty}\bar R^{-N}s_{t+N}\right]
		= \sum_{N=1}^{\infty}\bar R^{-N}\EE_{\mu_0}[s_{t+N}]
		= \EE_{\mu_0}[s_0]\sum_{N=1}^{\infty}\bar R^{-N}
		< \infty$
	as $\bar R>1$. Thus $\sum_{N=1}^{\infty}\bar R^{-N}s_{t+N}<\infty$
	almost surely, implying that $\bar R^{-N}s_{t+N}\to0$ almost surely 
	as $N\to\infty$. Since $\{z_{t+j}\}_{j\ge1}$ is bounded and $\bar R>1$, 
	the series $\sum_{j=1}^{\infty}\bar R^{-j}z_{t+j}$ converges absolutely, 
	hence almost surely. Letting $N\to\infty$ in \eqref{eq:s_iterate_N} yields
	$s_t = \bar c/(\bar R-1) - \bar w\sum_{j=1}^{\infty}\bar R^{-j}z_{t+j}$
	almost surely.
	Let $Y_t:=\sum_{j=1}^{\infty}\bar R^{-j}z_{t+j}$. Then 
	$Y_t= [\bar c/(\bar R-1)-s_t]/\bar w$.
	Since $s_t=a_t-c_0(a_t,z_t)$ and $\bar c=c_0(a_t,z_t)$, 
	we have $\Var(Y_t\mid a_t,z_t) = 0$ almost surely.
	
	\emph{Step 4. Impossibility of zero conditional variance}. Let
	$S_N:=\sum_{j=1}^{N}\bar R^{-j}$ and 
	$T_N:=\sum_{j=N+1}^{\infty}\bar R^{-j}$.
	Since $S_N\uparrow(\bar R-1)^{-1}$ and $T_N\downarrow 0$, 
	pick $N$ large enough so that $S_N>T_N$. 
	Define events $D_L:=\{z_{t+1}=\cdots=z_{t+N}=\uz\}$ and
	$D_H:=\{z_{t+1}=\cdots=z_{t+N}=\bar z\}$.
	As $Q(z,\hat z)\ge q_{\min}>0$ for all $z,\hat z\in \ZZ$, 
	with $q_{\min}$ given in \eqref{eq:q_min}, 
	$\PP(D_L\mid a_t,z_t)\ge q_{\min}^N>0$ and 
	$\PP(D_H\mid a_t,z_t)\ge q_{\min}^N>0$.
	
	On $D_L$, $Y_t\le \uz S_N+\bar z T_N =:A$, while on $D_H$,
	$Y_t\ge \bar z S_N+\uz T_N=:B.$
	The gap is
	$\delta := B-A = (\bar z-\underline z)(S_N-T_N)>0$.
	Let $m:=\EE(Y_t\mid a_t,z_t)$. If $m\ge(A+B)/2$, then on $D_L$,
	$Y_t-m\le A-m \le -\delta/2$, so $(Y_t-m)^2\ge\left(\delta/2\right)^2$.
	Since $\PP(D_L\mid a_t,z_t)\ge q_{\min}^N$, 
	$\Var(Y_t\mid a_t,z_t)
	\ge (\delta/2)^2 \PP(D_L\mid a_t,z_t)
	\ge q_{\min}^N (\delta/2)^2 > 0$.
	If $m\le(A+B)/2$, then on $D_H$, $Y_t-m\ge B-m \ge \delta/2$,
	so again $(Y_t-m)^2\ge\left(\delta/2\right)^2$, and  
	$\Var(Y_t\mid a_t,z_t)
	\ge (\delta/2)^2 \PP(D_H\mid a_t,z_t)
	\ge q_{\min}^N (\delta/2)^2>0$.
	Hence $\Var(Y_t\mid a_t,z_t) >0$ almost surely,
	contradicting $\Var(Y_t\mid a_t,z_t)=0$ in Step~3. Thus 
	no pair $(\mu_0, c_0)$ satisfies
	(i) $\EE_{\mu_0} a_t < \infty$, 
	(ii) $\mu_0$ is invariant for $P_{\bar R,0}$, and 
	(iii) $c_0$ satisfies the Euler inequality \eqref{eq:critical_euler_ineq}. 
	The stated claim then follows from Lemma~\ref{lm:c_ubdd}. 
	Finally, for any finite $\gamma\ge0$, 
	Proposition~\ref{pr:markov_small_gamma} yields $A(R,\gamma)\ge A(R,0)$.
	Therefore $A(R,\gamma)\to\infty$ as $R\uparrow\bar R$,
	completing the proof.
\end{proof}

\section{Proof of Lemma~\ref{lm:agg_cont} and Theorem~\ref{t:RCE}}\label{s:pr_GE}

Throughout, $\underline R$ and $\bar R:=1/\beta$ are as
in Section~\ref{s:GE}, and $K(R),w(R)$ are the firm's capital demand and 
induced wage defined in Section~\ref{ss:firm}. Recall that $c_{R,\gamma}$ 
and $v_{R,\gamma}$ are the optimal policy and value function of 
Theorem~\ref{t:opt_general}, $\mu_{R,\gamma}$ is the unique 
stationary distribution of Theorem~\ref{t:gs_rs}, and $P_{R,\gamma}$ is the 
stochastic kernel at $\theta=(R,\gamma)$. The aggregate wealth
$A(R,\gamma)$ and aggregate saving $S(R,\gamma)$ are 
defined in \eqref{eq:agg_wealth}. 


\begin{proof}[Proof of Lemma~\ref{lm:agg_cont}]
	Let $\theta_n=(R_n,\gamma_n)\to \theta_0=(R_0,\gamma_0)\in(\uR,\bar R) 
	\times[0,\infty)$, and choose a compact interval $\CC\subset(\uR,\bar R)$ 
	containing $R_0$ and $\{R_n\}$. Set $\TT:=\CC\times[0,\infty)$. 
	By Theorem~\ref{t:opt_general}, 
	$(a,z,\theta)\to c_\theta(a,z)$ is jointly continuous on $\XX\times\TT$.
	By Theorem~\ref{t:gs_rs}, $P_\theta$ has a unique stationary 
	measure $\mu_\theta$ for all $\theta\in \TT$. 
	By Proposition~\ref{pr:com_sup}, applied with the compact interval $\CC$, 
	there exists $a_\CC<\infty$ such that $\XX_\CC:=[0,a_\CC]\times\ZZ$ is 
	invariant under $P_\theta$ and the stationary distribution satisfies 
	$\mu_\theta (\XX_\CC)=1$ for all $\theta \in \TT$. Thus, the sequence 
	of stationary distributions $\{\mu_{\theta_n} \}_{n\geq 0}$ has common 
	compact support $\XX_\CC$. 	
	
	Now, pick any sequence $\{(a_n, z_n)\}_{n\ge 0} \subset \XX$ such that 
	$(a_n, z_n) \to (a, z) \in \XX$, and pick any bounded continuous function 
	$f:\XX \to \RR$. By definition, we have
	\begin{equation*}
		\int_\XX f(\hat a, \hat z) 
		P_{\theta_n}((a_n, z_n), \diff (\hat a, \hat z))
		= \sum_{\hat z\in \ZZ} 
		f\left(R_n(a_n-c_{\theta_n}(a_n,z_n))+w(R_n) \hat z, \hat z\right)
		Q(z_n, \hat z).
	\end{equation*}
	By the continuity of $f$, $c_\theta(a,z)$, $w(R)$ and finiteness of $\ZZ$, the right-hand side converges to 
	$\sum_{\hat z\in \ZZ} 
	f\left(R_0(a-c_{\theta_0}(a,z))+w(R_0) \hat z, \hat z\right)
	Q(z, \hat z)
	=\int_\XX f(\hat a, \hat z) P_{\theta_0}((a,z), \diff (\hat a,\hat z))$.
	Thus $\{P_{\theta_n}((a_n,z_n), \cdot )\}$ converges weakly to 
	$P_{\theta_0}((a,z), \cdot )$.
	
	Applying Theorem~12.13 in \cite{stokey1989}, we obtain: 
	$\{\mu_{\theta_n}\}$ converges weakly to $\mu_{\theta_0}$. Therefore,
	$A(\theta_n) = \int_{\XX} a  \mu_{\theta_n}[\diff (a, z)]
		= \int_ {\XX_\CC} a \mu_{\theta_n} [\diff (a, z)]$,
	which, by the weak convergence of $\{\mu_{\theta_n}\}$ to $\mu_{\theta_0}$ 
	and the boundedness and continuity of the integrand on the compact support 
	$\XX_\CC$, converges to 
	$\int_{\XX_\CC} a \mu_{\theta_0}(\diff a, \diff z)
		=\int_{\XX} a \mu_{\theta_0}(\diff a, \diff z)=A(\theta_0)$.	
	Thus $A(\theta_n)\to A(\theta_0)$, so $A(\theta)$ is continuous on 
	$(\uR,\bar R) \times [0,\infty)$.
	
	The continuity of $S(\theta)$ then follows immediately from the 
	continuity of $A(\theta)$, the accounting identity 
	$A(\theta)=A(R,\gamma)=R\,S(R,\gamma)+w(R)$ 
	established in Section~\ref{ss:equilibrium}, together with the 
	continuity of $w(R)$ and the fact that $R>0$ on $(\uR,\bar R)$.
\end{proof}

\begin{proof}[Proof of Theorem~\ref{t:RCE}]
	
	\emph{Step 1. Existence of a stationary equilibrium.}
	By stationarity, $\mu_\theta P_\theta = \mu_\theta$ for $\theta=(R,\gamma)$.
	As the $z$-marginal of $\mu_\theta$ is the stationary distribution $\pi$
	of $\{z_t\}$ and $\EE z=1$, we have
	$\int_\XX \sum_{\hat z\in \ZZ} Q(z,\hat z) \hat z\mu_\theta[\diff (a,z)]
		=\sum_{\hat z\in\ZZ}\hat z\,\pi(\hat z) =1$.
	Therefore
	\begin{align}\label{eq:stationary_accounting}
		A(\theta)
		&=A(R,\gamma) 
		= \int_{\XX} a (\mu_{\theta} P_{\theta})[\diff (a,z)] 
		= \int_{\XX}  (P_{\theta} \hat a)(a,z) \mu_\theta[\diff (a,z)] \notag \\
		&= \int_\XX\sum_{\hat z\in\ZZ}Q(z,\hat z)
		\bigl(R[a-c_\theta(a,z)]+w(R)\hat z\bigr)\,\mu_\theta[\diff (a,z)]
		=RS(\theta)+w(R).
	\end{align}    
	Since $\underline R=1-\delta+\ell<1$, take $R<1$ close to $\uR$, 
	the inequalities $S(R,\gamma)\le A(R,\gamma)$ and 
	\eqref{eq:stationary_accounting} imply $(1-R)S(R,\gamma)\le w(R).$ 
	Together with \eqref{eq:wage_capital_prop}, this yields
	\begin{equation}\label{eq:SK}
		0\le S(R,\gamma)\big/ K(R)
		\le w(R)\big/ [(1-R)K(R)]\to 0
		\quad \text{as }\; R\downarrow\underline R.
	\end{equation}
	Hence, we may choose $R_L\in(\uR,\bar R)$ sufficiently close to 
	$\uR$ such that  $S(R_L,\gamma)/K(R_L)<1.$ The excess capital 
	supply $\Psi(R,\gamma)$ then satisfies
	\begin{equation}\label{eq:Psi_negative}
		\Psi(R_L,\gamma)=S(R_L,\gamma)-K(R_L)<0.
	\end{equation}
	At the upper endpoint, together with Proposition~\ref{pr:agr_cap}, 
	\eqref{eq:wage_capital_prop} and \eqref{eq:stationary_accounting},
	\begin{equation*}
		S(R,\gamma)=[A(R,\gamma)-w(R)] \big/ R \to\infty
		\quad \text{as }\; R\uparrow\bar R.
	\end{equation*}
	Since $K(R)$ is decreasing and takes values in 
	$(0,\infty)$, we may choose $R_H\in(R_L,\bar R)$ sufficiently close to 
	$\bar R$ such that
	\begin{equation}\label{eq:Psi_positive}
		\Psi(R_H,\gamma) = S(R_H,\gamma)-K(R_H) >0.
	\end{equation}
	Consider the compact interval $\CC:=[R_L,R_H] \subset(\underline R,\bar 
	R).$ By Lemma~\ref{lm:agg_cont} and the continuity of $w(R)$, 
	\eqref{eq:stationary_accounting} implies that $S(R,\gamma)$ is 
	continuous on $\CC$. Together with the continuity of capital demand 
	$K(R)$, $\Psi(R,\gamma)$ is continuous on $\CC$. By \eqref{eq:Psi_negative},
	\eqref{eq:Psi_positive}, and the intermediate value theorem, there
	exists $R^*\in(R_L,R_H)$ such that
	\begin{equation}\label{eq:capital_market_clearing}
		\Psi(R^*,\gamma)=0
		\qquad \Longleftrightarrow \qquad
		S(R^*,\gamma)=K(R^*).
	\end{equation}
	This establishes the existence of a recursive competitive equilibrium
	$(R^*,w^*,v^*,c^*,\mu^*)$, where $w^*=w(R^*)$, $v^*=v_{R^*,\gamma}$, 
	$c^*=c_{R^*,\gamma}$, and $\mu^*=\mu_{R^*,\gamma}$.
	
	\emph{Step 2. Compactness and upper hemicontinuity of equilibrium interest rates.}
	The lower-endpoint $R_L$ in Step~1 is independent of $\gamma$. 
	Recall \eqref{eq:SK}. As $R\uparrow\bar R$, the expected-utility argument 
	in Appendix~\ref{s:pr_agg_wealth} gives $\Psi(R,0)\to\infty$.
	For each fixed $R$, applying Proposition~\ref{pr:markov_small_gamma}
	on a compact interval containing $R$ with $\gamma_1=0$ gives
	$\Psi(R,\gamma)\ge\Psi(R,0)$ for all $\gamma\ge0$.
	By \eqref{eq:Psi_positive}, we may choose $R_L, R_H \in (\uR,\bar R)$
	independent of $\gamma$ such that $R_L< R_H$,
	$\Psi(R,\gamma)<0$ for $R\in(\underline R,R_L]$, and 
	$\Psi(R,\gamma)>0$ for $R\in[R_H,\bar R)$
	for every $\gamma\ge0$. Consequently,
	\begin{equation*}
		\mathcal R(\gamma) = \{R\in[R_L,R_H]:\Psi(R,\gamma)=0\}.
	\end{equation*}
	By Lemma~\ref{lm:agg_cont} and continuity of $K$,
	$\Psi(\cdot,\gamma)$ is continuous. The intermediate
	value theorem and closedness of its zero set on
	$[R_L,R_H]$ imply that $\mathcal R(\gamma)$ is nonempty,
	compact, and contained in $(R_L,R_H)$ for every $\gamma\ge0$. 
	
	To establish upper hemicontinuity of $\mathcal R(\gamma)$, it suffices
	to show that $\mathcal R(\gamma)$ has a closed graph. Equivalently, 
	let $\gamma_n\to\gamma_0\ge0$ and $R_n\in\mathcal R(\gamma_n)$ with 
	$R_n\to R_0$, we need to verify $R_0\in \mathcal R(\gamma_0)$. 
	By the previous paragraph, the sequence $\{R_n\}$ lies in the common 
	compact set $[R_L,R_H]$. By the joint continuity of $\Psi$ 
	established in Lemma~\ref{lm:agg_cont}, we obtain 
	$\Psi(R_0,\gamma_0)=\lim_{n\to\infty}\Psi(R_{n},\gamma_{n})=0.$ 
	Hence $R_0\in \mathcal R(\gamma_0)$.
\end{proof}

\section{Proof of Section~\ref{s:comp_stats} Results}
\label{s:pr_comparative}

For fixed-price household analyses, let
$R\in(0,\bar R)$ and $w>0$. Equilibrium arguments evaluate household objects at
the production wage $w(R)$. We identify $\theta$ with $(R,\gamma)$.
We use the entropic risk expectation $\rho_{\gamma,z}$ of 
\eqref{eq:entropic_expectation}, the infinite-horizon tilt 
$f_\gamma(s,z,\hat z)$ of \eqref{eq:infh_tilt}, and the
finite-horizon objects of Lemma~\ref{lm:finite_horizon}: the 
iterates $v_{n,\theta}$, policies $c_{n,\theta},s_{n,\theta}$, 
and finite-horizon tilt $f_{n,\theta}(s,z,\hat z)$ defined in 
\eqref{eq:fh_tilt}. At fixed prices we abbreviate
\begin{equation*}
	v_{n,\gamma}:=v_{n,(R,\gamma)},\quad
	c_{n,\gamma}:=c_{n,(R,\gamma)},\quad
	s_{n,\gamma}:=s_{n,(R,\gamma)},\quad
	f_{n,\gamma}(s,z,\hat z):=f_{n,(R,\gamma)}(s,z,\hat z),
\end{equation*}
and similarly $v_\gamma:=v_{(R,\gamma)}$, $c_\gamma:=c_{(R,\gamma)}$,
$s_\gamma:=s_{(R,\gamma)}$. Both $f_{n,\theta}$ and $f_\gamma$ are
densities with respect to $Q(z,\cdot)$, equal to one at $\gamma=0$; 
hence the tilted probability mass attached to $\hat z=z_j$ is
$Q(z,z_j)f_{n,\gamma}(s,z,z_j)$. We use the shorthand
\begin{equation*}
	\rho_{n,\gamma}(s,z)
	:=\beta\rho_{\gamma,z}\bigl(v_{n,\gamma}(Rs+w\hat z,\hat z)\bigr)
	\quad\text{and}\quad 
	\rho_\gamma(s,z)
	:=\beta\rho_{\gamma,z}\bigl(v_\gamma(Rs+w\hat z,\hat z)\bigr),
\end{equation*}
as in the proof of Lemma~\ref{lm:finite_horizon}. Moreover, continuity, 
monotonicity, the finite-horizon Euler equation, and convergence of the 
finite-horizon objects are cited from Lemma~\ref{lm:finite_horizon} and 
the envelope derivative from Lemma~\ref{lm:T_diff}. Finally, under
Assumption~\ref{a:mlr}, the finite-horizon values are increasing in 
productivity, as established in Lemma~\ref{lm:mono_productivity}.

\subsection{Common comparison arguments}
\label{ss:common_comparison}
Fix $R\in(0,\bar R)$ and $w>0$. For all $\gamma\ge0$, 
let $T_\gamma$ be the Bellman operator~\eqref{eq:bo_rs} 
at the fixed prices $(R,w)$, and set $v_{0,\gamma}=0$ and 
$v_{n+1,\gamma}=T_\gamma v_{n,\gamma}$ for $n\ge0$.
In terms of saving, this recursion is
\begin{equation}\label{eq:iid_continuation}
	v_{n+1,\gamma}(a,z)
	=\max_{s\in[0,a]}\bigl\{u(a-s)+\rho_{n,\gamma}(s,z)\bigr\}.
\end{equation}
For $n\ge0$, let $s_{n+1,\gamma}(a,z)$ be the unique maximizer in 
\eqref{eq:iid_continuation} and 
$c_{n+1,\gamma}(a,z):=a-s_{n+1,\gamma}(a,z)$. By 
Lemma~\ref{lm:finite_horizon}\eqref{i:fh_to_infh},
$c_{n+1,\gamma}\to c_\gamma$ and $s_{n+1,\gamma}\to s_\gamma$.
By Lemma~\ref{lm:finite_horizon}\eqref{i:csn}, $c_{n+1,\gamma}$ and 
$s_{n+1,\gamma}$ are continuous increasing in $a$. 
By Lemma~\ref{lm:T_diff}, applied at each $n$,
\begin{equation}\label{eq:iid_iteration_envelope}
	v'_{n+1,\gamma}(a,z)
	:=\partial v_{n+1,\gamma}(a,z) /\partial a
	=u'(c_{n+1,\gamma}(a,z))
	\qquad(a>0,\ n\ge0).
\end{equation}
The derivatives here and below are with respect to wealth or saving, 
not with respect to $\gamma$. For $s\ge0$, put $x_j(s):=Rs+wz_j$. 
Since $f_{n,\gamma}$ is a density with respect to $Q(z,\cdot)$,
\begin{align}
	\rho_{n,\gamma}'(s,z):=
	\partial \rho_{n,\gamma}(s,z)/\partial s
	&=\beta R\,\EE_z\Bigl[
	f_{n,\gamma}(s,z,\hat z)v'_{n,\gamma}(Rs+w\hat z,\hat z)
	\Bigr] \notag \\
	&=\beta R\sum_{j=1}^{N}Q(z,z_j)
	f_{n,\gamma}(s,z,z_j)\,v'_{n,\gamma}(x_j(s),z_j). 
	\label{eq:comparison_continuation_derivative}
\end{align}
At $s=0$ the derivative is from the right; it is finite, as
$x_j(0)\ge w\underline z>0$. Fix $n\ge0$, $0\le\gamma_1<\gamma_2$, 
$(a,z)\in \XX$ with $a>0$, a feasible $s$, and $j\in\{1,\dots,N\}$.
Denote
\begin{align*}
	a_j:=Q(z,z_j)\,f_{n,\gamma_2}(s,z,z_j),&\qquad
	b_j:=Q(z,z_j)\,f_{n,\gamma_1}(s,z,z_j), \\
	A_j:=v'_{n,\gamma_2}(x_j(s),z_j),&\qquad 
	B_j :=v'_{n,\gamma_1}(x_j(s),z_j).
\end{align*}

\begin{lemma}[One-step comparison]\label{lm:comparison_step}
	Suppose that $A_j\ge B_j$ for all $j$,
	$j\to a_j/b_j$ is decreasing, and at least one of
	$j\to A_j$ and $j\mapsto B_j$ is decreasing. Then
	\begin{equation*}
		s_{n+1,\gamma_2}(a,z)\ge s_{n+1,\gamma_1}(a,z) 
		\quad \text{and} \quad 
		c_{n+1,\gamma_2}(a,z)\le c_{n+1,\gamma_1}(a,z).
	\end{equation*}
	Both inequalities are strict if
	$i_1:=s_{n+1,\gamma_1}(a,z)>0$ and, at $s=i_1$,
	the ratio $a_j/b_j$ and at least one of $A_j,B_j$
	are strictly decreasing in $j$.
	The same conclusions hold for the infinite-horizon problem
	under the corresponding assumptions.
\end{lemma}

\begin{proof}	
	If $j\to B_j$ is decreasing, since $\sum_j a_j=\sum_j b_j=1$, symmetry gives
	\begin{equation}\label{eq:abB_ineq}
		\sum_{j=1}^N(a_j-b_j)B_j
		=\sum_{j,k=1}^Na_jb_k(B_j-B_k)
		=\frac12\sum_{j,k=1}^Nb_jb_k
		\left(\frac{a_j}{b_j}-\frac{a_k}{b_k}\right)
		(B_j-B_k)\ge0.
	\end{equation}
	The inequality follows because  $a_j/b_j$ is decreasing in $j$. Together with $A_j\ge B_j$, this implies
	$\sum_j a_jA_j \ge\sum_j a_jB_j \ge\sum_j b_jB_j$.
	If instead $j\to A_j$ is decreasing, the same
	logic with $A_j$ in place of $B_j$ in \eqref{eq:abB_ineq} gives
	$\sum_j a_jA_j \ge\sum_j b_jA_j \ge\sum_j b_jB_j$.
	Thus, in either case,
	\eqref{eq:comparison_continuation_derivative} yields	
	$\rho'_{n,\gamma_2}(s,z) =\beta R\sum_j a_jA_j 
	\ge\beta R\sum_j b_jB_j =\rho'_{n,\gamma_1}(s,z)$.
	Thus $\rho_{n,\gamma_2}(s,z)-\rho_{n,\gamma_1}(s,z)$ is increasing
	in $s$. Suppose
	$i_2:= s_{n+1,\gamma_2}(a,z)<s_{n+1,\gamma_1}(a,z)=:i_1$.
	Uniqueness of maximizer at $\gamma_1$ gives
	$
		u(a-i_1)-u(a-i_2)
		>\rho_{n,\gamma_1}(i_2,z)-\rho_{n,\gamma_1}(i_1,z)
		\ge \rho_{n,\gamma_2}(i_2,z)-\rho_{n,\gamma_2}(i_1,z),
	$
	where the last inequality follows from the increasing
	difference of $\rho_{n,\gamma_2}(s,z)-\rho_{n,\gamma_1}(s,z)$. 
	Rearranging contradicts optimality of $s_{n+1,\gamma_2}(a,z)$. 
		
	For the strict inequalities, evaluate the preceding argument at $s=i_1$. The strict monotonicity assumptions, together with
        $b_j>0$, make the double sum in \eqref{eq:abB_ineq} strictly
        positive, using $A_j$ in place of $B_j$ if necessary. The preceding comparison therefore gives $\rho'_{n,\gamma_2}(i_1,z)>\rho'_{n,\gamma_1}(i_1,z).$
        Since optimal consumption is positive, $0<i_1<a$,
        and the Euler equality gives
        $
        -u'(a-i_1)+\rho'_{n,\gamma_2}(i_1,z)
        =\rho'_{n,\gamma_2}(i_1,z)-\rho'_{n,\gamma_1}(i_1,z)>0.
        $
        Thus $i_1$ cannot maximize the $\gamma_2$ objective.
        Together with the weak policy order already proved, this yields
        $s_{n+1,\gamma_2}(a,z)>i_1$.
        The same argument applies to the infinite-horizon problem, using  corresponding derivative formula and Euler equality.
\end{proof}

\begin{lemma}[From policy order to stationary distribution and aggregates]
	\label{lm:stationary_policy_order}
	Fix $R\in(0,\bar R)$ and
	$0\le\gamma_1<\gamma_2$. Let $\XX_R:=[0,a_R^*]\times\ZZ$ be a common support set
	provided by Proposition~\ref{pr:com_sup}. If
	$ s_{\gamma_2}(a,z)\ge s_{\gamma_1}(a,z)$ for all $(a,z)\in \XX_R,$
	then:
	\begin{enumerate}
		\item\label{i:agg_mono}
                 $A(R,\gamma_2)\ge A(R,\gamma_1)$
                and
                $S(R,\gamma_2)\ge S(R,\gamma_1)$.		
		\item\label{i:dist_mono}
		For all bounded $h:\XX\to\RR$ such that $h(a,z)$ 
		is increasing in $a$,
		\begin{equation}\label{eq:lm_cs_distribution}
			\int h\,\diff\mu_{R,\gamma_2}\;
			\ge\;\int h\,\diff\mu_{R,\gamma_1}.
		\end{equation}
		In particular, if $\{z_t\}$ is Markov, then conditional on each 
		productivity state $z\in \ZZ$, the stationary wealth
		distribution under $\gamma_2$ first-order stochastically dominates 
		the one under $\gamma_1$.
	\end{enumerate}
	If, in addition, $s_{\gamma_2}(a,z)>s_{\gamma_1}(a,z)$
        whenever $(a,z)\in\XX_R$ and $s_{\gamma_1}(a,z)>0$,
       then both inequalities in  claim~(\ref{i:agg_mono})  are strict unless $S(R,\gamma_1)=S(R,\gamma_2)=0$.
\end{lemma}

\begin{proof}
	Fix an arbitrary $z_0\in\ZZ$, and let
	$\mu_{\gamma_i}:=\mu_{R,\gamma_i}$ for $i\in\{1,2\}$. For each $i$, 
	define 
	\begin{equation*}
		a_{i,0}=0
		\qquad\text{and}\qquad 
		a_{i,t+1}
		=R s_{\gamma_i}(a_{i,t},z_t)+wz_{t+1},
		\quad t\ge0.
	\end{equation*}
	Let $P_{\gamma_i}$ be the stochastic kernel of 
	$\{(a_{i,t},z_t)\}_{t\ge0}$. Since $(0,z_0)\in\XX_R$ and $\XX_R$ is invariant under
	both kernels, we have $(a_{i,t},z_t)\in\XX_R$
	almost surely for $i=1,2$ and all $t\ge0$.
	Moreover, $\mu_{\gamma_i}(\XX_R)=1$ for $i=1,2$.  We first verify that
	\begin{equation}\label{eq:iid_pathwise_order}
		a_{2,t}\ge a_{1,t}
		\quad \text{and} \quad 
		s_{\gamma_2}(a_{2,t},z_t)
		\ge s_{\gamma_1}(a_{1,t},z_t)
		\quad\text{almost surely for all } \; t\ge0.
	\end{equation}
	By Theorem~\ref{t:opt_general}\eqref{i:c*}, each saving policy is 
	increasing in wealth, while the hypothesis gives
	$s_{\gamma_2}(a,z)\ge s_{\gamma_1}(a,z)$ for all $(a,z)\in \XX_R$.
	The initial wealth levels coincide.
	Suppose $a_{2,t}\ge a_{1,t}$ at some date $t$.
	Then $s_{\gamma_2}(a_{2,t},z_t)\ge s_{\gamma_2}(a_{1,t},z_t)
	\ge s_{\gamma_1}(a_{1,t},z_t).$
	Since the two processes receive the same labor income at date $t+1$, 
	it follows that $a_{2,t+1}-a_{1,t+1} = R\bigl[
	    s_{\gamma_2}(a_{2,t},z_t)-s_{\gamma_1}(a_{1,t},z_t)
	\bigr] \ge0$. 
	Induction therefore yields \eqref{eq:iid_pathwise_order}.
	
	\emph{Step 1. Ordering of stationary aggregates}.
	Let $M:=\max\{1,a_R^*\}<\infty.$
	Feasibility of the saving policies implies
	$0\le s_{\gamma_i}(a_{i,t},z_t)\le a_{i,t}\le M$ almost surely for 
	$i=1,2$ and all $t\ge0$.
	Let $\nu_{i,t}$ be the distribution of $(a_{i,t},z_t)$.
	Then $\nu_{i,t}=\delta_{(0,z_0)}P_{\gamma_i}^{\,t}$, where
	$\delta_{(0,z_0)}$ is the point mass at $(0,z_0)$.
	By Theorem~\ref{t:gs_rs},
	$\|\nu_{i,t}-\mu_{\gamma_i}\|_{TV} \to 0$ as $t \to \infty$,
	and both $\nu_{i,t}$ and $\mu_{\gamma_i}$ assign probability one
	to $[0,M]\times\ZZ$. For aggregate comparison, by 
	\eqref{eq:agg_wealth}, 
	\begin{align*}
		\bigl|\EE a_{i,t}-A(R,\gamma_i)\bigr|
		= \left|
		\int_{\XX}a
		(\nu_{i,t}-\mu_{\gamma_i})[\diff(a,z)]
		\right|
		\le M\|\nu_{i,t}-\mu_{\gamma_i}\|_{TV}
		\to 0 \quad \text{and} \qquad \\
		\bigl|\EE s_{\gamma_i}(a_{i,t},z_t)-S(R,\gamma_i)\bigr|
		= \left|
		\int_{\XX} s_{\gamma_i}(a,z)
		(\nu_{i,t}-\mu_{\gamma_i})[\diff(a,z)]
		\right|
		\le
		M\|\nu_{i,t}-\mu_{\gamma_i}\|_{TV}
		\to0.
	\end{align*}
	Taking expectations in \eqref{eq:iid_pathwise_order}
	and then letting $t\to\infty$ gives 
	$A(R,\gamma_2)\ge A(R,\gamma_1)$ and
	$S(R,\gamma_2)\ge S(R,\gamma_1)$, so claim~\eqref{i:agg_mono} is verified.
	
	\emph{Step 2. Ordering of stationary distributions.}
	Fix $h$ as in claim~\eqref{i:dist_mono}. Since the two processes
	in \eqref{eq:iid_pathwise_order} share the same $\{z_t\}$ path
	and $h(a,z)$ is increasing in $a$, \eqref{eq:iid_pathwise_order} 
	implies $h(a_{2,t},z_t)\ge h(a_{1,t},z_t)$ almost surely for all $t\ge0$.
	So $\EE[h(a_{2,t},z_t)]\ge\EE[h(a_{1,t},z_t)]$ for all $t\ge0$.
	As boundedness of $h$ and the total variation convergence give
	$\int h\diff\nu_{i,t}\to\int h\diff\mu_{\gamma_i}$ for $i=1,2$,
	letting $t\to\infty$ in $\EE[h(a_{2,t},z_t)]\ge\EE[h(a_{1,t},z_t)]$ 
	verifies \eqref{eq:lm_cs_distribution}. 
	
	\emph{Step 3. Ordering of stationary wealth distribution.}
	Fix $z_j\in \ZZ$ and a bounded increasing
	$\varphi:\RR_+\to\RR$, and apply claim~\eqref{i:dist_mono} to
	$h(a,z)=\varphi(a)\mathbf 1\{z=z_j\}$. Both $\mu_{R,\gamma_1}$ and
	$\mu_{R,\gamma_2}$ have the productivity marginal $\pi$, the stationary
	distribution of $Q$, and $\pi(z_j)>0$ under the maintained assumptions.
	Dividing by $\pi(z_j)$ implies
	\begin{equation*}
		\int_{\RR_+}\varphi(a)\,\mu_{R,\gamma_2}(\diff a\mid z=z_j)
		\ge \int_{\RR_+}\varphi(a)\,\mu_{R,\gamma_1}(\diff a\mid z=z_j),
	\end{equation*}
	which is the stated conditional first-order stochastic dominance in
	claim~\eqref{i:dist_mono}.
	
	\emph{Step 4. Strict aggregate comparisons.}
	If $S(R,\gamma_1)=0<S(R,\gamma_2)$, the saving comparison is already strict.	
	Consider $S(R,\gamma_1)>0$. The function $h(a,z):=s_{\gamma_2}(\min\{a,a_R^*\},z)$ is bounded, 
	increasing in wealth, and equals $s_{\gamma_2}$ on $\XX_R$. Hence
    \eqref{eq:lm_cs_distribution} yields
    \begin{equation}\label{eq:agg_diff}
    S(R,\gamma_2)-S(R,\gamma_1)
    \ge \int_{\XX_R}s_{\gamma_2}(a,z)\,\diff\mu_{R,\gamma_1}
       -\int_{\XX_R}s_{\gamma_1}(a,z)\,\diff\mu_{R,\gamma_1}.
    \end{equation}
    Since $s_{\gamma_1}(a,z)\ge0$, $S(R,\gamma_1)>0$ implies
    $\mu_{R,\gamma_1}\{s_{\gamma_1}(a,z)>0\}>0$. The integrand is strictly positive on this
    set and nonnegative elsewhere, so  \eqref{eq:agg_diff} is strictly positive.
    Thus $S(R,\gamma_2)>S(R,\gamma_1)$ unless $S(R,\gamma_1)=S(R,\gamma_2)=0$.
    For $i\in\{1,2\}$, the identity $A(R,\gamma_i)=RS(R,\gamma_i)+w$ 
    gives $A(R,\gamma_2)>A(R,\gamma_1)$ with the same exception.	
\end{proof}

\begin{lemma}[From aggregate order to extremal equilibrium sets]
	\label{lm:extremal_equilibrium_order}
	There exists a compact interval $\CC=[R_L,R_H]\subset(\underline R,\bar R)$
	containing $\mathcal R(\gamma)$ for every $\gamma\ge0$.
	Fix $0\le\gamma_1<\gamma_2$ and let
        $r_1:=R_{\min}(\gamma_1)$ and $r_2:=R_{\max}(\gamma_2)$.
        If $S(r_i,\gamma_2)> S(r_i,\gamma_1)$ for $i=1,2$, then
        $R_{\min}(\gamma_2)< R_{\min}(\gamma_1)$ and 
        $R_{\max}(\gamma_2)< R_{\max}(\gamma_1).$
\end{lemma}

\begin{proof}
	By Step~2 of the proof of Theorem~\ref{t:RCE} in
	Appendix~\ref{s:pr_GE}, there exist $R_L<R_H$, independent of
	$\gamma$, such that $\CC:=[R_L,R_H]\subset(\underline R,\bar R)$,
	$\Psi(R_L,\gamma)<0<\Psi(R_H,\gamma)$, and
	$\mathcal R(\gamma)=\{R\in\CC:\Psi(R,\gamma)=0\}$
	for all $\gamma\ge0$. By Lemma~\ref{lm:agg_cont} and continuity
	of $K(R)$, $\Psi(\cdot,\gamma)$ is continuous on $\CC$. So
	$\mathcal R(\gamma)$ is nonempty, compact, and both extremal 
	equilibrium rates in \eqref{eq:iid_extremal_rates} are attained.
	
	Fix $0\le\gamma_1<\gamma_2$. 
	Suppose $S(r_i,\gamma_2)>S(r_i,\gamma_1)$ for $i=1,2$.
        By the definition of $\Psi$,
        $\Psi(r_i,\gamma_2)-\Psi(r_i,\gamma_1)
        =S(r_i,\gamma_2)-S(r_i,\gamma_1)>0$.
        At $r_1$, we have
        $\Psi(r_1,\gamma_2)>\Psi(r_1,\gamma_1)=0$.
        Since $\Psi(R_L,\gamma_2)<0$, continuity gives an equilibrium
        for $\gamma_2$ in $(R_L,r_1)$, so
        $R_{\min}(\gamma_2)<r_1=R_{\min}(\gamma_1)$.
        For the largest equilibria, we have
        $\Psi(r_2,\gamma_1)<\Psi(r_2,\gamma_2)=0$.
        Together with $\Psi(R_H,\gamma_1)>0$, continuity yields an
        equilibrium for $\gamma_1$ in $(r_2,R_H)$, hence
        $R_{\max}(\gamma_1)>r_2=R_{\max}(\gamma_2)$.
        This proves the lemma.
        \end{proof}

\subsection{Proof under IID productivity}
\label{ss:pr_comparative}

\begin{proof}[Proof of Proposition~\ref{pr:iid_gamma_saving}]
	Fix $R\in(0,\bar R)$ and $0\le\gamma_1<\gamma_2$.
	As the Bellman iterates in \eqref{eq:iid_continuation} are 
	independent of current
	productivity, we suppress $z$ in the notation
	of Section~\ref{ss:common_comparison} and write
	$p_j:=Q(z,z_j)$.
	We prove by induction that
	\begin{equation}\label{eq:iid_marginal_value_order}
		v'_{n,\gamma_2}(a)\ge v'_{n,\gamma_1}(a)
		\qquad\text{for all }\; a>0.
	\end{equation}
	The claim holds for $n=0$. Suppose it holds at iteration
	$n$ and define
	$D_n(x):=\gamma_2v_{n,\gamma_2}(x)-\gamma_1v_{n,\gamma_1}(x)$.
	Since marginal values are nonnegative,
	\begin{equation}\label{eq:iid_scaled_value_order}
		D'_n(x)
		=\gamma_2[v'_{n,\gamma_2}(x)-v'_{n,\gamma_1}(x)]
		+(\gamma_2-\gamma_1)v'_{n,\gamma_1}(x)
		\ge0.
	\end{equation}
	In the {\sc iid} case, the finite-horizon tilt is
	$
		f_{n,\gamma}(s,z_j)
		=\me^{-\gamma v_{n,\gamma}(x_j(s))}
		/\sum_k p_k\me^{-\gamma v_{n,\gamma}(x_k(s))},
	$
	so the density ratio satisfies
	\begin{equation}\label{eq:iid_tilted_ratio}
		\frac{f_{n,\gamma_2}(s,z_j)}{f_{n,\gamma_1}(s,z_j)}
		=C_n(s)\,\me^{-D_n(x_j(s))}
		\quad\text{with}\quad
		C_n(s):=\frac{\sum_k p_k\me^{-\gamma_1v_{n,\gamma_1}(x_k(s))}}
		{\sum_k p_k\me^{-\gamma_2v_{n,\gamma_2}(x_k(s))}}.
	\end{equation} 
	Here $C_n(s)>0$ is independent of $j$.
	Since $x_j(s)$ is increasing in $j$,
	\eqref{eq:iid_scaled_value_order} implies that the ratio
	in \eqref{eq:iid_tilted_ratio} is decreasing in $j$.
	Concavity also makes $v'_{n,\gamma_1}(x_j(s))$ decreasing
	in $j$. All the conditions of Lemma~\ref{lm:comparison_step}
	are therefore satisfied. Its policy comparison and
	\eqref{eq:iid_iteration_envelope} establish
	\eqref{eq:iid_marginal_value_order} at iteration $n+1$.
	In particular, for all $n\ge1$,
	\begin{equation}\label{eq:iid_iteration_policy_order}
		s_{n,\gamma_2}(a)\ge s_{n,\gamma_1}(a)
		\quad\text{and}\quad
		c_{n,\gamma_2}(a)\le c_{n,\gamma_1}(a).
	\end{equation}
	By Lemma~\ref{lm:finite_horizon}\eqref{i:fh_to_infh}, passing
	\eqref{eq:iid_iteration_policy_order} to the limit
	proves \eqref{eq:iid_policy_order}. Applying 
	Lemma~\ref{lm:stationary_policy_order} with the policy order
	\eqref{eq:iid_policy_order} yields \eqref{eq:iid_aggregate_order}
	together with the distributional comparison. 
	
	To establish strict policy ordering, we verify the strict 
	likelihood ratio condition in Lemma~\ref{lm:comparison_step}. 
	Note that $c_{\gamma_2}\le c_{\gamma_1}$ and Lemma~\ref{lm:T_diff} 
	imply $v'_{\gamma_2}\ge v'_{\gamma_1}>0$.
    Define $D(x):=\gamma_2v_{\gamma_2}(x)-\gamma_1v_{\gamma_1}(x)$.
    Then 
    $
    D'(x) =\gamma_2[v'_{\gamma_2}(x)-v'_{\gamma_1}(x)]
     +(\gamma_2-\gamma_1)v'_{\gamma_1}(x)>0.
    $
    For any feasible saving choice $s$, $x_j(s)=Rs+wz_j$
    is strictly increasing in $j$. Accordingly,
    $
    f_{\gamma_2}(s,z_j)/f_{\gamma_1}(s,z_j)
    \propto \me^{-D(x_j(s))}
    $
    is strictly decreasing in $j$.
    By Theorem~\ref{t:os_rs_opt} and Lemma~\ref{lm:T_diff},
    $v_{\gamma_1}$ is differentiable and strictly concave, so
    $j\mapsto v'_{\gamma_1}(x_j(s))$ is strictly decreasing.
    Since $p_j>0$ for all $j$, the infinite-horizon
    version of Lemma~\ref{lm:comparison_step} gives
    $s_{\gamma_2}(a)>s_{\gamma_1}(a)$ and 
    $c_{\gamma_2}(a)<c_{\gamma_1}(a)$
    whenever $s_{\gamma_1}(a)>0$.
    Lemma~\ref{lm:stationary_policy_order} yields the distributional
    and aggregate comparisons; both aggregate inequalities are
    strict unless $S(R,\gamma_1)=S(R,\gamma_2)=0$.
\end{proof}

\begin{proof}[Proof of Theorem~\ref{t:iid_gamma_equilibrium}]
        Fix $0\le\gamma_1<\gamma_2$ and let
        $r_1:=R_{\min}(\gamma_1)$ and $r_2:=R_{\max}(\gamma_2)$.
        Since market clearing and positivity of capital demand imply $S(r_1,\gamma_1)=K(r_1)>0$ and $S(r_2,\gamma_2)=K(r_2)>0$,
        Proposition~\ref{pr:iid_gamma_saving} implies
        $S(r_i,\gamma_2)>S(r_i,\gamma_1)$ for $i=1,2$.
        Lemma~\ref{lm:extremal_equilibrium_order} then yields
        \eqref{eq:iid_equilibrium_order}.
\end{proof}

\subsection{Proof under Markov productivity}
\label{ss:pr_markov_comparative}

Fix a compact set $\CC\subset(0,\bar R)$ and a continuous
wage schedule $w:\CC\to(0,\infty)$. Let $a_\CC$ be the
common bound provided by Proposition~\ref{pr:com_sup}.
Write $q_{ij}:=Q(z_i,z_j)$ and $\underline q_1:=\min_iq_{i1}>0$. Set
\begin{equation}\label{eq:markov_comparison_constants}
	\begin{gathered}
		\uw_\CC:=\min_{R\in\CC}w(R),\quad
		\ua:=\uw_\CC \uz,\\
		U_\CC:=u(a_\CC)-u(\ua),\quad 
		B_\CC:= U_\CC/(1-\beta),\quad
		\Lambda_\CC:=\beta B_\CC^2/[8(1-\beta)],\\
		\Delta z:=\min_{j<N}(z_{j+1}-z_j),\quad
		d_\CC:=\uw_\CC\Delta z\,u'(a_\CC)>0,  \quad
		\chi:=\beta \log\big(1/\underline q_1\big)/(1-\beta),
	\end{gathered}
\end{equation}
\vspace{-.5cm}
\begin{align}\label{eq:markov_utility_modulus}
	\varepsilon_\CC
	&:=\min\left\{(a_\CC-\ua)/2, \;
	(1-\beta)d_\CC /[2u'(\ua)]\right\}>0,\notag\\
	\zeta_\CC
	&:=\min_{\substack{c,\tilde c\in[\ua,a_\CC]\\
			|c-\tilde c|\ge\varepsilon_\CC}}
	\left\{u\left[(c+\tilde c)/2\right]
	- [u(c)+u(\tilde c)]/2\right\}>0.
\end{align}
The feasible set in \eqref{eq:markov_utility_modulus} is nonempty 
compact; strict concavity makes $\zeta_\CC$ positive. Let
\begin{equation}\label{eq:markov_gamma_bound}
	J_\CC:=\zeta_\CC d_\CC/(2U_\CC),\quad
	\gamma_\CC^{-}:=J_\CC/\Lambda_\CC>0,
	\quad \text{and} \quad 
	\gamma_\CC^+ :=\max\left\{2\gamma_\CC^-, \chi /J_\CC \right\}.
\end{equation}
Since $\zeta_\CC\le U_\CC$, we have
$\gamma_\CC^{-}\le d_\CC/(2\Lambda_\CC)$.
All constants in 
\eqref{eq:markov_comparison_constants}--\eqref{eq:markov_gamma_bound} 
depend only on the primitives and the common bound of 
Proposition~\ref{pr:com_sup}.

We establish the policy comparison on $\XX_\CC=[0,a_\CC]\times\ZZ$.
For $0\le a<\ua$, feasibility and the consumption floor
\eqref{eq:markov_consumption_lower} imply 
$a\ge c_\gamma(a,z)\ge\min\{a,\ua\}=a$, hence $c_\gamma(a,z) = a$.
Similarly, $c_{n,\gamma}(a,z)= a$ for all $0\le a< \ua$.
Thus optimal saving is zero in this region, uniformly in $\gamma$, 
and the policy comparison holds trivially there. 
It remains to consider $a\in I_\CC:=[\ua,a_\CC].$ 
Let $H_R:=[a_\CC-w(R)\bar z]/R>0.$ By \eqref{eq:markov_strict_saving_cap} 
in the proof of Proposition~\ref{pr:com_sup}, 
for $(a,z)\in I_\CC\times\ZZ$, $n\ge1$, and $\gamma\ge0$,
$
	s_{n,\gamma}(a,z)\le H_R- 1/R<H_R
	$ and $ 
	s_{\gamma}(a,z)\le H_R- 1/R<H_R.
$
By the normalization argument in Online Appendix~\ref{ss:opt_utility_ext}, 
we may subtract $u(\ua)$ from period utility without changing
policies, marginal values, or tilted probabilities. 
The optimal saving function then has domain $I_\CC\times\ZZ$ and 
range $[0,\min\{a-\ua, H_R\}]$. We henceforth write $u$ for the 
normalized utility $u-u(\ua)$ and retain the notation for the 
associated values.

Starting from $v_{0,\gamma}=0$, normalized current utility is at most
$(1-\beta)B_\CC$, while the certainty equivalent of a
continuation value in $[0,B_\CC]$ also lies in $[0,B_\CC]$. 
Induction shows that the transformed objects satisfy 
$0\le v_{n,\gamma}(a,z)\le B_\CC$ for all $(a,z)\in I_\CC\times\ZZ$.

\begin{proof}[Proof of Proposition~\ref{pr:markov_small_gamma}]
	We prove a more general result. Let $\CC$ be any compact
	subset of $(0, \bar R)$. We show that there exist 
	$0<\gamma_\CC^{-}<\gamma_\CC^{+}$ such that for all 
	$0\le\gamma_1<\gamma_2$ with $\gamma_1\le \gamma_\CC^{-}$ 
	or $\gamma_2\ge \gamma_\CC^{+}$, the stated claims hold uniformly
	for $R\in \CC$.
	
	\emph{Step 1. A criterion for the finite-horizon comparison.}
	Define $D_n(a,z):=\gamma_2v_{n,\gamma_2}(a,z)
	-\gamma_1v_{n,\gamma_1}(a,z)$. We show that
	$s_{n,\gamma_2}(a,z) \ge s_{n,\gamma_1}(a,z)$ for all 
	$(a,z) \in I_\CC\times \ZZ$
	if for all $n\ge1$ and $0\le s\le \min\{a-\ua, H_R\}$,
	\begin{equation}\label{eq:markov_comparison_criterion}
		\begin{gathered}
			j\to v'_{n,\gamma}(x_j(s),z_j)
			\quad\text{is decreasing for some }
			\gamma\in\{\gamma_1,\gamma_2\},\\
			j\to D_n(x_j(s), z_j) 
			\quad\text{is increasing}.
		\end{gathered}
	\end{equation}
	To that end, we apply Lemma~\ref{lm:comparison_step}.
	Note that the first condition of \eqref{eq:markov_comparison_criterion} 
	indicates $A_j$ or $B_j$ is decreasing in $j$. Since
	$a_j/b_j =f_{n,\gamma_2}(s,z_i,z_j)/f_{n,\gamma_1}(s,z_i,z_j)
	=C_{n,i}(s)\me^{-D_n(x_j(s), z_j)}$,
	where $C_{n,i}(s)>0$ is independent of $j$,
	\eqref{eq:markov_comparison_criterion} 
	implies that $a_j/b_j$ is decreasing in $j$. 
	
	Next, we show that $v_{n,\gamma_2}' \ge v_{n,\gamma_1}'$ on 
	$I_\CC\times \ZZ$ for all $n$.
	By \eqref{eq:iid_continuation},
	$v_{1,\gamma}'(a,z) = u'(a)$.
	Now suppose $v'_{n,\gamma_2}\ge v'_{n,\gamma_1}$ on
	$I_\CC\times\ZZ$. By Lemma~\ref{lm:comparison_step}, 
	$s_{n+1,\gamma_2}\ge s_{n+1,\gamma_1}$ and 
	$ c_{n+1,\gamma_2}\le c_{n+1,\gamma_1}$.
	The envelope identity \eqref{eq:iid_iteration_envelope}
	then gives $v'_{n+1,\gamma_2}\ge v'_{n+1,\gamma_1}$,
	closing the induction. This gives $A_j \ge B_j$.
	The stated claim then follows immediately from 
	Lemma~\ref{lm:comparison_step}.
	
	\emph{Step 2. Uniform value estimates.} 
	First, Lemma~\ref{lm:mono_productivity}, the envelope identity,
	and feasibility give, for $k>j$ and $n\ge1$,
	\begin{align}\label{eq:markov_diagonal_value_gap}
		&v_{n,\gamma}(x_k(s),z_k)-v_{n,\gamma}(x_j(s),z_j)
		\ge v_{n,\gamma}(x_k(s),z_j)-v_{n,\gamma}(x_j(s),z_j)\notag\\
		&\quad=\int_{x_j(s)}^{x_k(s)}u'(c_{n,\gamma}(a,z_j))\,\diff a
		\ge\underline w_\CC(z_k-z_j)u'(a_\CC).
	\end{align}
	For a fixed random variable $X\in[0,B_\CC]$, write
	$f(t):=\log\EE_z\me^{-tX}.$
	Differentiating with $X$ held fixed gives
	$f'(t)=-\EE_z X\me^{-tX} / \EE_z\me^{-tX}$
	and $f''(t)=\Var_{t,z}(X)\le B_\CC^2/4$,
	where $\Var_{t,z}$ denotes variance under the tilted
	probability $\me^{-tX}/\EE \me^{-tX}$. As $f(0)=0$,
	\begin{equation*}
		0\le-\frac{\partial\rho_{\gamma,z}(X)}{\partial\gamma}
		=\frac{\gamma f'(\gamma)-f(\gamma)}{\gamma^2}
		=\frac1{\gamma^2}\int_0^\gamma t\Var_{t,z}(X)\,\diff t
		\le\frac{B_\CC^2}{8}.
	\end{equation*}
	Integrating between $\gamma_1$ and $\gamma_2$, and using
	continuity at zero when $\gamma_1=0$, yields
	\begin{equation}\label{eq:bd_rho}
		0\le\rho_{\gamma_1,z}(X)-\rho_{\gamma_2,z}(X)
		\le B_\CC^2(\gamma_2-\gamma_1)/8.
	\end{equation}
	Monotonicity therefore gives $v_{n,\gamma_1}\ge v_{n,\gamma_2}$
	by induction from zero. Define
	\begin{equation*}
		\delta_n:=\sup_{(a,z)\in I_\CC\times\ZZ}
		|v_{n,\gamma_1}(a,z)-v_{n,\gamma_2}(a,z)|.
	\end{equation*}
	Recall that $T_\gamma$ uses the common saving restriction
	$0\le s\le\min\{a-\ua,H_R\}$. For each such choice,
	$\hat a=Rs+w(R)\hat z$ satisfies
	$(\hat a,\hat z)\in I_\CC\times\ZZ$ by the definition of $H_R$.
	First, since $0\le v_{n,\gamma_1}-v_{n,\gamma_2}\le\delta_n$
	on $I_\CC\times\ZZ$, monotonicity and cash invariance
	(Lemma~\ref{lm:entropic}\eqref{i:mono}
	and \eqref{i:drift_inv}) imply
	$\rho_{\gamma_1,z} (v_{n,\gamma_1}(\hat a,\hat z))
	\le \rho_{\gamma_1,z} (v_{n,\gamma_2}(\hat a,\hat z)+\delta_n)
	= \rho_{\gamma_1,z} (v_{n,\gamma_2}(\hat a,\hat z))+\delta_n$.
	Multiplying by $\beta$, adding current utility,
	and maximizing over the same saving set gives
	\begin{equation}\label{eq:contract_bd}
		0\le T_{\gamma_1}v_{n,\gamma_1}(a,z) -T_{\gamma_1}v_{n,\gamma_2}(a,z)
		\le \beta\delta_n.
	\end{equation}
	Second, hold $v_{n,\gamma_2}$ fixed. Since its successor
	values lie in $[0,B_\CC]$, the preceding
	certainty equivalent bound \eqref{eq:bd_rho} yields
	$0 \le \rho_{\gamma_1,z} (v_{n,\gamma_2}(\hat a,\hat z))
	- \rho_{\gamma_2,z}(v_{n,\gamma_2}(\hat a,\hat z))
	\le B_\CC^2(\gamma_2-\gamma_1)/8$.
	As this bound is uniform over saving choices, the same
	maximization argument gives
	\begin{equation}\label{eq:diff_bd}
		0\le T_{\gamma_1}v_{n,\gamma_2}(a,z)-T_{\gamma_2}v_{n,\gamma_2}(a,z)
		\le\beta B_\CC^2(\gamma_2-\gamma_1)/8.
	\end{equation}
	By \eqref{eq:contract_bd} and \eqref{eq:diff_bd}, 
	$
		\delta_{n+1} = \sup_{(a,z)\in I_\CC\times\ZZ} \left|
		T_{\gamma_1}v_{n,\gamma_1}(a,z)
		-T_{\gamma_2}v_{n,\gamma_2}(a,z)
		\right|
		\le \beta\delta_n +\beta B_\CC^2 (\gamma_2-\gamma_1) \big/ 8.
	$
	As $\delta_0=0$, this implies
	$\delta_n 
		\le \beta B_\CC^2(1-\beta^n)(\gamma_2-\gamma_1) /[8(1-\beta)]
		\le\Lambda_\CC(\gamma_2-\gamma_1)$,	
	so
	\begin{equation}\label{eq:markov_value_loss}
		0\le v_{n,\gamma_1}(a,z)-v_{n,\gamma_2}(a,z)
		\le\Lambda_\CC(\gamma_2-\gamma_1)
		\quad\text{on }\;I_\CC\times\ZZ.
	\end{equation}
	\emph{Step 3. A uniform strict concavity bound.}
	Consider expected utility first. Fix $n\ge1$, $z\in\ZZ$, and $\ua\le a<b\le a_\CC$
	with $b-a\ge \uw_\CC\Delta z$. Feasibility and Lemma~\ref{lm:T_diff} give
	\begin{equation}\label{eq:diff_value}
		v_{n,0}(b,z)-v_{n,0}(a,z)
		=\int_a^b u'(c_{n,0}(x,z))\,\diff x
		\ge(b-a)u'(a_\CC)\ge d_\CC.
	\end{equation}
	Use the same productivity path with $z_0=z$. For $q\in\{a,b\}$ and 
	$t< n$, set
	$a_0^q=q$, $c_t^q:=c_{n-t,0}(a_t^q,z_t)$, and $a_{t+1}^q=R(a_t^q-c_t^q)+w(R)z_{t+1}$.
	By Proposition~\ref{pr:com_sup} and
	\eqref{eq:markov_consumption_lower},
	$\ua\le c_t^q\le a_t^q\le a_\CC$.
	Put $E_t:=\{|c_t^b-c_t^a|\ge\varepsilon_\CC\}$.
	Since
	$
		|u(c_t^b)-u(c_t^a)|
		\le u'(\ua)\varepsilon_\CC\1_{E_t^c}
		+U_\CC\1_{E_t},
	$
	optimality of the two paths, the definition of $\epsilon_\CC$ in
	\eqref{eq:markov_utility_modulus}, and \eqref{eq:diff_value} yield
	\begin{equation*}
		d_\CC
		\le\EE_z\sum_{t=0}^{n-1}\beta^t[u(c_t^b)-u(c_t^a)]
		\le\frac{u'(\ua)\varepsilon_\CC}{1-\beta}
		+U_\CC\sum_{t=0}^{n-1}\beta^t\PP_z(E_t)
		\le\frac{d_\CC}{2}
		+U_\CC\sum_{t=0}^{n-1}\beta^t\PP_z(E_t).
	\end{equation*}
	Hence, $\sum_{t=0}^{n-1}\beta^t\PP_z(E_t)\ge d_\CC/(2U_\CC)$.
	The midpoint plan $a_t^m:=(a_t^a+a_t^b)/2$ and $c_t^m:=(c_t^a+c_t^b)/2$ 
	is adapted and satisfies $a_0^m=(a+b)/2$,
	$0\le c_t^m\le a_t^m$, and 
	$a_{t+1}^m=R(a_t^m-c_t^m)+w(R)z_{t+1}$.
	By feasibility, optimality of $v_{n,0}$, and definition
	of $\zeta_\CC$ in \eqref{eq:markov_utility_modulus},
	\begin{align}\label{eq:markov_eu_midpoint_gap}
		&\mathcal J_{n,R}(a,b,z)
		:=v_{n,0}\left((a+b)/2,z\right)
		-[v_{n,0}(a,z)+v_{n,0}(b,z)]/2\notag\\
		&\ge\EE_z\sum_{t=0}^{n-1}\beta^t
		\left[u\!\left(\frac{c_t^a+c_t^b}{2}\right)
		-\frac{u(c_t^a)+u(c_t^b)}{2}\right]
		\ge\zeta_\CC\sum_{t=0}^{n-1}\beta^t\PP_z(E_t)
		\ge J_\CC>0.
	\end{align}
	The bound is uniform over the stated choices of $n,R,z,a,b$.
	
	Alongside the expected-utility problem, consider the deterministic
	problem with constant income $w(R)z_1$. Let its $n$-period
	value function and consumption policy be $v_n^D$ and $c_n^D$,
	respectively. Its normalized values satisfy $v_0^D=0$ and
	\begin{equation*}
		v_{n+1}^D(a)
		=\max_{0\le s\le\min\{a-\ua,H_R\}}
		\bigl\{u(a-s)+\beta v_n^D(x_1(s))\bigr\}.
	\end{equation*}
	The same argument applies to this special deterministic problem: take
	$z_t\equiv z_1$, replace $c_{n-t,0}(a_t^q,z_t)$ by
	$c_{n-t}^D(a_t^q)$, and omit expectations. The common consumption
	floor and wealth bound ensure that the same constants apply.
	Thus
	\begin{equation}\label{eq:markov_de_midpoint_gap}
		v_n^D\left((a+b)/2\right)
		-[v_n^D(a)+v_n^D(b)]/2
		\ge J_\CC>0.
	\end{equation}
	Passing to the infinite horizon limit gives the same bound for
	$v^D:=\lim_{n\to\infty}v_n^D$, uniformly over the stated choices
	of $R,a,b$.
	
	\emph{Step 4. The zero-gap region near expected utility.}
	Fix $0\le\gamma_1<\gamma_2$ with $\gamma_1\le\gamma_\CC^{-}$. 
	We first show that $c_{n,0}(a,z)$ is increasing in $z$ for all $n$.
	Since $c_{1,0}(a,z)=a$, suppose inductively that $c_{n,0}(a,z)$ is 
	increasing in $z$. The monotonicity of $x_j(s)$ in $j$
	then makes $j\mapsto u'(c_{n,0}(x_j(s),z_j))$ decreasing.
	For $i<k$, Assumption~\ref{a:mlr} therefore gives
	\begin{align*}
		\rho'_{n,0}(s,z_k)
		&\le\beta R\sum_j Q(z_i,z_j)
		u'(c_{n,0}(x_j(s),z_j))
		=\rho'_{n,0}(s,z_i).
	\end{align*}
	Thus $\rho_{n,0}(s,z_i)-\rho_{n,0}(s,z_k)$ is
	increasing in $s$. The maximizer argument in the proof of
	Lemma~\ref{lm:comparison_step} yields
	$s_{n+1,0}(a,z_i)\ge s_{n+1,0}(a,z_k)$, completing
	the induction. By \eqref{eq:iid_iteration_envelope},
	\begin{equation}\label{eq:markov_eu_marginal_order}
		v'_{n,0}(a,z_{j+1})
		=u'(c_{n,0}(a,z_{j+1}))
		\le u'(c_{n,0}(a,z_j))=v'_{n,0}(a,z_j)
		\qquad(n\ge1).
	\end{equation}
	Fix $n\ge1$, $0\le s\le \min\{a-\ua,H_R\}$, and $j<N$. Put
	$\xi_0:=x_j(s)$, $\xi_1:=x_{j+1}(s)$, $h:=\xi_1-\xi_0$, and $m:=(\xi_0+\xi_1)/2$.
	Then $\ua\le \xi_0<m<\xi_1\le a_\CC$ and
	$h=w(R)(z_{j+1}-z_j)\ge \uw_\CC\Delta z$.
	For $\gamma\ge0$, concavity gives
	\begin{align}\label{eq:marginal_est}
		v'_{n,\gamma}(\xi_0,z_j)
		&\ge 2[v_{n,\gamma}(m,z_j)-v_{n,\gamma}(\xi_0,z_j)]\big/ h,\notag\\
		v'_{n,\gamma}(\xi_1,z_{j+1})
		&\le 2[v_{n,\gamma}(\xi_1,z_{j+1})-v_{n,\gamma}(m,z_{j+1})] \big/ h.
	\end{align}
	At $\gamma=0$, integrating \eqref{eq:markov_eu_marginal_order}
	over $[m,\xi_1]$ yields
	$
		v_{n,0}(\xi_1,z_{j+1})-v_{n,0}(m,z_{j+1})
		\le v_{n,0}(\xi_1,z_j)-v_{n,0}(m,z_j).
	$
	Moreover, \eqref{eq:markov_value_loss} gives
	$0\le v_{n,0}-v_{n,\gamma}\le\Lambda_\CC\gamma$ on
	$I_\CC\times\ZZ$. Applying this bound to the two positive terms
	below and using $-v_{n,\gamma}\ge-v_{n,0}$ for the other
	two terms, 
	\begin{align*}
		&v_{n,\gamma}(m,z_j)-v_{n,\gamma}(\xi_0,z_j)
		-v_{n,\gamma}(\xi_1,z_{j+1})+v_{n,\gamma}(m,z_{j+1})\\
		&\ge v_{n,0}(m,z_j)-v_{n,0}(\xi_0,z_j)
		-v_{n,0}(\xi_1,z_{j+1})+v_{n,0}(m,z_{j+1})
		-2\Lambda_\CC\gamma\\
		&\ge2v_{n,0}(m,z_j)-v_{n,0}(\xi_0,z_j)-v_{n,0}(\xi_1,z_j)
		-2\Lambda_\CC\gamma
		=2\mathcal J_{n,R}(\xi_0,\xi_1,z_j)-2\Lambda_\CC\gamma.
	\end{align*}
	Combining these estimates with \eqref{eq:markov_eu_midpoint_gap}, \eqref{eq:marginal_est}
	and $\gamma\le\gamma_\CC^{-}$ gives
	\begin{align}\label{eq:markov_diagonal_marginal_order}
		&v'_{n,\gamma}(x_j(s),z_j)
		-v'_{n,\gamma}(x_{j+1}(s),z_{j+1})\notag\\
		&\quad\ge4
		[\mathcal J_{n,R}(\xi_0,\xi_1,z_j)-\Lambda_\CC\gamma]\big/h
		\ge 4(J_\CC-\Lambda_\CC\gamma)/h
		\ge0.
	\end{align}
	Hence
	$j\to v'_{n,\gamma}(x_j(s),z_j)$ is decreasing, establishing the first condition in \eqref{eq:markov_comparison_criterion}.
	For the second condition, recall that
	$D_n=(\gamma_2-\gamma_1)v_{n,\gamma_2}
	-\gamma_1(v_{n,\gamma_1}-v_{n,\gamma_2}).$
	The one-sided estimate \eqref{eq:markov_value_loss} and
	the diagonal value bound \eqref{eq:markov_diagonal_value_gap}
	established in Step~2 give
	\begin{align}\label{eq:markov_scaled_value_order}
		D_n(x_{j+1}(s),z_{j+1})-D_n(x_j(s),z_j)
		\ge(\gamma_2-\gamma_1)(d_\CC-\gamma_1\Lambda_\CC)\ge0,
	\end{align}
	where the last inequality follows from
	$\gamma_1\le\gamma_\CC^{-}\le d_\CC/(2\Lambda_\CC)$.
	Thus both conditions in
	\eqref{eq:markov_comparison_criterion} hold for every
	$n\ge1$ and $0\le s\le \min \{a-\ua, H_R\}$.
	
	\emph{Step 5. The zero-gap region at high risk sensitivity.} 
	For any vector $y=(y_1,\dots,y_N)$ and $\gamma>0$,
	$q_{i1}\me^{-\gamma y_1}
	\le \sum_jq_{ij}\me^{-\gamma y_j}
	\le \me^{-\gamma\min_jy_j}$.
	Applying $-\gamma^{-1}\log$ gives
	\begin{equation}\label{eq:markov_soft_min_bounds}
		\min_jy_j\le\rho_{\gamma,z_i}(y)
		\le y_1 + \log(1/q_{i1}) / \gamma.
	\end{equation}
	We first compare stochastic and deterministic values.
	Observe $v_{1,\gamma}(a,z_i)=v_1^D(a)=u(a)$. 
	Suppose $0\le v_{n,\gamma}-v_n^D\le \chi/\gamma$.
	As $v_n^D$ is increasing, by monotonicity, cash invariance,
	and \eqref{eq:markov_soft_min_bounds}, 
	$v_n^D(x_1(s))
		\le \rho_{\gamma,z_i}
		\bigl(v_{n,\gamma}(x_j(s),z_j)\bigr)
		\le v_n^D(x_1(s))
		+ [\chi+\log(1/\underline q_1)] \big/\gamma$.
	Since $\beta \chi+\beta\log(1/\underline q_1)=\chi$, 
	maximizing over the common saving set yields
	\begin{equation}\label{eq:markov_uniform_value_bound}
		0\le v_{n,\gamma}(a,z_i)-v_n^D(a)\le \chi/\gamma
		\qquad (a\in I_\CC,\, n\ge1,\,\gamma>0).
	\end{equation}      
	Fix $n\ge1$, $0\le s\le\min\{a-\ua, H_R\}$, and $j<N$.
	Concavity, \eqref{eq:markov_de_midpoint_gap}, 
	\eqref{eq:marginal_est}, and \eqref{eq:markov_uniform_value_bound} give 
	\begin{align}\label{eq:markov_high_diagonal_marginals}
		&v'_{n,\gamma}(\xi_0,z_j)-v'_{n,\gamma}(\xi_1,z_{j+1})\notag\\
		&\ge\frac2{\xi_1-\xi_0}
		\bigl[v_{n,\gamma}(m,z_j)-v_{n,\gamma}(\xi_0,z_j)
		-v_{n,\gamma}(\xi_1,z_{j+1})
		+v_{n,\gamma}(m,z_{j+1})\bigr]\notag\\
		&\ge\frac4{\xi_1-\xi_0}
		\left[v_n^D(m)-\frac{v_n^D(\xi_0)+v_n^D(\xi_1)}2-\frac \chi\gamma\right]
		\ge\frac4{\xi_1-\xi_0}\left(J_\CC-\frac \chi\gamma\right).
	\end{align}
	We next establish the second condition in \eqref{eq:markov_comparison_criterion}. 
	Fix $0<\gamma_1<\gamma_2$ and a vector $y$ with
	$y_1=\min_jy_j$. Define
	$Z_\ell:=\sum_jq_{ij}\me^{-\gamma_\ell(y_j-y_1)}, \ell=1,2$.
	Since $y_j-y_1\ge0$ and the $j=1$ term equals $q_{i1}$,
	we have $q_{i1}\le Z_2\le Z_1\le1$.
	Jensen's inequality, applied to
	$t\mapsto t^{\gamma_2/\gamma_1}$, also gives
	$Z_2\ge Z_1^{\gamma_2/\gamma_1}$.
	Using
	$\rho_{\gamma_\ell,z_i}(y)=y_1-\gamma_\ell^{-1}\log Z_\ell$,
	we therefore obtain
	\begin{align}\label{eq:markov_entropic_inverse_bound}
		0
		&\le \rho_{\gamma_1,z_i}(y)-\rho_{\gamma_2,z_i}(y)
		=-\log Z_1/\gamma_1 + \log Z_2/\gamma_2  \notag\\
		&\le
		\left(1/\gamma_1-1/\gamma_2\right)(-\log Z_1)
		\le
		\left(1/\gamma_1-1/\gamma_2\right)
		\log (1/q_{i1}).
	\end{align}
	The lower bound uses $Z_2\ge Z_1^{\gamma_2/\gamma_1}$; the two upper bounds
	follow from $Z_2\le Z_1$ and $Z_1\ge q_{i1}$, respectively. By an argument similar to the proof of \eqref{eq:markov_value_loss} in Step 2, 
	\begin{equation}\label{eq:markov_inverse_parameter_bound}
		0\le v_{n,\gamma_1}(a,z)-v_{n,\gamma_2}(a,z)
		\le \chi\left(1/\gamma_1-1/\gamma_2\right),
		\qquad (a,z)\in I_\CC\times\ZZ.
	\end{equation}
	Recall that
	$D_n=(\gamma_2-\gamma_1)v_{n,\gamma_2}
	-\gamma_1(v_{n,\gamma_1}-v_{n,\gamma_2})$.
	By \eqref{eq:markov_diagonal_value_gap} and \eqref{eq:markov_inverse_parameter_bound}, we obtain
	\begin{align}\label{eq:markov_high_scaled_value_order}
		&D_n(x_{j+1}(s),z_{j+1})-D_n(x_j(s),z_j)\notag\\
		&\ge
		(\gamma_2-\gamma_1)d_\CC
		-\gamma_1\chi\left(1/\gamma_1-1/\gamma_2\right)
		=(\gamma_2-\gamma_1)\left(d_\CC-\chi/\gamma_2\right).
	\end{align}
	Fix $\gamma_{\CC}^{-}\le\gamma_1<\gamma_2$ with
	$\gamma_2\ge\gamma_\CC^+$.
	For every $n\ge1$ and $0\le s\le \min\{a-\ua, H_R\}$,
	\eqref{eq:markov_high_diagonal_marginals} and
	$\gamma_2\ge \chi/J_\CC$ show that
	$j\to v'_{n,\gamma_2}(x_j(s),z_j)$
	is decreasing. Thus the first condition in
	\eqref{eq:markov_comparison_criterion} holds at $\gamma_2$.
	\eqref{eq:markov_high_scaled_value_order}
	and $\chi/\gamma_2\le J_\CC\le d_\CC/2$ imply that
	$j\mapsto D_n(x_j(s),z_j)$ is increasing.
	Hence the second condition also holds.
	
	\emph{Step 6. Passage to the infinite horizon and stationary aggregates.}
	For each $\gamma\in\{\gamma_1,\gamma_2\}$,
	Lemma~\ref{lm:finite_horizon}\eqref{i:fh_to_infh} gives
	$s_{n+1,\gamma}(a,z)\to s_{\gamma}(a,z)$ and 
	$c_{n+1,\gamma}(a,z)\to c_{\gamma}(a,z).$
	Passing to the limit in the policy order from Step~4 and Step~5 proves
	\eqref{eq:markov_policy_order} for $0<a\le a_\CC$; at $a=0$,
	both consumption and saving equal zero.
	The common support in Proposition~\ref{pr:com_sup}
	then permits an application of Lemma~\ref{lm:stationary_policy_order},
	which yields \eqref{eq:markov_aggregate_order} together with the 
	distributional comparison stated in the proposition.
	
	To establish strict ordering, fix $(a,z)\in\XX_\CC$ with
    $i_1:=s_{\gamma_1}(a,z)>0$, and put
    $x_j:=Ri_1+w(R)z_j$.
    The common invariant set ensures $(x_j,z_j)\in\XX_\CC$.
    Thus the above weak policy ordering and Lemma~\ref{lm:T_diff} imply
    $v'_{\gamma_2}(x_j,z_j)\ge v'_{\gamma_1}(x_j,z_j)$
    for every $j$.   
    First, value convergence preserves the weak bounds used to
    derive \eqref{eq:markov_diagonal_marginal_order} and
    \eqref{eq:markov_high_diagonal_marginals}.
    Next, the strict concavity of $v^*$ in Theorem~\ref{t:os_rs_opt} 
    makes the inequalities in \eqref{eq:marginal_est} strict for the 
    infinite-horizon value functions.
    Combining these strict inequalities with the limiting value
    bounds gives, for $j<N$ and $h_j:=x_{j+1}-x_j>0$,       
    \begin{align*}
    v'_{\gamma_1}(x_j,z_j)-v'_{\gamma_1}(x_{j+1},z_{j+1})
    &>4(J_\CC-\Lambda_\CC\gamma_1)/h_j\ge0
    &&\text{if }\gamma_1\le\gamma_\CC^-,\\
    v'_{\gamma_2}(x_j,z_j)-v'_{\gamma_2}(x_{j+1},z_{j+1})
    &>4\left(J_\CC-\chi/\gamma_2\right)/h_j\ge0
    &&\text{if }\gamma_2\ge\gamma_\CC^+.
    \end{align*}
     Hence $j\mapsto v'_{\gamma_1}(x_j,z_j)$ is strictly decreasing
    in the low-parameter region, and
    $j\mapsto v'_{\gamma_2}(x_j,z_j)$ is strictly decreasing
    in the high-parameter region.
    
    By \eqref{eq:markov_gamma_bound},
    $\gamma_1\Lambda_\CC\le J_\CC\le d_\CC/2$
    in the low-parameter region, and
    $\chi/\gamma_2\le J_\CC\le d_\CC/2$
    in the high-parameter region.
    Thus the lower bounds in
    \eqref{eq:markov_scaled_value_order} and
    \eqref{eq:markov_high_scaled_value_order} are at least
    $(\gamma_2-\gamma_1)d_\CC/2>0$, uniformly in $n$. Taking 
    the limit shows that
    $j\mapsto D(x_j,z_j)$ is strictly increasing, where
    $D:=\gamma_2v_{\gamma_2}-\gamma_1v_{\gamma_1}$. Hence
    $
    f_{\gamma_2}(i_1,z,z_j)/f_{\gamma_1}(i_1,z,z_j)
    \propto \me^{-D(x_j,z_j)}
    $
    is strictly decreasing in $j$.
    These inequalities verify the strict comparison conditions
    of Lemma~\ref{lm:comparison_step} at $s=i_1$.
    Applying its infinite-horizon version yields
    $s_{\gamma_2}(a,z)>s_{\gamma_1}(a,z).$  
    Lemma~\ref{lm:stationary_policy_order} then implies the
    distributional and aggregate comparisons, with both aggregate
    inequalities strict unless
    $S(R,\gamma_1)=S(R,\gamma_2)=0$.
\end{proof}

\begin{proof}[Proof of Theorem~\ref{t:markov_gamma_equilibrium}]
	Let $\CC=[R_L,R_H]$ be the common equilibrium interval from 
	Lemma~\ref{lm:extremal_equilibrium_order}, and set
	$\gamma_*^{-}:=\gamma_\CC^{-}$ and $\gamma_*^{+}:=\gamma_\CC^{+}$ defined in
	\eqref{eq:markov_gamma_bound}. For $0\le\gamma_1<\gamma_2$ with
	$\gamma_1\le\gamma_*^{-}$ or $\gamma_2\ge \gamma_*^{+}$, the comparison established in the proof of Proposition~\ref{pr:markov_small_gamma}
	gives $S(R,\gamma_2)\ge S(R,\gamma_1)$ for every $R\in\CC$.
	Let $r_1:=R_{\min}(\gamma_1)$ and $r_2:=R_{\max}(\gamma_2)$.
    Since $r_1,r_2\in\CC$, and market clearing and positivity of capital demand imply $S(r_1,\gamma_1)=K(r_1)>0$ and $S(r_2,\gamma_2)=K(r_2)>0,$
    the aggregate comparisons at both prices are strict:
    $S(r_i,\gamma_2)>S(r_i,\gamma_1)$ for $i=1,2$.
    Lemma~\ref{lm:extremal_equilibrium_order} then yields
    \eqref{eq:markov_equilibrium_order}.		
\end{proof}

\begin{proof}[Proof of Proposition~\ref{pr:decomp}]
	For all bounded $X$, Lemma~\ref{lm:entropic}\eqref{i:mono_gam}
	gives $\rho_{\gamma_1}(X)\ge\rho_{\gamma_2}(X)$. So, for any fixed
	policy, the recursive criterion is decreasing in $\gamma$. Taking  
	supremum over policies then gives
	$v_{R,\gamma_1}\ge v_{R,\gamma_2}$, hence
	$L_j:=V_{1j}-V_{2j}\ge0$ for all $j$. 
	
	Fix $(a,z)\in\XX$, $R \in (0,\bar R), w>0$, $0\le\gamma_1<\gamma_2$, 
	and suppose $s_1:=s_{R,\gamma_1}(a,z)>0$. For $\gamma\ge0$ and
	$s\in[0,a]$, define 
	$\Phi_\gamma(s) 
		:= u(a-s) + \beta \rho_{\gamma,z}\bigl(
		v_{R,\gamma}(Rs+w\hat z,\hat z)
	\bigr)$. At $\gamma_1$, the choice $s_1$ is an interior maximizer, 
	so $\Phi'_{\gamma_1}(s_1)=0$. By the derivative formula 
	\eqref{eq:part_vhat} of Appendix~\ref{s:pr_euler},
	for $M_k:=\sum_j p_{kj} m_{kj}, k=1,2$, we have
	\begin{equation*}
		\Phi'_{\gamma_1}(s_1)
		= -u'(a-s_1)+\beta R\sum_j p_{1j} m_{1j}
		= -u'(a-s_1)+\beta R M_1
		= 0,
	\end{equation*}
	At $\gamma_2$, evaluating the derivative at the same point gives
	\begin{equation*}
		\Phi'_{\gamma_2}(s_1)
		= -u'(a-s_1)+\beta R M_2
		= \beta R (M_2-M_1).
	\end{equation*}
	Note that $\Phi_{\gamma_2}$ is strictly concave in $s$: $u$ is
	strictly concave, $s\mapsto v_{R,\gamma_2}(Rs+w \hat z,\hat z)$
	is concave for each $\hat z$ and $\rho_{\gamma_2,z}$ is monotone and
	concave by Lemma~\ref{lm:entropic}\eqref{i:mono}--\eqref{i:concave}. 
	Hence
	\begin{equation}\label{eq:s_mono_equiv}
		s_{R,\gamma_2}(a,z)\ge s_1
		\quad \iff \quad 
		\Phi'_{\gamma_2}(s_1)\ge0
		\quad \iff \quad 
		M_2-M_1\ge0,
	\end{equation}
	and $s_{R,\gamma_2}(a,z)<s_1$ if and only if $M_2-M_1<0$. 
	It remains to decompose $M_2-M_1$. Adding and subtracting
	$\sum_j\tilde p_j\,m_{1j}$,
	\begin{equation}\label{eq:decomp_algebra}
		M_2-M_1
		=\sum_j\bigl(\tilde p_j-p_{1j}\bigr)m_{1j}
		+\sum_j\bigl(p_{2j}-\tilde p_j\bigr)m_{1j}
		+\sum_j p_{2j}\bigl(m_{2j}-m_{1j}\bigr).
	\end{equation}
	By the definition of $p_1$ and $\tilde p$,
	\begin{equation*}
		p_{1j} = \frac{Q(z,z_j)\me^{-\gamma_1 V_{1j}}}
		{\sum_\ell Q(z,z_\ell)\me^{-\gamma_1 V_{1\ell}}}
		\quad \text{and} \quad 
		\tilde p_j
		= \frac{Q(z,z_j)\me^{-\gamma_2 V_{1j}}}
		{\sum_\ell Q(z,z_\ell)\me^{-\gamma_2 V_{1\ell}}}
		= \frac{\,p_{1j}\me^{-(\gamma_2-\gamma_1)V_{1j}}\,}
		{\EE_{p_1}\me^{-(\gamma_2-\gamma_1)V_1}}.
	\end{equation*}
	\begin{equation*}
		\Longrightarrow \quad 
		\sum_j\bigl(\tilde p_j-p_{1j}\bigr)m_{1j}
		= \frac{\EE_{p_1}m_1 \me^{-(\gamma_2-\gamma_1)V_1}
		}{\EE_{p_1}\me^{-(\gamma_2-\gamma_1)V_1}} 
		- \EE_{p_1} m_1
		= \frac{\Cov_{p_1}\bigl(m_1, \me^{-(\gamma_2-\gamma_1)V_1}\bigr)}
		{\EE_{p_1} \me^{-(\gamma_2-\gamma_1)V_1}}
		= T.
	\end{equation*}
	Since $V_{2j}=V_{1j}-L_j$, we have
	$p_{2j} \propto Q(z,z_j)\me^{-\gamma_2V_{2j}}
	= Q(z,z_j)\me^{-\gamma_2V_{1j}} \me^{\gamma_2L_j}
	\propto \tilde p_j \me^{\gamma_2L_j}$, which implies
	$p_{2j} = \tilde p_j\me^{\gamma_2L_j}/\EE_{\tilde p}\me^{\gamma_2L}$.
	Consequently,
	\begin{equation*} 
		\sum_j\bigl(p_{2j}-\tilde p_j\bigr)m_{1j}
		= \frac{\,\EE_{\tilde p} \, m_1 \me^{\gamma_2L}\,
		}{\EE_{\tilde p} \me^{\gamma_2L}} - \EE_{\tilde p} \, m_1
		= \frac{\Cov_{\tilde p}\bigl(m_1,\me^{\gamma_2L}\bigr)
		}{\EE_{\tilde p} \me^{\gamma_2L}}
		=W.
	\end{equation*}
	The envelope identity of Lemma~\ref{lm:T_diff} gives
	$m_{kj} = v'_{R,\gamma_k}(x_j,z_j)
	= u'\bigl(c_{R,\gamma_k}(x_j,z_j)\bigr)$ for $k=1,2$,
	so $\sum_j p_{2j}(m_{2j}-m_{1j})=\EE_{p_2}(m_2-m_1) = B$ as stated.
	Hence \eqref{eq:decomp_algebra} implies $M_2-M_1 = T+W+B$.
	The equivalence results then follow from \eqref{eq:s_mono_equiv}.
\end{proof}

\bibliographystyle{ecta}
\bibliography{os}

\clearpage

\setcounter{section}{0}
\setcounter{subsection}{0}
\renewcommand{\sectionname}{Appendix}
\renewcommand{\thesection}{O.\Alph{section}}

\makeatletter
\let\OA@oldseccntformat\@seccntformat
\def\@seccntformat#1{%
	\ifstrequal{#1}{section}
	{O.\Alph{section}:\enspace}  
	{\OA@oldseccntformat{#1}}%
}
\makeatother

%
%

\setcounter{equation}{0}
\renewcommand{\theequation}{O.\arabic{equation}}

\setcounter{figure}{0}
\renewcommand{\thefigure}{O.\arabic{figure}}

\setcounter{table}{0}
\renewcommand{\thetable}{O.\arabic{table}}

\renewcommand{\theHsection}{O.\arabic{section}}
\renewcommand{\theHequation}{O.\arabic{equation}}
\renewcommand{\theHfigure}{O.\arabic{figure}}
\renewcommand{\theHtable}{O.\arabic{table}}

\begin{center}
	\LARGE Online Appendix to ``Risk Sensitivity and Stationary Equilibrium in Incomplete Markets''
	\medskip
	
	\normalsize Qingyin Ma, 
	\quad John Stachurski,
	\quad Bin Wu
\end{center}

\section{Counterexamples}\label{oa:counterexamples}
This appendix provides two counterexamples for
Remarks~\ref{rm:agg_reversal} and \ref{rm:eq_reversal}, respectively. 
The first shows that, without stochastic monotonicity of $Q$, 
risk-sensitive households can save less than expected-utility households, 
i.e., $A(R,\gamma)<A(R,0)$ for some $\gamma> 0$. The second shows that, 
even when $Q$ is monotone with strictly
positive entries, the saving comparison and ordering of both
extremal equilibrium interest rates can reverse between intermediate
risk-sensitivity values. Therefore, the restrictions on 
$(\gamma_1,\gamma_2)$ in Proposition~\ref{pr:markov_small_gamma}
and Theorem~\ref{t:markov_gamma_equilibrium}
cannot be dispensed with in general.

\subsection{Common Setup and a Zero-Saving Criterion}
\label{ss:oa_verification}

Throughout, we consider $u(c)=\sqrt c$.
Let $\ZZ=\{z_1,\ldots,z_N\}$, with
$0<z_1<\cdots<z_N$, and write $q_{ij}=Q(z_i,z_j)$.
Let $\pi$ be the stationary distribution of the strictly
positive transition matrix $Q$, normalized so that
$\sum_i\pi_i z_i=1$.
Fix $w_0>0$ and define
\begin{equation}\label{eq:oa_normalization}
	y_i:=w_0z_i,\qquad i=1,\ldots,N.
\end{equation}
Thus $\sum_i\pi_i y_i=w_0$.
At wage $w$, productivity remains fixed and
labor income is $wz_i$. At fixed prices $(R,w_0)$, the notation
$v_{R,\gamma},c_{R,\gamma},s_{R,\gamma},\mu_{R,\gamma},A(R,\gamma)$,
and $S(R,\gamma)$ has the same meaning as in the main text, with the wage
held at $w_0$. Theorem~\ref{t:os_rs_opt} supplies the household solution, 
and Proposition~\ref{pr:compact} and Theorem~\ref{t:gs_rs} supply a unique
stationary distribution with compact wealth support. Their conditions hold 
whenever $Q$ is strictly positive and $0<R<1/\beta$, since
$-u''(c)/u'(c)=1/(2c)\to0$. 

We begin with the same feasible plan: consume current income
and save zero whenever wealth equals income. Let $\varrho(\gamma)=(\varrho_i(\gamma))_{i=1}^N$ be its value vector, defined by
\begin{equation}\label{eq:oa_endowment}
	\varrho_i(\gamma)=\sqrt{y_i}+\beta\rho_{\gamma,z_i}(\varrho(\gamma)).
\end{equation}
By Lemma~\ref{lm:entropic}\eqref{i:lips}, the right-hand side of
\eqref{eq:oa_endowment} is a contraction with modulus $\beta$.
Hence the system has a unique solution $\varrho(\gamma)$. It is the value of a specified plan, not yet the optimal value
at the income nodes.
Define
\begin{equation}\label{eq:oa_marginals}
	d_i=u'(y_i)=\frac1{2\sqrt{y_i}} 
	\qquad\text{and} \qquad
	M_i(\gamma)=
	\frac{\sum_jq_{ij}\me^{-\gamma \varrho_j(\gamma)}d_j}
	{\sum_jq_{ij}\me^{-\gamma \varrho_j(\gamma)}}.
\end{equation}
The ratio is well defined also at $\gamma=0$, when it reduces to
$\sum_jq_{ij}d_j$. As in the Euler equation of
Section~\ref{ss:2_period}, $d_i$ is the marginal cost of saving and
$\beta RM_i$ is a tilted marginal continuation benefit. The following
lemma makes this comparison an optimality test.

\begin{lemma}[Zero saving at the income nodes]
	\label{lm:oa_zero_saving}
	Suppose $Q$ is strictly positive and $0<R<1/\beta$. For each
	$\gamma\ge0$, the following statements are equivalent:
	\begin{enumerate}[label=(\roman*)]
		\item $S(R,\gamma)=0$.
		\item $s_{R,\gamma}(y_i,z_i)=0$ for all $i$.
		\item $\beta RM_i(\gamma)\le d_i$ for all $i$.
	\end{enumerate}
	When these statements hold,
	$v_{R,\gamma}(y_i,z_i)=\varrho_i(\gamma)$ and
	$\mu_{R,\gamma}=\sum_i\pi_i\delta_{(y_i,z_i)}$.
\end{lemma}

\begin{proof}
	Fix $\gamma$ and abbreviate $\varrho=\varrho(\gamma)$. First suppose (iii) holds.
	For $a\ge0$, define $W(a,z_i)=\varrho_i+d_i(a-y_i).$
	Since $\varrho_i\ge\sqrt{y_i}$ and $d_i y_i=\sqrt{y_i}/2$, $W$ is nonnegative.
	Concavity of utility $u$ and of $\rho_{\gamma,z_i}$ implies, for every
	$s\in[0,a]$,
	\begin{align*}
		u(a-s)+\beta\rho_{\gamma,z_i}
		\bigl(W(Rs+w_0 \hat z, \hat z)\bigr)
		&\le \sqrt{y_i}+d_i(a-s-y_i)+\beta\rho_{\gamma,z_i}(\varrho)+\beta RsM_i(\gamma)\\
		&=W(a,z_i)+s(\beta RM_i(\gamma)-d_i)\le W(a,z_i).
	\end{align*}
	Thus the Bellman operator satisfies $TW\le W$ pointwise. Monotonicity 
	of $T$ gives $T^n0\le W$, and value iteration convergence delivers
	$v_{R,\gamma}\le W$. At $(y_i,z_i)$, the zero saving plan attains
	$\varrho_i=W(y_i,z_i)$, so it is optimal there. This establishes (ii). Let $\nu:=\sum_i\pi_i\delta_{(y_i,z_i)}$. Under (ii), a household
	at $(y_i,z_i)$ moves to $(y_j,z_j)$ with probability $q_{ij}$.
	Thus, if the current state has distribution $\nu$, its next-period
	distribution is
	$$
	\sum_i\pi_i\sum_jq_{ij}\delta_{(y_j,z_j)}
	=
	\sum_j(\pi Q)_j\delta_{(y_j,z_j)}
	=
	\nu,
	$$
	where the last equality uses $\pi Q=\pi$. Hence $\nu$ is invariant.
	By Theorem~\ref{t:gs_rs}, $\mu_{R,\gamma}=\nu$, and therefore
	$
	S(R,\gamma)
	=\sum_i\pi_i s_{R,\gamma}(y_i,z_i)=0.
	$
	This proves that (ii) implies (i).
	
	Conversely, if (i) holds, then saving is zero
	$\mu_{R,\gamma}$-almost surely. By \eqref{eq:evol_w},
	$a_{t+1}=w_0z_{t+1}$ almost surely.
	Stationarity and the productivity marginal $\pi$ give
	$\mu_{R,\gamma}=\sum_i\pi_i\delta_{(y_i,z_i)}$.
	Since $\pi_i>0$ for all $i$, (ii) follows. 
	Under (ii), the Bellman equation implies that
	$(v_{R,\gamma}(y_i,z_i))_{i=1}^N$ satisfies
	\eqref{eq:oa_endowment}. Uniqueness of its solution then gives
	$v_{R,\gamma}(y_i,z_i)=\varrho_i$ for all $i$. To prove (iii), take a 
	feasible deviation: save $s\in[0,y_i]$ today, consume all wealth 
	next period, and save zero thereafter. Conditional on next-period productivity $z_j$, its
	continuation value is
	$
	u(y_j+Rs)+\beta\rho_{\gamma,z_j}(\varrho)
	=\varrho_j+u(y_j+Rs)-u(y_j).
	$
	Thus the deviation has value
	\begin{equation*}
		D_i(s)
		=u(y_i-s)+\beta\rho_{\gamma,z_i}
		\left(\bigl(\varrho_j+u(y_j+Rs)-u(y_j)\bigr)_{j=1}^N\right).
	\end{equation*}
	Since $D_i(s)\le v_{R,\gamma}(y_i,z_i)=\varrho_i=D_i(0)$,
	the right derivative at zero must be nonpositive:
	$
	D_i'(0+)=-d_i+\beta R M_i(\gamma)\le0.
	$
	This proves (iii).
\end{proof}

The lemma implies that if all income nodes save zero,
their stationary distribution is known. If one of the inequalities in
(iii) fails, stationary saving must instead be positive. Integrating
\eqref{eq:evol_w} then gives the same accounting identity as in the main
text,
\begin{equation}\label{eq:oa_accounting}
	A(R,\gamma)=R S(R,\gamma)+w_0.
\end{equation}

\subsection{Counterexample I}
\label{ss:oa_nonmonotone}
This example illustrates Remark~\ref{rm:agg_reversal}. When $Q$ fails 
stochastic monotonicity, households with positive risk sensitivity can exhibit
lower aggregate saving and wealth than their expected-utility counterparts.
Consider $N=3$, hence $\ZZ=\{z_1,z_2,z_3\}$, and fix
\begin{equation}\label{eq:oa_nm_primitives}
	\begin{gathered}
		u(c)=\sqrt c,\qquad \beta=9/10,\qquad
		R_0=58/57>1,\qquad \beta R_0=87/95<1,\\
		y=\left(81,\frac{1369}{16},100\right),\qquad
		Q=\frac1{20}\begin{pmatrix}1&1&18\\1&18&1\\10&9&1\end{pmatrix}.
	\end{gathered}
\end{equation}
Here $y_i=w_0z_i$. To impose the normalization of Section~\ref{ss:firm},
let $\pi$ denote the stationary distribution of productivity and set
\begin{equation}\label{eq:oa_nm_normalization}
	\pi=\frac1{247}(29,181,37),\qquad
	w_0=\sum_i\pi_i y_i=\frac{344573}{3952},\qquad z_i=y_i/w_0.
\end{equation}
Direct multiplication verifies $\pi Q=\pi$, and hence
$\sum_i\pi_i z_i=1$. The matrix $Q$ is strictly positive but violates the
monotonicity requirement in Assumption~\ref{a:mlr}: for the increasing
function $h(z)=\mathbf 1\{z=z_3\}$, one has
$Qh(z_1)=18/20>1/20=Qh(z_2)$.

\begin{proposition}[Reversal relative to expected utility]
	\label{pr:oa_nonmonotone}
	At prices $(R_0,w_0)$,
	\begin{align}
		s_{R_0,4}(100,z_3)&=0<s_{R_0,0}(100,z_3),
		\label{eq:oa_nm_policy}\\
		S(R_0,4)&=0<S(R_0,0),
		\label{eq:oa_nm_saving}\\
		A(R_0,4)&=w_0<A(R_0,0).
		\label{eq:oa_nm_wealth}
	\end{align}
\end{proposition}

\begin{proof}
	\emph{Step 1. Order the zero-saving values.}
	Let $\varrho=\varrho(4)$ be the income-node value vector of the zero-saving plan.
	Its defining contraction is
	\begin{equation}\label{eq:oa_nm_endowment}
		\varrho_i=\sqrt{y_i}-\frac9{40}\log\sum_jq_{ij}\me^{-4\varrho_j}.
	\end{equation}
	We first show that its fixed point satisfies
	\begin{equation}\label{eq:oa_nm_gaps}
		\varrho_1-\varrho_2\ge1/5 
		\quad \text{and} \quad
		\varrho_3-\varrho_2\ge 3/4.
	\end{equation}
	Choose $\varrho^{(0)}$ satisfying
	$\varrho_1^{(0)}-\varrho_2^{(0)}\ge1/5$ and
	$\varrho_3^{(0)}-\varrho_2^{(0)}\ge3/4$.
	Iterate \eqref{eq:oa_endowment} at $\gamma=4$:
	\begin{equation*}
		\varrho_i^{(n+1)}
		=
		\sqrt{y_i}
		-\frac9{40}\log\left(
		\sum_{j=1}^3 q_{ij}\exp[-4\varrho_j^{(n)}]
		\right),
		\qquad i=1,2,3.
	\end{equation*}
	Suppose both inequalities hold at iteration $n$, and set
	$r_n:=\exp[-4(\varrho_1^{(n)}-\varrho_2^{(n)})]$, $r:=\exp[-4(\varrho_1-\varrho_2)]$,
	$t_n:=\exp[-4(\varrho_3^{(n)}-\varrho_2^{(n)})]$, and $t:=\exp[-4(\varrho_3-\varrho_2)]$.
	Then $0<r_n\le\exp(-4/5)<9/20$ and
	$0<t_n\le\exp(-3)<1/20$, so
	\begin{equation*}
		\frac{r_n+1+18t_n}{r_n+18+t_n}
		<
		\frac{9/20+1+18/20}{18}
		=\frac{47}{360}
		<\frac2{15}
		<\exp(-2).
	\end{equation*}
	Moreover,
	$(10r_n+9+t_n)-(r_n+18+t_n)=9(r_n-1)<0$.
	Substituting the rows of $Q$ therefore gives
	\begin{equation*}
		\begin{aligned}
			\varrho_1^{(n+1)}-\varrho_2^{(n+1)}
			&=-\frac14-\frac9{40}
			\log\frac{r_n+1+18t_n}{r_n+18+t_n}
			>-\frac14+\frac9{20}
			=\frac15,\\
			\varrho_3^{(n+1)}-\varrho_2^{(n+1)}
			&=\frac34-\frac9{40}
			\log\frac{10r_n+9+t_n}{r_n+18+t_n}
			>\frac34.
		\end{aligned}
	\end{equation*}
	By induction, both bounds hold for every $n$.
	The contraction property implies that $\varrho^{(n)}$ converges to the
	unique fixed point $\varrho$. Passing to the limit yields \eqref{eq:oa_nm_gaps}.
	
	\emph{Step 2. Verify zero saving under risk sensitivity.}
	For these incomes, the marginal utility vector is
	$d=\left(1/18,2/37,1/20\right)$.
	Abbreviate $M_i=M_i(4)$. Since  $d_1>d_2>d_3$, 
	\begin{equation}\label{eq:oa_nm_low_rows}
		\beta R_0M_i\le\beta R_0d_1 = 29/570<2/37\le d_i,
		\qquad i=1,2.
	\end{equation}
	At the third node, dropping the term with lowest marginal utility
	and using $r<9/20$:
	\begin{equation*}
		M_3=\frac{10r d_1+9d_2+t d_3}{10r+9+t}
		<\frac{10r d_1+9d_2}{10r+9}
		<\frac{d_1+2d_2}{3}=\frac{109}{1998}.
	\end{equation*}
	So $\beta R_0M_3 <(87/95) (109/1998) =d_3-1/25308<d_3$.
	Lemma~\ref{lm:oa_zero_saving} now verifies the zero-saving plan at
	every income node: $v_{R_0,4}(y_i,z_i)=\varrho_i$ and
	$s_{R_0,4}(y_i,z_i)=0$ for all $i$.
	
	\emph{Step 3. Establish the expected-utility policy comparison.}
	Under expected utility, the corresponding marginal comparison has the
	opposite sign:
	\begin{equation*}\label{eq:oa_nm_eu_marginal}
		\beta R_0\sum_jq_{3j}d_j
		=(87/95)\left(1/36+9/370+1/400\right)
		=d_3+17/4218000>d_3.
	\end{equation*}
	To relate this inequality to the optimal expected-utility policy, apply
	Lemma~\ref{lm:T_diff} at the income nodes. Since consumption cannot
	exceed available wealth,
	\begin{equation*}
		\partial v_{R_0,0}(y_j,z_j)/\partial a
		=u'(c_{R_0,0}(y_j,z_j))\ge u'(y_j)=d_j.
	\end{equation*}
	The right derivative at zero of the saving objective at
	$(y_3,z_3)$ is therefore
	\begin{equation*}
		-d_3+\beta R_0\sum_jq_{3j}
		\partial v_{R_0,0}(y_j,z_j)/\partial a
		\ge 17/4218000>0.
	\end{equation*}
	Thus zero saving is not optimal there, proving
	\eqref{eq:oa_nm_policy}.
	
	\emph{Step 4. Compare stationary aggregates.}  By Lemma~\ref{lm:oa_zero_saving},
	$s_{R_0,4}(y_i,z_i)=0$ for every $i$ implies
	$S(R_0,4)=0$, whereas
	$s_{R_0,0}(y_3,z_3)>0$ implies $S(R_0,0)>0$.
	The accounting identity
	\eqref{eq:oa_accounting} then gives
	$A(R_0,4)=w_0<A(R_0,0)$, proving \eqref{eq:oa_nm_saving} and 
	\eqref{eq:oa_nm_wealth}.    
\end{proof}

\subsection{Counterexample II}
\label{ss:oa_equi_reversal}
The example illustrates Remark~\ref{rm:eq_reversal} by showing 
that the saving comparison in Proposition~\ref{pr:markov_small_gamma} 
and equilibrium interest rate comparison in 
Theorem~\ref{t:markov_gamma_equilibrium} need not extend to
the intermediate region of risk sensitivity. We construct
a five-state economy with strictly positive, stochastically
monotone income transitions and two positive values
of $\gamma$ such that the larger value delivers lower
aggregate saving and mean stationary wealth at fixed
prices. We then introduce production and show that,
for the same parameter pair, both the smallest and the
largest equilibrium interest rates are higher at the
larger value of $\gamma$.

\subsubsection{Saving reversal at fixed prices}
\label{ss:oa_monotone}
We first construct an example with two positive parameters
$\gamma_1<\gamma_2$ such that, at the same prices, stationary
aggregate saving is positive under $\gamma_1$ and zero under
$\gamma_2$. Let $\ZZ=\{z_1,\ldots,z_5\}$ and fix
\begin{equation}\label{eq:oa_m_primitives}
	\begin{gathered}
		\beta=3/4,\qquad
		y=\left(64,81,100,(161/16)^2,169\right),\\
		R_0=1564/1521>1,\qquad
		\gamma_1=\log4<\gamma_2=\log16,\qquad \epsilon=2^{-32}.
	\end{gathered}
\end{equation}
Define the matrices $Q$ and $\bar Q = (q^0_{ij})_{1\le i,j\le 5}$ by
\begin{equation}\label{eq:oa_m_kernel}
	Q=(1-\epsilon)\bar Q+\epsilon U,\qquad U_{ij}=\frac15,
	\qquad
	\bar Q=\begin{pmatrix}
		1&0&0&0&0\\
		1&0&0&0&0\\
		1/4&3/4&0&0&0\\
		1/4&0&3/4&0&0\\
		0&0&1/2&1/2&0
	\end{pmatrix}.
\end{equation}
Let $\pi$ be the stationary distribution of $Q$ and define $w_0$ and
$z_i$ by \eqref{eq:oa_normalization}. The matrix $\bar Q$ is monotone:
each row first-order stochastically dominates the preceding row.
Mixing every row with the same distribution preserves this order, so
$Q$ is monotone and every entry is at least $\epsilon/5$.
Also $\beta R_0=391/507<1$.

\begin{proposition}[Reversal under monotone productivity]
	\label{pr:oa_monotone}
	At prices $(R_0,w_0)$,
	\begin{align}
		s_{R_0,\gamma_1}(169,z_5)&>0=s_{R_0,\gamma_2}(169,z_5),
		\label{eq:oa_m_policy}\\
		S(R_0,\gamma_1)&>0=S(R_0,\gamma_2),
		\label{eq:oa_m_saving}\\
		A(R_0,\gamma_1)&>A(R_0,\gamma_2)=w_0.
		\label{eq:oa_m_wealth}
	\end{align}
	Moreover, for every bounded $h:\XX\to\RR$ that is
	increasing in $a$,
	\begin{equation}\label{eq:oa_m_stationary_order}
		\int_\XX h\,\diff \mu_{R_0,\gamma_1}
		\ge
		\int_\XX h\,\diff \mu_{R_0,\gamma_2},
	\end{equation}
	with strict inequality for $h(a,z)=a/(1+a)$.
	
\end{proposition}

\begin{proof}
	\emph{Step 1. Calculate the value gap under the auxiliary matrix.}
	Let $\varrho$ and $\varrho^0$ denote the corresponding zero-saving value vectors under $Q$ and $\bar Q$ respectively. The first two rows of $\bar Q$ assign probability  
	one to state $1$, so both certainty equivalents equal $\varrho_1^0$. Hence
	$\varrho_1^0=8+3\varrho_1^0/4$ and $\varrho_2^0=9+3 \varrho_1^0/4$.
	Solving these equations gives $\varrho_1^0=32$ and $\varrho_2^0=33$,
	independently of $\gamma$.
	Define $A_\gamma=(1+3\me^{-\gamma})/4$ and 
	$C_\gamma=(1+3\me^{-2\gamma}A_\gamma^{3/4})/4$.
	Factoring $\me^{-\gamma \varrho_1^0}$ out of the exponential sums in rows
	3 and 4 yields
	\begin{equation}\label{eq:oa_m_values}
		\varrho_3^0=34-(3\log A_\gamma)/(4\gamma)
		\quad \text{and} \quad 
		\varrho_4^0=545/16-(3\log C_\gamma)/(4\gamma).
	\end{equation}
	Let $B_\gamma=\me^{\gamma(\varrho_4^0-\varrho_3^0)}$, so 
	$B_\gamma^4=\me^{\gamma/4}(A_\gamma/C_\gamma)^3$.
	We claim that
	\begin{equation}\label{eq:oa_m_B_order}
		B_{\gamma_1}>3/2>10/7>11/8>B_{\gamma_2}.
	\end{equation}
	At $\gamma_1=\log4$, $A_{\gamma_1}=7/16$. Since $16\cdot7^3<9^4$,  $A_{\gamma_1}^{3/4}<9/16$.
	So $C_{\gamma_1}<283/1024 $ and 
	$A_{\gamma_1}/C_{\gamma_1}>448/283>11/7$.
	Using $\sqrt2>7/5$ gives
	$B_{\gamma_1}^4>(7/5)\left(11/7\right)^3
	=1331/245>\left(3/2\right)^4$.
	At $\gamma_2=\log16$, $A_{\gamma_2}=19/64$ and
	$C_{\gamma_2}>1/4$, hence
	$B_{\gamma_2}^4<2\left(19/16\right)^3
	=13718/4096<\left(11/8\right)^4$.
	This proves \eqref{eq:oa_m_B_order} by elementary rational comparisons.
	
	\emph{Step 2. Turn the value-gap comparison into a marginal comparison.}
	Write $M_i(\gamma)$ and $M_i^0(\gamma)$ for the tilted marginal
	means computed using $(Q,\varrho)$ and $(\bar Q,\varrho^0)$, respectively. Since $q^0_{53}=q^0_{54}=1/2$, the tilted probability of state $3$ is
	$\me^{-\gamma \varrho_3^0}/(\me^{-\gamma \varrho_3^0}+\me^{-\gamma \varrho_4^0})
	=B_\gamma/(1+B_\gamma)$.
	Using $d_3=1/20$ and $d_4=8/161$, we obtain
	\begin{equation}\label{eq:oa_m_M5}
		M_5^0(\gamma)
		=(B_\gamma d_3+d_4)/(1+B_\gamma)
		=8/161 + B_\gamma/(1+B_\gamma)/3220.
	\end{equation}
	Thus a fall in $B_\gamma$ reduces the weight on the lower-income
	state and lowers the tilted marginal mean. Choose the intermediate mean
	$M_*:=8/161+(10/17)/3220 =39/782$.
	The return in \eqref{eq:oa_m_primitives} satisfies
	$\beta R_0M_*=d_5$.
	Since $B/(1+B)$ is strictly increasing for $B>0$,
	\eqref{eq:oa_m_B_order} gives
	\begin{align}
		M_5^0(\gamma_1)-M_*
		&>\left(3/5-10/17\right)/3220
		=1/273700>2^{-19},
		\label{eq:oa_m_margin1}\\
		M_*-M_5^0(\gamma_2)
		&>\left(10/17-11/19\right)/3220
		=3/1040060>2^{-19}.
		\label{eq:oa_m_margin2}
	\end{align}
	Consequently, $\beta R_0M_5^0(\gamma_1)>d_5>\beta R_0M_5^0(\gamma_2)$.
	
	\emph{Step 3. Control the positive perturbation.}
	We show that, for $\gamma\in\{\gamma_1,\gamma_2\}$,
	\begin{equation}\label{eq:oa_m_perturbation}
		\max_i|M_i(\gamma)-M_i^0(\gamma)|<2^{-20}.
	\end{equation}
	Fix $\gamma\in\{\gamma_1,\gamma_2\}$ and set
	$E_j:=\me^{-\gamma \varrho_j^0}$,
	$L_i:=\sum_j q_{ij}^0E_j$, and $L_U:=\sum_jE_j/5$.
	Since income and $\bar Q$ are monotone, the map in
	\eqref{eq:oa_endowment} preserves increasing vectors.
	Iterating from zero and passing to the limit gives
	$\varrho_1^0\le\cdots\le \varrho_5^0$.
	Thus $E_j\le E_1$ for every $j$, and hence $L_U\le E_1$.
	The first four rows satisfy $q_{i1}^0\ge1/4$, while
	$q_{53}^0=1/2$. Therefore, for $i\le 4$,
	$L_i\ge E_1/4$ and $L_5\ge E_3/2$.
	Moreover, \eqref{eq:oa_m_values}, $\gamma\le\log16$, and
	$A_\gamma\ge1/4$ imply
	$E_1/E_3
	=\me^{\gamma(\varrho_3^0-\varrho_1^0)}
	=\me^{2\gamma}/A_\gamma^{3/4}
	\le256\cdot4^{3/4}<1024.$
	Hence, for $i\le 4$, $L_U/L_i\le4$ and $L_U/L_5\le2 E_1/E_3<2^{11}.$
	Thus $L_U/L_i\le2^{11}$ for every row.
	Since $q_{ij}=(1-\epsilon)q_{ij}^0+\epsilon/5$,
	evaluating the exponential sums at $\varrho^0$ gives
	\begin{equation*}
		\sum_jq_{ij}E_j
		=(1-\epsilon)\sum_jq_{ij}^0E_j
		+(\epsilon/5) \sum_jE_j
		=L_i\left(1-\epsilon+\epsilon L_U/L_i\right), 
	\end{equation*}
	which implies that
	$
	1-\epsilon
	\le1-\epsilon+\epsilon L_U/L_i
	\le1-\epsilon+2^{-21}
	\le1+2^{-21}.
	$
	Thus, for every row $i$, the ratio of the exponential sum under $Q$
	to that under $\bar Q$, with $E_j=\me^{-\gamma \varrho_j^0}$ fixed, satisfies
	$(\sum_j q_{ij}E_j) \big/
	(\sum_j q_{ij}^0E_j)
	=1-\epsilon+\epsilon L_U/L_i
	\in[1-\epsilon,1+2^{-21}]$.
	Let $T_Q$ and $T_{\bar Q}$
	denote the maps in \eqref{eq:oa_endowment}, with fixed points
	$\varrho$ and $\varrho^0$, respectively. Evaluating both maps at $\varrho^0$ gives
	$(T_Q\varrho^0)_i-(T_{\bar Q}\varrho^0)_i
	=-(\beta/\gamma) \log[(\sum_jq_{ij}E_j)/L_i]$.
	The ratio inside the logarithm lies in $[1-\epsilon,1+2^{-21}]$.
	Since $\log(1+2^{-21})\le2^{-21}$ and
	$-\log(1-\epsilon)\le\epsilon/(1-\epsilon)<2\epsilon$, we have $\|T_Q\varrho^0-T_{\bar Q}\varrho^0\|_\infty
	\le\beta 2^{-20}/\gamma.$
	Using the fixed-point identities and the contraction property,
	\begin{equation*}
		\|\varrho-\varrho^0\|_\infty
		\le\|T_Q\varrho-T_Q\varrho^0\|_\infty
		+\|T_Q\varrho^0-T_{\bar Q}\varrho^0\|_\infty
		\le\beta\|\varrho-\varrho^0\|_\infty+\beta 2^{-20}/\gamma.
	\end{equation*}
	\begin{equation*}\label{eq:oa_m_value_error}
		\Longrightarrow \quad 
		\Delta:=\gamma\|\varrho-\varrho^0\|_\infty
		\le\beta 2^{-20}/(1-\beta)
		=3\cdot2^{-20}.
	\end{equation*}
	To separate the effects of changing the matrix and the values,
	introduce
	$\widehat M_i(\gamma):=(\sum_jq_{ij}E_jd_j)/(\sum_jq_{ij}E_j)$,
	which uses $Q$ together with the old values $\varrho^0$.
	Since $q_{ij}E_j=(1-\epsilon)q_{ij}^0E_j+\epsilon E_j/5,$
	normalizing these weights and taking the mean of $d_j$ gives
	\begin{equation*}
		\widehat M_i(\gamma)-M_i^0(\gamma)
		=
		\frac{\epsilon L_U}{(1-\epsilon)L_i+\epsilon L_U}
		\left[
		\left(\sum_jE_jd_j\right)/(5L_U)-M_i^0(\gamma)
		\right].
	\end{equation*}
	Since $d_j\in(0, 1/16]$, both means in brackets lie in $(0,1/16]$, so their absolute
	difference is at most $1/16$. Moreover, $L_U/L_i\le2^{11}$
	and $\epsilon=2^{-32}$ imply $\epsilon L_U/\bigl((1-\epsilon)L_i+\epsilon L_U\bigr)
	\le2^{-21}/(1-\epsilon)<2^{-20}.$
	Consequently, replacing $\bar Q$ by $Q$ while holding $\varrho^0$ fixed
	changes each tilted mean by less than $2^{-24}$:
	$|\widehat M_i(\gamma)-M_i^0(\gamma)|<2^{-20}/16=2^{-24}$.
	
	Next, keep $Q$ fixed and replace $\varrho^0$ by $\varrho$.
	The definition of $\Delta$ and summation against row $i$ give
	$\me^{-\Delta}E_j
	\le \me^{-\gamma \varrho_j}\le \me^\Delta E_j,$ which implies that
	$\me^{-\Delta}\sum_kq_{ik}E_k
	\le\sum_kq_{ik}\me^{-\gamma \varrho_k}
	\le \me^\Delta\sum_k q_{ik}E_k.$
	Thus, in row $i$ and column $j$, the tilted probability computed
	at $\varrho$, divided by that computed at $\varrho^0$, equals
	$\me^{-\gamma(\varrho_j-\varrho_j^0)}
	\left(\sum_kq_{ik}E_k\right)/
	\left(\sum_kq_{ik}\me^{-\gamma \varrho_k}\right)
	\in[\me^{-2\Delta},\me^{2\Delta}].$
	Each probability therefore changes in absolute value by at most
	$\me^{2\Delta}-1$ times its value at $\varrho^0$.
	Since $d_j\le1/16$, the resulting change in the mean is at most
	$(\me^{2\Delta}-1)/16$.
	Moreover, $2\Delta\le6\cdot2^{-20}<1$, so $\me^x-1\le2x$ yields
	$|M_i(\gamma)-\widehat M_i(\gamma)|
	\le(\me^{2\Delta}-1)/16
	\le\Delta/4
	\le3\cdot2^{-22}$.
	The triangle inequality now gives
	\begin{equation*}\label{eq:oa_m_perturbation_bound}
		\begin{aligned}
			\max_i|M_i(\gamma)-M_i^0(\gamma)|
			&\le\max_i|\widehat M_i(\gamma)-M_i^0(\gamma)|
			+\max_i|M_i(\gamma)-\widehat M_i(\gamma)|\\
			&<2^{-24}+3\cdot2^{-22}
			=13\cdot2^{-24}<2^{-20}.
		\end{aligned}
	\end{equation*}
	This estimate holds for both parameters. Combining it with
	\eqref{eq:oa_m_margin1}--\eqref{eq:oa_m_margin2} gives
	$M_5(\gamma_1)>M_*>M_5(\gamma_2)$.
	Since $\beta R_0M_*=d_5$,
	\begin{equation}\label{eq:oa_m_crossing}
		\beta R_0M_5(\gamma_1)>d_5>
		\beta R_0M_5(\gamma_2).
	\end{equation}
	For either parameter, every tilted mean is at most
	$\max_jd_j=1/16$. Hence
	\begin{equation}\label{eq:oa_m_low_rows}
		\beta R_0M_i(\gamma)
		\le391/8112<8/161\le d_i,
		\qquad i\le4.
	\end{equation}
	Together with \eqref{eq:oa_m_crossing}, this verifies
	$\beta R_0M_i(\gamma_2)\le d_i$ for every $i$.
	Lemma~\ref{lm:oa_zero_saving} therefore implies
	$s_{R_0,\gamma_2}(y_i,z_i)=0$ at every income node.
	
	\emph{Step 4. Locate the policy reversal.}
	It remains to identify the fifth income node as the point of reversal.
	We first establish that, for the actual matrix $Q$ and every $\gamma\ge0$,
	\begin{equation}\label{eq:oa_m_zero_region}
		s_{R_0,\gamma}(a,z)=0\quad\text{if}\quad
		a\le L:=\left[8/(\beta R_0)\right]^2
		=\left(4056/391\right)^2.
	\end{equation}
	Use the finite-horizon policies of
	Lemma~\ref{lm:finite_horizon}.
	Since $c_{1,\gamma}(a,z)=a$, the induction starts with
	$c_{1,\gamma}(64,z)=64$ for every $z$.
	Suppose $c_{n,\gamma}(64,z)=64$ for every $z$.
	Monotonicity in wealth and $y_j\ge64$ imply
	$c_{n,\gamma}(y_j+R_0s,z_j)\ge64$.
	If $s_{n+1,\gamma}(a,z)>0$ for some $0<a\le L$,
	the finite-horizon Euler equation gives
	$u'(a)<u'(c_{n+1,\gamma}(a,z))\le\beta R_0/16\le u'(a)$,
	a contradiction. Thus $c_{n+1,\gamma}(a,z)=a$
	on $[0,L]$. Since $64<L$, induction and policy
	convergence in Lemma~\ref{lm:finite_horizon}\eqref{i:fh_to_infh}
	prove \eqref{eq:oa_m_zero_region}.
	Since $4056\cdot16>391\cdot161$, we have $L>y_4$. The first four
	income nodes therefore save zero under both parameters. If the fifth
	node also saved zero under $\gamma_1$, Lemma~\ref{lm:oa_zero_saving}
	would imply $\beta R_0M_5(\gamma_1)\le d_5$, contradicting
	\eqref{eq:oa_m_crossing}. This proves \eqref{eq:oa_m_policy}.
	Finally, \eqref{eq:oa_m_crossing}--\eqref{eq:oa_m_low_rows}
	and Lemma~\ref{lm:oa_zero_saving} give
	\eqref{eq:oa_m_saving}. The accounting identity \eqref{eq:oa_accounting}
	then proves \eqref{eq:oa_m_wealth}.	
	
	Under $\mu_{R_0,\gamma_1}$, stationarity implies
	$a\ge w_0z$ almost surely, and the productivity marginal
	is $\pi$. Since
	$\mu_{R_0,\gamma_2}=\sum_i\pi_i\delta_{(y_i,z_i)}$,
	every bounded $h$ increasing in $a$ satisfies
	\begin{equation*}
		\int_\XX h(a,z)\, \diff\mu_{R_0,\gamma_1}
		\ge
		\int_\XX h(w_0z,z)\,\diff \mu_{R_0,\gamma_1}
		=
		\sum_i\pi_i h(y_i,z_i)
		=
		\int_\XX h\,\diff \mu_{R_0,\gamma_2}.
	\end{equation*}
	Moreover, stationary accounting and
	\eqref{eq:oa_m_saving} give
	$\int_\XX(a-w_0z)\,\diff \mu_{R_0,\gamma_1}=R_0S(R_0,\gamma_1)>0$.
	Thus $a>w_0z$ on a set of positive
	$\mu_{R_0,\gamma_1}$ probability. For the bounded,
	strictly increasing function $h(a,z)=a/(1+a)$,
	the first inequality is therefore strict.
	This proves \eqref{eq:oa_m_stationary_order}
	and its strictness assertion.        
\end{proof}

\subsubsection{Reversal of equilibrium interest rates}
\label{ss:oa_equilibrium}
We show that the ordering of extremal equilibrium interest
rates can also reverse in the intermediate region of risk
sensitivity: both the smallest and the largest equilibrium
rates are higher under $\gamma_2$ than $\gamma_1$.
This reversal emerges when we introduce a family of production
technologies into the five-state example.
For arbitrary wage $w>0$, define $\varrho$, $d_i$, and $M_i$ as in
\eqref{eq:oa_endowment}--\eqref{eq:oa_marginals}, replacing $y_i$ by
$wz_i$, and write
$\tau(\gamma;w)=\min_i d_i/[\beta M_i(\gamma)]$.
Here the wage dependence of $\varrho$, $d_i$, and $M_i$ is understood. By
Lemma~\ref{lm:oa_zero_saving},
\begin{equation}\label{eq:oa_threshold_criterion}
	S(R,\gamma;w)=0\quad\Longleftrightarrow\quad R\le\tau(\gamma;w),
	\qquad 0<R<\bar R=4/3.
\end{equation}
Set depreciation to $\delta=9/10$, so that $\uR=1/10$.
For each $\alpha\in(0,1)$, define the Cobb--Douglas technology
\begin{equation*}\label{eq:oa_technology}
	k_\alpha=\frac{\alpha w_0}{(1-\alpha)(R_0-\uR)}
	\quad \text{and} \quad 
	F_\alpha(K,L)=\frac{w_0}{(1-\alpha)k_\alpha^\alpha}
	K^\alpha L^{1-\alpha}.
\end{equation*}
The firm conditions \eqref{eq:pro_opt} give
\begin{equation}\label{eq:oa_price_schedules}
	\begin{aligned}
		w_\alpha(R)=w_0
		\left(\frac{R_0-\uR}{R-\uR}\right)^{\alpha/(1-\alpha)}
		\; \text{ and} \quad 
		K_\alpha(R)=\frac{\alpha w_\alpha(R)}{(1-\alpha)(R-\uR)}.
	\end{aligned}
\end{equation}
In particular, $w_\alpha(R_0)=w_0$. Denote the equilibrium set by
\begin{equation*}
\mathcal R_\alpha(\gamma)=
\{R\in(\uR,\bar R):S(R,\gamma;w_\alpha(R))=K_\alpha(R)\}.
\end{equation*}
Assumption~\ref{a:prod_func} holds for every $\alpha$, and the
five-state matrix satisfies Assumption~\ref{a:mlr}. Hence
Theorem~\ref{t:RCE} makes these sets nonempty and compact.

\begin{proposition}[Equilibrium reversal]
	\label{pr:oa_equilibrium}
	For the five-state economy,
	\begin{equation}\label{eq:oa_equilibrium_limit}
		\lim_{\alpha \downarrow 0} 
		\sup_{R\in\mathcal R_\alpha(\gamma)}
		|R-\tau(\gamma;w_0)| = 0,
		\qquad \gamma\in\{\gamma_1,\gamma_2\}.
	\end{equation}
	For every sufficiently small positive $\alpha$,
	\begin{equation}\label{eq:oa_equilibrium_order}
		1<\min\mathcal R_\alpha(\gamma_1)
		\le\max\mathcal R_\alpha(\gamma_1)
		<R_0<\min\mathcal R_\alpha(\gamma_2).
	\end{equation}
	Thus both extremal equilibrium returns increase between the two parameters.
\end{proposition}

\begin{proof}
	\emph{Step 1. Bound the two zero-saving thresholds.}
	At $w_0$, the first four component thresholds satisfy
	$d_i/(\beta M_i(\gamma))\ge d_4/(\beta d_1) =512/483>1$.
	Equations \eqref{eq:oa_m_M5} and \eqref{eq:oa_m_perturbation} imply
	$5/117<8/161-2^{-20}<M_5(\gamma,Q)<1/20+2^{-20}<2/39$.
	Since $d_5/\beta=2/39$, the fifth component threshold lies in
	$(1,6/5)$. At $w_0$, \eqref{eq:oa_m_crossing}--\eqref{eq:oa_m_low_rows}
	show that $\tau(\gamma_1;w_0)<R_0<\tau(\gamma_2;w_0)$. Consequently,
	\begin{equation}\label{eq:oa_threshold_order}
		1<\tau(\gamma_1;w_0)<R_0<\tau(\gamma_2;w_0)<6/5<\bar R.
	\end{equation}
	For any $w>0$, we have $M_i\le u'(wz_1)$ and
	$d_i\ge u'(wz_5)$. Hence
	\begin{equation*}
		\tau(\gamma;w)
		\ge u'(wz_5)/(\beta u'(wz_1))
		=(1/\beta)\sqrt{z_1/z_5}
		=32/39=:r_*>\underline R.
	\end{equation*}
	At equilibrium, capital demand is positive, so
	\eqref{eq:oa_threshold_criterion} implies
	$R>\tau(\gamma;w_\alpha(R))\ge r_*$.
	On $[r_*,\bar R]$, the denominator $R-\underline R$ is
	bounded away from zero, and
	$\log((R_0-\underline R)/(R-\underline R))$ is bounded.
	Since $\alpha/(1-\alpha)\to0$,
	\eqref{eq:oa_price_schedules} therefore yields, as $\alpha\downarrow0$,
	\begin{equation}\label{eq:oa_uniform_prices}
		w_\alpha(R)\to w_0
		\quad \text{and} \quad 
		K_\alpha(R)\to0
		\quad\text{uniformly on }[r_*,\bar R].
	\end{equation}
	Finally, for fixed $\gamma$, comparing the zero-saving
	fixed points at wages $w$ and $\widetilde w$ gives
	\begin{equation*}
		\|\varrho(w)-\varrho(\widetilde w)\|_\infty
		\le
		\sqrt{z_5}\,|\sqrt w-\sqrt{\widetilde w}|/(1-\beta).
	\end{equation*}
	Thus $\varrho(w)$ is continuous. The formulas for $d_i$ and $M_i$
	then imply that both are continuous and positive for $w>0$.
	Their ratios, and hence
	$\tau(\gamma;w)=\min_i\{d_i/(\beta M_i)\}$, are therefore
	continuous.
	
	\emph{Step 2. Identify every equilibrium limit.}
	Fix $\gamma\in \{\gamma_1, \gamma_2\}$ and a sequence $\alpha_n\downarrow0$ with
	$R_n\in\mathcal R_{\alpha_n}(\gamma)$.
	Passing to a subsequence, let
	$R_n\to R^*\in[r_*,\bar R]$, and set
	$w_n:=w_{\alpha_n}(R_n)$ and $K_n:=K_{\alpha_n}(R_n)$.
	By \eqref{eq:oa_uniform_prices},
	$w_n\to w_0$ and $K_n\to0$.
	To justify the limit of saving policies,
	fix a compact wage interval $[w_-,W]\subset(0,\infty)$
	and set $\kappa(a):=\sqrt{1+a+3W\bar z}$.
	For $R\in[r_*,\bar R]$, $w\in[w_-,W]$, and $s\in[0,a]$,
	we have $u(a-s)\le\kappa(a)$ and
	$
	\kappa(Rs+wz_j)
	\le \sqrt{1+\bar R a+4W\bar z}
	\le \sqrt{\bar R}\,\kappa(a).
	$
	Since $\beta\sqrt{\bar R}=\sqrt3/2<1,$ the weighted contraction and verification arguments
	of Theorem~\ref{t:opt_general} apply uniformly on this
	parameter set, including $R=\bar R$. 
	Bellman iteration from zero preserves joint continuity
	and concavity in wealth and converges locally uniformly
	to the value function.
	Strict concavity of the Bellman objective and the maximum
	theorem yield continuity of the unique saving policy.
	Since the wage interval is arbitrary, for each $z\in \ZZ$,
	$(a,R,w)\mapsto s_{R,\gamma}(a,z;w)$
	is continuous on $[0,\infty)\times[r_*,\bar R]\times(0,\infty).$
	
	Write $s_n(a,z):=s_{R_n,\gamma}(a,z)$ and let
	$\mu_n$ be its stationary distribution at prices $(R_n,w_n)$.
	Stationarity implies $a\ge w_nz$, $\mu_n$-almost surely,
	and the productivity marginal is $\pi$.
	Since saving is increasing in wealth by
	Theorem~\ref{t:opt_general}\eqref{i:c*}, market clearing gives
	\begin{equation*}
		0\le\sum_i\pi_i s_n(w_nz_i,z_i)
		\le\int_\XX s_n\, \diff \mu_n=K_n\to0.
	\end{equation*}
	Joint policy continuity and $\pi_i>0$ therefore imply
	$s_{R^*,\gamma}(y_i,z_i;w_0)=0$ for every $i$.
	The income-node Bellman equations reduce to
	\eqref{eq:oa_endowment}, so the node values equal $\varrho_i$.
	The feasible saving deviation used in the proof of
	Lemma~\ref{lm:oa_zero_saving} has right derivative
	$-d_i+\beta R^*M_i(\gamma)$ at zero.
	Optimality of zero saving therefore implies
	$
	-d_i+\beta R^*M_i(\gamma)\le0$
	for every $i$, and 
	$R^*\le\tau(\gamma;w_0).$
	Since \eqref{eq:oa_threshold_order} gives
	$\tau(\gamma;w_0)<\bar R$, we obtain $R^*<\bar R$.
	If $R^*<\tau(\gamma;w_0)$, continuity of the threshold gives
	$R_n<\tau(\gamma;w_n)$ for all large $n$.
	Equation~\eqref{eq:oa_threshold_criterion} would then imply
	$S(R_n,\gamma;w_n)=0$, contradicting market clearing
	and $K_n>0$. Hence $R^*=\tau(\gamma;w_0).$
	Since the equilibrium sequence was arbitrary and all
	returns lie in the common compact interval $[r_*,\bar R]$,
	this proves \eqref{eq:oa_equilibrium_limit}.
	Combining it with \eqref{eq:oa_threshold_order} gives
	\eqref{eq:oa_equilibrium_order}.
\end{proof}

\section{Extension to Utilities Unbounded Below}
\label{ss:opt_utility_ext}
The baseline argument assumes nonnegative period utility.
In this appendix, we show that bounded below utilities 
can be treated by normalization. Moreover, we extend 
the argument to utilities unbounded below when labor
income has a uniform positive lower bound, as stated in
Footnote~\ref{fn:utility}.

\subsection{Bounded below utilities}\label{sss:bdd_below}
Suppose $u$ is bounded below on $\RR_+$, say
$\inf u=:m>-\infty$, and replace $u$ by $\tilde u:=u-m\ge0$.
By cash invariance of the entropic certainty equivalent
(Lemma~\ref{lm:entropic}\eqref{i:drift_inv}), if $\tilde v^*$ denotes the 
recursive value under $\tilde u$, then the original value is
$v^*=\tilde v^*+m/(1-\beta)$. The two problems have the same
optimal policy, and the bounds in \eqref{eq:u_kappa_C_G} remain valid
with $\tilde u(1)$ and $\tilde u'(1)=u'(1)$ in place of $u(1)$
and $u'(1)$. Hence all results extend verbatim.

\subsection{Utilities unbounded below}\label{sss:unbdd_bwlow}
Suppose $u$ satisfies the maintained assumptions on $(0,\infty)$
and $u(c)\to-\infty$ as $c\downarrow0$, and set $u(0):=-\infty$.
Assume that the continuous wage schedule is positive on $\CC$, and
define
\begin{equation*}
	\underline a:=\uz\min_{R\in\CC}w(R)>0
	\qquad \text{and} \qquad 
	\XX^0_\CC:=[\underline a,\infty)\times\ZZ.
\end{equation*}
The set $\XX^0_\CC$ is invariant because next period wealth is at least 
$w(R)\uz\ge\underline a$ under every feasible action. 
On $\XX^0_\CC\times\TT$, retain the candidate space of
Section~\ref{ss:opt_setup}, dropping only the nonnegativity requirement. 
This space remains complete. 

At $c=a$, the continuation wealth is $w(R) \hat z$ with $\hat z\in\ZZ$, 
a finite set contained in $[\ua,\bar a]$ for 
$\bar a:=\bar z \max_{R\in\CC}w(R)$. Since $\|v\|_{\kappa_\CC}<\infty$ 
gives $v(w(R)\hat z,\hat z)\ge-\|v\|_{\kappa_\CC}\kappa_\CC(\bar a)=:m_v$ 
for all $\hat z$, choosing $c=a$ yields
\begin{equation*}
	(T_\TT v)_\theta(a,z)\ge u(\underline a)+
	\beta \rho_{\gamma,z} \big(v(w(R) \hat z, \hat z)\big)
	\ge u(\underline a)+\beta m_v,
\end{equation*}
a uniform lower bound independent of $(a,z)$. Moreover, the linear 
growth bound on $u$ from \eqref{eq:u_kappa_C_G} and the definition of 
$\|\cdot\|_{\kappa_\CC}$ provide the upper bound, exactly as in the 
bounded below case. Since $u(c)\to-\infty$ as $c\downarrow0$ while
the continuation term is bounded on compact subsets of feasible
consumption, the maximand tends to $-\infty$ as $c\downarrow0$ and is
finite at $c=a$; hence the maximizer lies away from zero, uniformly on
compact sets of $(a,z)$.
The monotonicity, concavity, and continuity arguments of the 
finite-horizon iteration are unchanged and pass to the limit, so $v^*$ 
retains these properties. The contraction estimate in 
Lemma~\ref{lm:contr_C} is unchanged, yielding a unique fixed point $v^*$ 
and a continuous greedy policy $c^*>0$.

For verification, set $r:=\beta M_\CC<1$. By concavity of $u$, there exists
$B\ge0$ such that $u(c)\le B\kappa_\CC(a)$ whenever $0\le c\le a$ and
$a\ge\underline a$. For policy evaluation, extend the certainty equivalent 
to $[-\infty,\infty)$-valued successor payoffs, ignoring successors of 
zero transition probability; monotonicity and cash invariance remain valid.
Accordingly, the operators $L^\theta_{\pi_t}$ are well defined on
extended-real history functions with a weighted upper bound. By
\eqref{eq:pointwise_drift_C_G}, for every $d\ge0$,
\begin{equation*}
	f(h_{t+1})\le g(h_{t+1})+d\kappa_\CC(a_{t+1})
	\ \Longrightarrow\
	(L^\theta_{\pi_t}f)(h_t)\le (L^\theta_{\pi_t}g)(h_t)+rd\kappa_\CC(a_t).
\end{equation*}
Since $L^\theta_{\pi_{n+1}}0=u(\pi_{n+1})\le B\kappa_\CC$, iterating
this estimate backward gives
\begin{equation*}
	J_{n+1,\theta}(a,z,\pi)
	\le J_{n,\theta}(a,z,\pi)+Br^{n+1}\kappa_\CC(a).
\end{equation*}
Thus, the sequence
$J_{n,\theta}(a,z,\pi) + Br^{n+1} \kappa_\CC(a)/(1-r)$
is decreasing, hence has a limit in $[-\infty,\infty)$; since the
correction vanishes, $J_\theta(a,z,\pi):=\lim_n J_{n,\theta}(a,z,\pi)$
exists in $[-\infty,\infty)$. To prove optimality, let
$\varepsilon_n(a):=r^{n+1}\lVert v^*\rVert_{\kappa_\CC}\kappa_\CC(a)$.
Since $|v_\theta^*|\le\lVert v^*\rVert_{\kappa_\CC}\kappa_\CC$, the
terminal bound $0\le v_\theta^*+\lVert v^*\rVert_{\kappa_\CC}\kappa_\CC$
holds; applying the estimate with $f=0$, $g=v_\theta^*$, and
$d=\lVert v^*\rVert_{\kappa_\CC}$ backward over $n+1$ periods, and then
using the Bellman inequality $L^\theta_{\pi_t}v_\theta^*\le v_\theta^*$, 
we obtain 
\begin{equation*}
	J_{n,\theta}(a,z,\pi)
	\le\bigl[(L^\theta_{\pi_0}\circ\cdots\circ
	L^\theta_{\pi_n})v_\theta^*\bigr](a,z)+\varepsilon_n(a)
	\le v_\theta^*(a,z)+\varepsilon_n(a).
\end{equation*}
For the greedy policy $c_\theta^*$ every Bellman inequality is an
equality, and the same terminal-value estimate in both directions gives
\begin{equation*}
	v_\theta^*(a,z)-\varepsilon_n(a)
	\le J_{n,\theta}(a,z,c_\theta^*)
	\le v_\theta^*(a,z)+\varepsilon_n(a).
\end{equation*}
Letting $n\to\infty$ yields
$v_\theta^*(a,z)=\sup_{\pi\in\Pi}J_\theta(a,z,\pi)
=J_\theta(a,z,c_\theta^*).
$
The monotonicity of optimal consumption and saving does not depend on the
sign of utility and remains valid. Thus Theorem~\ref{t:opt_general}
extends to $\XX^0_\CC$, with real-valued candidate functions.

\section{Properties of the Production Side}
\label{s:pf_firm_opt}

This appendix proves the firm's capital demand and wage results stated in Section~\ref{ss:firm}.

Under Assumption~\ref{a:prod_func}, for each
$R\in(\underline R,\bar R)$, \eqref{eq:pro_opt} admits a unique 
positive solution $K(R)=F_1^{-1}(R-1+\delta,1)$, continuous and 
strictly decreasing in $R$. Since $R-1+\delta\to\ell$ as 
$R\downarrow\underline R$ while $\bar R-1+\delta > \delta >\ell$, 
\eqref{eq:K_lim} follows immediately from Assumption~\ref{a:prod_func}.

The first-order condition \eqref{eq:pro_opt}  defines $w(R):=F_2(K(R),1)$, 
which is continuous by Assumption~\ref{a:prod_func} and the continuity of 
$K(R)$. By constant returns to scale of $F$, $F(tK,t)=tF(K,1)$. 
Differentiating with respect to $t$ gives $KF_1(tK,t)+F_2(tK,t)=F(K,1)$, 
hence $F(K,1)=KF_1(K,1)+F_2(K,1)$. Consequently,
\begin{equation}\label{eq:wage_euler_identity}
	w(R)/K(R) = F(K(R),1)/K(R) - F_1(K(R),1).
\end{equation}
Fix $K_0>0$. For $K>K_0$, the fundamental
theorem of calculus gives
\begin{equation*}
	\frac{F(K,1)}{K} = \frac{F(K_0,1)}{K} 
	+ \frac{1}{K} \int_{K_0}^{K}F_1(x,1) \diff x.
\end{equation*}
Since $F_1(x,1)\to\ell$ as $x\to\infty$, the continuous Cesàro mean 
theorem implies that 
\begin{equation*}
	\lim_{K\to\infty} \frac{1}{K-K_0}\int_{K_0}^{K}F_1(x,1) \diff x =\ell
	\quad \Longrightarrow \quad 
	\lim_{K\to\infty} \frac{1}{K}\int_{K_0}^{K}F_1(x,1) \diff x =\ell.
\end{equation*}
So $\lim_{K\to\infty} F(K,1) / K=\ell.$
Combined with Assumption~\ref{a:prod_func}, \eqref{eq:pro_opt} and 
\eqref{eq:wage_euler_identity}, 
\begin{equation*}
	w(R)/K(R) \to \ell - (\underline R-1+\delta) = 0
	\qquad \text{as }\; R\downarrow\underline R.
\end{equation*}
At the upper endpoint, \eqref{eq:K_lim} and the continuity of
$F_2$ imply
\begin{equation*}\label{eq:wage_upper_limit}
	w(R)\to w(\bar R)=F_2(K(\bar R),1) \in (0,\infty)
	\qquad \text{as }\;R\uparrow\bar R.
\end{equation*}
All claims from Section~\ref{ss:firm} are therefore established.

\end{document}